Camelia Oprean-Stan

# A Novel Approach to Assessing Corporate Sustainable Economic Value

## 1. Introduction

The core objective of any manager's activity is to maximize the value of the organization s/he leads, as value creation is essential to the operations and success of the business. Companies need customers and employees to willingly create new values for them. They also need innovative ways to increase profits through cost-cutting strategies. Ultimately, there is no limit to what companies can do when thinking about how they can have a positive impact on society through their creations – they just need to think outside the box.

An important way to achieve this goal is to efficiently combine available – tangible and intangible – resources in the production and marketing of goods or services with the aim of gaining a significant share of the market in order to increase profitability and increase the value of the organization, implicitly of the owners' wealth. Companies create new value through innovation efforts such as inventing new products, expanding markets, or building better organizational structures. They also create new value by contracting – that is, by cutting costs to increase profits. Other ways to create value include protecting intellectual property and recruiting top talent. Ultimately, companies are responsible for creating value for themselves and for the people who work for them.

According to the perspective supported by endogenous growth theory and other approaches, traditional production factors (natural resources, labor and capital) have gradually diminished their importance. At the same time, the importance of intangible assets – such as information, knowledge and creativity – has increased. Knowledge becomes a new resource, a true capital of modern economies.

There are many ways in which companies can create value beyond what their customers and employees create. For example, companies can adopt sustainability practices, such as using environmentally friendly materials or manufacturing processes that use minimal energy (Oprean-Stan et al., 2020). In addition, companies can raise awareness of societal issues such as poverty through philanthropy or charitable causes. Essentially, there is no limit to what a company can do when it gives back to the people it serves.

Although the process of knowing, understanding, determining, and measuring economic value in the context of sustainable development, while creating profit, are the main objectives of modern business, at present the determinants of corporate value are purely financial and do not take into account non-financial factors, such as those related to sustainability orientation.

The study questions are as follows: How is economic value reinterpreted in the context of sustainable development? Alternatively, what is sustainable business value and what factors contribute to it? This study's major hypotheses are that corporate economic value in the context of sustainable development is influenced by both financial and non-financial factors. The goal of this study is to propose a new concept, Sustainable Economic Value, to define it logically, and to build a simplified model for its evaluation.

## 2. General Understandings of Value in a Company: From Shareholder Value to Shared Value

Table 1 presents the evolution of the major value theories over the last decades, showing the shift from the traditional financial system, where the goal is to maximize shareholder value by seeking the optimal combination of financial return and risk, to the financial system based on stakeholder value theory and then to the financial system based on shared value theory. The last stage concerns a financial system capable of producing value for the common good, first and foremost in response to societal and environmental challenges. Importantly, as each step proceeds, the time horizon broadens from short to long term.

**Table 1.** Main theories of corporate value

| Type of Value | Value Created | Optimization Method | Time Horizon |
|---|---|---|---|
| **Traditional value** | Present shareholder value | Maximizing net profits or cash flows | Short term |
| **Contemporary value** | Future shareholder value | Maximizing shareholder value:<br>• Book value<br>• Market value<br>• Financial value | Short and medium term |
| **Stakeholder value** | Shared value | Maximizing shareholder value:<br>• Financial value<br>• Social value<br>• Environmental value | Long term |

**Source:** in author's view

The traditional shareholder value strategy, promoted by Milton Friedman in the 1970s (Friedman 1970), has placed a strong emphasis on maximizing net profits or cash flows in order to increase shareholder present value. This strategy takes a short-term time perspective into account. However, the singular focus on shareholder value has come under intense criticism (Tirole 2001; Aglietta & Reberioux, 2006), notably in the wake of the financial crisis of the late 2000s. Although a corporation's owners may financially gain from a focus on shareholder value, it does not give a clear indicator of social aspects like employment, environmental problems, or ethical business practices. A managerial choice may increase shareholder wealth at the expense of other people's welfare.

The contemporary approach to value aims to create future shareholder value by maximizing the value of the firm measured by: book value, market value, and financial value. In this value framework, objectives regarding the social and environmental effects of the firm are not taken into account or are incorporated to a rather small extent, as the creation of financial value is put first.

The most prevalent foundation for alternative frameworks is the theory of stakeholder value, which emphasizes that a company's value is determined by a mix of its financial, social, and environmental performance and has a long-term meaning. Redefining the corporation's mission to include developing viable solutions to social challenges is the goal of creating shared value (CSV) (Porter & Kramer, 2011; Menghwar & Daood, 2021). The idea behind generating shared value is that a company's competitiveness and the wellbeing of the communities in which it operates are interdependent.

Thus, by understanding the drivers of value creation, companies will be able to create better products, provide better service to their customers, have engaged employees, and help create shareholder value, in other words create sustainable value.

## 3. The Rationale behind the Concept of Sustainable Economic Value

In this study, a new type of value is proposed: Sustainable Economic Value, which is conceptually formalized below. This approach to value maximizes the total value of a firm in the long run, taking into account moral implications. The concept of sustainable finance thus underlines the importance of the behavioral premises of modern economic agents and explicitly expands the company's objectives. In addition to the necessary risk/reward objective, based on the performance generated by all available capital at the firm level (tangible and intangible), a company should, in its strategy, aim for future environmental and social requirements to be at the core of the company's business. Optimizing the firm's orientation toward social and environmental aspects will also lead to an increase in multi-capital performance, as this approach is long term. In a sense it represents a reintegration of social values into economic theory.

In this chapter, the concept of Sustainable Economic Value is defined from a logical perspective, by identifying logical attributes, hereafter called sufficiency predicates, which need to be fully verified. Thus, for the definition of the concept of Sustainable Economic Value, the following steps are considered:

– punctual identification of sufficiency predicates;
– logical analysis of the concept by examining the following conditions of the identified sufficiency predicates: independence – no sufficiency predicate is a logical result of another sufficiency predicate; consistency – no sufficiency predicate is contradictory to another sufficiency predicate; completeness – refers to the concurrent verification of sufficiency predicates, which provides the intended qualification;
– defining the concept based on the identified sufficiency predicates.

We will use the following notations:

– P – the identified sufficiency predicates;
– M – the sufficiency predicates, where: $M = \{P_1, P_2, \ldots, P_n\}$, n – natural number.

Consider the following sufficiency predicates for the concept of Sustainable Economic Value:

– $P_1$. It is a concept that can be used and evaluated at corporate level;
– $P_2$. Determined by financial performance;
– $P_3$. In direct dependence with the share of intangible resources in total resources used by companies;
– $P_4$. It is maximized by the company's focus on sustainability;
– $P_5$. Generates the firm's ability to be resilient and survive in the long term.

## 3.1  Logical Analysis of Sufficiency Predicates

### 3.1.1  Independence Analysis

The identified sufficiency predicates were compared two by two and the number of analyzed cases was $C_n^2$, according to the following formula:

$$C_n^2 = \frac{n!}{2! * (n-2)!}, n \geq 2 \tag{1}$$

$C_5^2$ possible cases were analyzed, which means 10 cases. None of the five identified sufficiency predicates is the logical result of another.

### 3.1.2  Consistency Analysis

The identified sufficiency predicates were compared two by two. As in the case of the independence condition, $C_n^2$ cases were subjected to analysis. None of the sufficiency predicates identified is contradictory to another predicate, especially since we are referring to the sustainable finance framework.

### 3.1.3  Completeness Analysis

With regard to the completeness condition, the simultaneous relevance of the identified sufficiency predicates to the concept of Sustainable Economic Value is observed.

Thus, the sufficiency predicates identified above meet the conditions of independence, consistency, and completeness.

In light of the foregoing, Sustainable Economic Value can be defined as: *The overall value created by a company when it achieves corporate performance, i.e., financial performance correlated with the share of intangible resources in total resources used, but also the performance generated by the company's orientation toward sustainability, which increases the company's ability to be resilient and survive in the long term.*

From a formal point of view, the following logical expression of Sustainable Economic Value can be written:

$$M(SEV) = \{P1, P2, P3, P4, P4\} \qquad (2)$$

where M(SEV) is the crowd of sufficiency predicates of Sustainable Economic Value.

## 4.  Proposal of a Mathematical Model to Assess Sustainable Economic Value

### 4.1.  Substantiation of Variables

Alongside measuring Sustainable Economic Value, identifying the factors influencing this value is equally necessary. The construction of the system of indicators to measure the value created at firm level (Sustainable Economic Value) divided according to the three dimensions (financial-intangible-sustainable) is described below (and can be found in Table 2).

For the purpose of this research, an organization's Sustainable Economic Value represents a dynamic system influenced at each moment by several factors, such as:

– financial performance as measured by: Enterprise Value, EBITDA, Net income, PE ratio, and long-term returns score. These indicators have the ability to measure the organizational performance most accurately;
– performance generated by human capital. Human capital is one of the most important parts of intangible assets (Oprean-Stan, Stan, & Brătian, 2020), namely those resources created by employees. In this context, it is considered critical to assess the effectiveness of the staff and management by measuring the outcomes. We have therefore used three indicators from this perspective: productivity, one of the most commonly used indicators in analyses, number of employees and ROA, and ROE perceived as indicators of management performance.
– performance generated by the company's orientation toward sustainability, assessed through pillars of Environmental, Social and Governance (ESG): governance score, environment score, social score.

**Table 2.** Indicators proposed for analysis

| No. | Proposed Indicators | Description | Category |
|---|---|---|---|
| **FINANCIAL PERFORMANCE** | | | |
| 1 | **EV** | Alternative to conventional market capitalization, enterprise value (EV) measures the market value of the complete company. Enterprise value encompasses all expenses associated with a company, including debt and equity. *Enterprise Value = Market Capitalization + Current Portion of Long Term Debt + Non-Current Portion of Long Term Debt + Book Value of Preferred Stock + Book Value of Minority Interest – Cash and Short Term Investments* | Profitability Market Value |
| 2 | **EBITDA** | EBITDA is frequently used by investors to conduct discounted cash flow analyses or to evaluate a company's performance independently of its capital structure (debt vs. equity financing structure). *EBITDA = Net Income + Interest Expense + Tax Expense + Depreciation Expense + Amortization Expense* | Profitability |
| 3 | **Net_income** | Traditional quantitative indicator of profitability | Profitability |

**Table 2.**   Continued

| No. | Proposed Indicators | Description | Category |
|---|---|---|---|
| 4 | **PE ratio** | The price to earnings ratio is a current stock price over its earnings per share. *PE Ratio = price / Earnings per share* | Profitability Market Value |
| 5 | **LTR_score** | Long-term returns score is based on the following indicators: <br> – earnings quality region rank, which shows the degree to which past earnings are secure and likely to persist; <br> – long-term orientation score, which shows the company's ability to manage its long-term financial sustainability; <br> – credit combined region rank, which measures a company's credit risk. | Profitability |
| PERFORMANCE GENERATED BY HUMAN CAPITAL | | | |
| 6 | **Number of employees** | Quantitative indicator of human resources | Human Capital |
| 7 | **Productivity** | Indicator of the efficiency of human resources, calculated as the ratio of sales to the number of employees | Human Capital |
| 8 | **ROA** | Return on Assets (ROA) is a measure of how efficiently the company is using all stakeholders' assets to earn returns. *ROA = (Net Profit) / (Total Assets) * 100* | Management effectiveness |
| 9 | **ROE** | Return on equity (ROE) demonstrates a company's ability to generate profits from shareholders' equity. *ROE = (Net Profit) / Equity * 100* | |
| PERFORMANCE GENERATED BY SUSTAINABILITY ORIENTATION | | | |
| 10 | **GOV_score** | Governance pillar score based on: <br> – shareholders Score; <br> – management Score; <br> – CSR Strategy Score. | Sustainability |
| 11 | **ENV_score** | Environment pillar score based on: <br> – resource use Score; <br> – emissions Score management; <br> – environmental innovation Score. | Sustainability |
| 12 | **SOC_score** | Social pillar score based on: <br> – workforce Score; <br> – community Score; <br> – human rights Score; <br> – product responsibility Score. | Sustainability |

**Source:** In author's view

## 4.2. Methodological Aspects of Research and Findings

There are a large number of factors or variables with different influences on Sustainable Economic Value of an organization, so it is useful to find those factors/variables with significant determinant role in optimizing their level. Principal Component Analysis, also called *Hotteling Transformation* or *Karhunen-Loeve Transformation*, actually uses a mathematical concept, namely, "own vector" (solution of a matrix equation of form $(A - \lambda I) x = 0, x \neq 0$), and it aims to reduce the number of variables initially used by taking into account a small number of representative variables, i.e.., reducing large data volume to a structure that retains as much common variability as possible (Bucur, 2015).

For the analysis, the values for the 12 indicators of Sustainable Economic Value were extracted for the companies included in the calculation of the STOXX Europe 50 Index, Europe's leading blue chip index, which offers a measure of Europe's supersector members. It represents around 50 % of the European stock market capitalization. The main reason for choosing this index is the need to improve the generalization of the results obtained, as this study covers a wide range of economic sectors. Data are for 2021 and were collected from the Refinitiv Eikon Thompson Reuters database.

This study uses IBM SPSS (version 25) software to analyze the main components influencing Sustainable Economic Value, by which we shall track the key, representative factors that can be constituted in a decision maker to prioritize specific measures and actions to ensure an optimal level of Sustainable Economic Value. The solution to the raised problem consists in identifying the orthogonal factors out of the available ones and subjected to the factorial analysis, following several steps outlined below.

By accessing Extraction option, in Method list, we have selected the factorial analysis method, Principal components. The following indicators were obtained: a Kaiser-Meyer-Olkin Measure of Sampling Adequacy of 0.629 and Bartlett's Test of Sphericity Approx. Chi-Square (df=66) was 213.205 (sig. .000). The results of these tests indicate a meaningful analysis. The same results were obtained by performing the communalities analysis, where all values were greater than 0.5, showing that the test is valid.

The criterion of Eigenvalues larger than 1 (Aleshinloye et al., 2021; Field, 2013) was used to determine the number of factors to be retained and to "identify a latent factor structure" (Watkins, 2018). In the case of this study, five components contributed to the structure of the latent variables and had Eigenvalues larger than 1 (Table 3). Furthermore, 75.772 % of the variation was explained by the five components. In accordance with suggestions from Hair et al. (2010), the threshold of the variance explained at 70 % was thus reached. The lowest result was 0.656 (Table 4), indicating that all factor loadings were greater than the suggested limit of 0.6. Accordingly, the exploratory factor analysis is regarded pertinent based on input from earlier research on this subject (Field, 2013; Goretzko et al., 2021; Vinerean et al., 2021).

**Table 3.** Total variance explained table

| Compo-nent | Initial Eigenvalues | | | Extraction Sums of Squared Loadings | | | Rotation Sums of Squared Loadings | | |
|---|---|---|---|---|---|---|---|---|---|
| | Total | % of Variance | Cumulative % | Total | % of Variance | Cumulative % | Total | % of Variance | Cumulative % |
| 1 | 3.127 | 26.056 | 26.056 | 3.127 | 26.056 | 26.056 | 2.020 | 16.832 | 16.832 |
| 2 | 2.348 | 19.569 | 45.625 | 2.348 | 19.569 | 45.625 | 1.984 | 16.534 | 33.366 |
| 3 | 1.389 | 11.572 | 57.198 | 1.389 | 11.572 | 57.198 | 1.886 | 15.721 | 49.087 |
| 4 | 1.131 | 9.421 | 66.619 | 1.131 | 9.421 | 66.619 | 1.868 | 15.566 | 64.653 |
| 5 | 1.098 | 9.154 | 75.772 | 1.098 | 9.154 | 75.772 | 1.334 | 11.119 | 75.772 |
| 6 | .837 | 6.978 | 82.751 | | | | | | |
| 7 | .708 | 5.897 | 88.648 | | | | | | |
| 8 | .491 | 4.093 | 92.740 | | | | | | |
| 9 | .271 | 2.255 | 94.995 | | | | | | |
| 10 | .259 | 2.162 | 97.158 | | | | | | |
| 11 | .208 | 1.734 | 98.891 | | | | | | |
| 12 | .133 | 1.109 | 100.000 | | | | | | |

Extraction Method: Principal Component Analysis
Source: Displayed by SPSS Software

**Table 4.** Solution after factor rotation

| Rotated Component Matrix[a] | | | | | |
|---|---|---|---|---|---|
| | Component | | | | |
| | 1 | 2 | 3 | 4 | 5 |
| SOC_score | **.863** | | −.132 | .113 | −.127 |
| ENV_score | **.827** | | | .131 | |
| ROA | −.172 | **.909** | .106 | | |
| ROE | | **.893** | .158 | | |
| LTR_score | | .155 | **.833** | .134 | .233 |
| EP | −.304 | .199 | **.709** | | |
| Net_income | .229 | .432 | −.549 | .471 | .323 |
| EV | | .241 | .223 | **.719** | |
| EBITDA | .351 | −.194 | −.325 | **.656** | .419 |
| No_employees | .228 | | −.249 | .594 | −.258 |
| GOV_score | .452 | | −.263 | −.532 | .259 |
| Productivity | −.178 | | .212 | | **.908** |

Extraction Method: Principal Component Analysis
Rotation Method: Varimax with Kaiser Normalization
[a] Rotation converged in 9 iterations
**Source:** Displayed by SPSS Software

In summary, by using the factorial method, respectively, the analysis of the 12 indicators of Sustainable Economic Value level processed by the SPSS software, a structure of five orthogonal factors, was obtained as a final solution which has accumulated the greatest amount of variability common to Sustainable Economic Value (presented in Table 5).

**Table 5.** Presentation of the final solution, consisting of five-basic factors

| Factor Number | Constituent Variables | The Percentage of Variation that Can Be Explained | Definition |
|---|---|---|---|
| 1 | SOC_score (0.863), ENV_score (0.827) | 26 % | *Sustainability performance* |
| 2 | ROA (0.909), ROE (0.893) | 19 % | *Management effectiveness* |
| 3 | LTR_score (0.833), PE (0.709) | 11 % | *Future financial performance* |
| 4 | EV (0.719), EBITDA (0.656) | 9 % | *Current financial performance* |
| 5 | productivity (0.908) | 9 % | *Human resources performance* |

**Source:** in author's evaluations

## 5. Conclusions

The proposal of a simplified model for the evaluation of Sustainable Economic Value by the factorial method refers to a final solution consisting of five-basic factors: sustainability performance, management effectiveness, future financial performance, present financial performance, human resources performance. We mark these factors with $f_1$, $f_2$, $f_3$, $f_4$, $f_5$. This sequence is a new indicator of Sustainable Economic Value, a key, representative indicator, which incorporates the greatest amount of variability common to other available factors in the factorial method.

We mark the new indicator with $SEV_t$. Then, mathematically, it can be written as a vector of $R^5$, as follows:

$$SEV_t = (f_1, f_2, f_3, f_4, f_5) \tag{3}$$

This type of factorial analysis, Principal Component Analysis, provides useful and practical information to researchers; it provides statisticians and all managers the possibility to follow the ascending or descending evolution of Sustainable Economic Value and to take the appropriate corrective measures. In today's ever-changing world, the challenge for business is to produce and maximize long-term value by developing effective strategies, managing risks effectively and seizing opportunities. For sustainable growth, the private sector needs clear global rules, effective laws, and government support.

## References

Aglietta, M., & Reberioux, A. (2006). Corporate Governance Adrift. A Critique of Shareholder Value. *Journal of Economics*, *88*(3), 307–311. https://doi.org/10.1007/s00712-006-0204-8.

Aleshinloye, K., Woosnam, K., Tasci, A., & Ramkissoon, H. (2021). Antecedents and Outcomes of Resident Empowerment through Tourism. *Journal of Travel Research*, *61*(3), 1–18. https://doi.org/10.1177/0047287521990437

Bucur, A. (2015). *Contributions to the Scientific Approach to Quality and Quality Management Through Modelling and Simulation* (in Romanian). Sibiu: University of Sibiu Publishing House.

Field, A. (2013). *Discovering Statistics Using IBM SPSS Statistics*, 4th ed. London, UK: SAGE Publications.

Friedman, M. (1970). A Friedman Doctrine: The Social Responsibility of Business is to Increase Its Profits. *New York Times Magazine, 33*, 122–124.

Goretzko, D., Pham, T. T. H., & Bühner, M. (2021). Exploratory Factor Analysis: Current Use, Methodological Developments, and Recommendations for Good Practice. *Current Psychology*, *40*, 3510–3521.

Hair, J. F., Black, W. C., Babin, B. J., Anderson, R. E., & Tatham, R. L. (2010). *Multivariate Data Analysis*, 7th ed. NJ, USA: Prentice Hall, Upper Saddle River.

Menghwar P. S., & Daood, A. (2021). Creating Shared Value: A Systematic Review, Synthesis and Integrative Perspective. *International Journal of Management Reviews, 23*(4). https://doi.org/10.1111/ijmr.12252.

Oprean-Stan, C., Oncioiu, I., Iuga, I. C., & Stan, S. (2020). Impact of Sustainability Reporting and Inadequate Management of ESG Factors on Corporate Performance and Sustainable Growth. *Sustainability*, *12*(20), 8536.

Oprean-Stan, C., Stan, S., & Brătian, V. (2020). Corporate Sustainability and Intangible Resources Binomial: New Proposal on Intangible Resources Recognition and Evaluation. *Sustainability*, *12*, 4150. https://doi.org/10.3390/su12104150.

Porter, M. E., & Kramer, M. R. (2011). Creating Shared Values: How to Reinvent Capitalism and Unleash a Wave of Innovation and Growth. *Harvard Business Review, 1*, 1–5.

Tirole, J. (2001). Corporate Governance. *Econometrica*, *69(1)*, 1–35.

Vinerean, S., Opreana, A., Tileagă, C., & Popșa, R. E. (2021). The Impact of COVID-19 Pandemic on Residents' Support for Sustainable Tourism Development. *Sustainability*, *13*, 12541. https://doi.org/10.3390/su132212541.

Watkins, M. W. (2018). Exploratory Factor Analysis: A Guide to Best Practice. *Journal of Black Psychology*, *44*, 219–246.

Juliya S. Tsertseil, Tatiana G. Bondarenko,
Nadezhda M. Tchuykova, and Aleksei I. Bolvachev

# Prospects for the Development of Small and Medium-Sized Businesses: The Case of Russia

## 1. Introduction

According to Federal Law No. 209-FZ of July 24, 2007 (amended on July 2, 2021) "On the Development of Small and Medium Enterprises in the Russian Federation", all SMEs must register in accordance with the legislation of the Russian Federation and meet the conditions presented below.

There are three categories of SMEs:

– micro-enterprise – up to 15 employees,
– small enterprise – up to 100 employees,
– medium enterprise – from 101 to 250 employees.

The term "SMEs" also covers enterprises, unions, partnerships, corporations, cooperatives, farms, and individual entrepreneurs. The Federal State Statistics Service published report on SMEs contribution to the federal 2017–2019 GDP. The numbers stay on the same level: 22 %, 20.4 % and 20.8 %, respectively. In 2020 Russian Federation national GDP has 22.5 % SMEs share, the Central Federal District – 23.2 %, the Northwestern Federal District – 24.7 %, the Southern Federal District – 29.1 %, the North Caucasian Federal District – 31, 4 %, Volga Federal District – 25.3 %, Ural Federal District – 13.6 %, Siberian Federal District – 21.2 %, Far Eastern Federal District – 19.8 %.

Russian Federation actively implements multiple national and regional projects to support SMEs. This chapter examines the dynamic development of SMEs within "Small and Medium Enterprises and Support for Individual Entrepreneurial Initiative" (2019 – 2020). The main goal of the project is to expand SMEs access to all types of financing, including concessional financing. The project has an ambitious target to increase the number of the SMEs' employees up to 25 million people including individual entrepreneurs.

According to the website of the Federal State Statistics Service, the share of SMEs in the GDP of the Russian Federation for the period from 2017 to 2020 were: 22.0, 20.4, 20.7, and 20.3 %, respectively, which is shown in Figure 1.

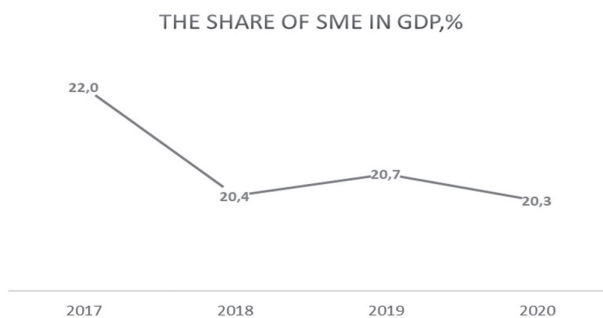

**Figure 1.** The share of small and medium-sized businesses in the gross domestic product, %

(Compiled according to the official website of the Federal State Statistics Service of the Russian Federation dated December 29, 2021)

At present, the key tool for the development of the national economy is the national projects implemented in the country based on Decree of the President of the Russian Federation No. 13 of 16.01.2017. On Approval of the Fundamentals of the State Policy of Regional Development of the Russian Federation for the Period Up to 2025. (2017), formed in each direction of the development of society. In particular, the presented research will examine the dynamics of the development of small and medium-sized enterprises (SMEs) within the framework of the Passport of the National Project "Small and Medium-Sized Entrepreneurship and Support for Individual Entrepreneurial Initiative". (2020).

This project aims to expand the access of small and medium-sized enterprises (SMEs) to financial resources, including concessional financing.

It provides for the achievement of the projected target: an increase in the number of employees in the field of small and medium-sized businesses, including individual entrepreneurs, to 25 million people.

## 2. Literature Review

The self-evident significance of the topic explains why so many researchers study the role of SMEs in the development of the national economy from practical and theoretical points of views. The literature review provides a comprehensive analytical summary of publications on the topics of SMEs' advantages, disadvantages, trends, classifications, characteristics, and comparisons in developed and developing economies.

In their work, Keskin, Sentürk, Sungur, and Kiris (2010) describe and compare large companies and SMEs by marketing, management, international cooperation, application for R&D patents, and additional costs.

In their study, Mari, Nebhwani, and Sohag (2011) describe the role of SMEs as entities contributing to competitive market. The authors identify the following SME support measures: tax breaks, state support, access to financing including loans with preferential conditions. In their study, Herr and Nettekoven (2017) assess the influence of the SMEs in the economic development of the different countries. They focus on Germany and discuss the employment in the SMEs. In her work, Cibela Neagu (2016) studies the SMEs and R&D investments. She identifies bank loans (80 %) as the main source for R&D financing and states that state authorities have little R&D financing for SMEs.

In his research, Hobohm (2001) explores how SME clusters can promote economic growth, build cooperation, and create new business opportunities. A brief literature review highlights the importance of well-thought and well-executed state policy for SME development: Bouri et al. (2011), Herr and Nettekoven (2017) contribute to the theory and practice of SMEs support state policy's design and execution with emphasis on financing.

In 2011, the collective of authors (Bouri et al) published the Report on Support to SMEs in Developing Countries Through Financial Intermediaries (2011) highlighting SMEs share in the GDP and the innovations.

The studies on SMEs in different economic sectors are worth attention. Here authors agree that SMEs activities depend on large companies in the real sector of economy. Therefore, Bartlett (2001), McIntyre (2001), Tewar et.al (2013), Jamieson et al.(2012), M. Nureldin Hussain (2000), Pech and Vrchota (2020) pay close attention to the cooperation between SMEs and large businesses.

Bartlett (2001) identifies SMEs' key role for developing economies. He pays attention to the analysis of the impediments for SMEs, including financial and institutional barriers. McIntyre (2001) is interested in new forms of SMEs. In his paper, the author reviews the dynamics of the SMEs in connection to the share in total employment for the period from 1937 to 1988. He indicates that during that period SME employment share reached 88 % in Portugal and 68 % in Southern Italy.

Dougal Jamieson et al. (2012) study the connection and influence of large business on the SMEs. They highlight positive indicators such as contractual relationship and cooperation. Contacts with large business support growth, promote capital investment, diversify activities, and build and improve reputations for SMEs. Similar conclusions were reached by M. Nureldin Hussain (2000) in his studies on emerging African markets.

In their article, Tewar et al. (2013) describe the essential characteristics of SMEs and emphasize their different contribution to industrial growth and direct economic growth. The authors also note that most of the SMEs shut down after seven years. Only 20 % SMEs grow at normal rate (15 %) and high growth rate (5 %).

In their study Pech and Vrchota (2020) examine SMEs potential to implement technological processes and digital transformation. They conclude that although SMEs have financial and production opportunities, they still need state support. The authors use statistical research methods to study SME clusters in different economic sectors.

Employment is often used for the evaluation of SMEs' role in economy. Kok et al. (2011) published a report on European SMEs in 2010/2011. They noted that in some European countries, the employment growth rate was slightly higher and amounted to 1.2 % compared to 0.8 % of the EU growth rate. They assessed it as a positive trend.

SMEs are often seen as a driver for the regional economic strategies; that is why, authors often focus on recommendations, legislative problems, and impediments for SMEs. Many authors focus on the regional growth due to SMEs clusters. The founder of the clustering theory M. E. Porter published extensively on the issue (1998, 2000) as well as such authors as Ketels (2017), Jon Swords (2013), Enright (2003), Barkley and Henry (2005), Asheim, Cooke, and Martin (2009), Tsertseil and Kookueva (2017), Tsertseil (2015) and others.

Based on the presented literature reviews, one can make a conclusion that SMEs are major contributors to the national economy. Our paper offers analysis of trends and development of SMEs in the federal districts of the Russian Federation in the first half of 2021 with focus on investments per capita and R&D expenditures to assess the long-term prospects.

## 3. Methodology

### 3.1. Theoretical and Methodological Approaches

In the first section, the authors apply general scientific methods of learning in relation to the study of the nature and content of the process of implementing a national project in the field of entrepreneurship, tools for implementing state policy, and factors of the external and internal environment of the functioning of SMEs.

The methodological basis of the research is based on the principles of dialectics, induction, and deduction, which allow us to identify the main characteristics of phenomena and processes in their interrelation with respect to SMEs, to determine the key trends of their formation and development in the global economic environment.

At this stage the results of the study are expected to systematize existing research on the problems of the formation and implementation of the state sectoral policy in the institutional economy in the field of entrepreneurship, clarification on the formation and implementation of tools for the implementation of the state sectoral policy in the field of entrepreneurship, taking into account the institutional changes that lead to an increase in the competitiveness of both industries and the national economy as a whole.

### 3.2. Empirical Approaches

The comparative method is supposed to be implemented using a group of analytical methods of statistics and economics, since the examination of the quantitative trends in the implementation and development of economic processes in small and medium-sized businesses within the framework of the implemented state industry policy in Russia provides for the processing of extensive empirical material. The results obtained will form the basis for a comparative analysis of the current situation of the stated issues:

The grouping method, being a part of analytical and statistical methods, will allow to form the directions of implementation of state support measures using various forms of funding sources with different degrees of innovation and investment activity of small and medium-sized entrepreneur entities.

Regression analysis. This method is used to identify the relationship between quantitative and qualitative variables that have an impact on the resulting indicator: the value of the turnover of SMEs.

## 4. Conclusion

According to the Passport of the national project, the following amounts of funding for state support measures for SMEs are expected in relation to the formation of a system that ensures the availability of funding sources. It should be noted that the law provides for the provision of subsidies to JSC "Corporation "SME"" at the expense of the federal budget for the financial support of the fulfilment of the obligations of JSC "Corporation "SME"" on guarantees provided to SME entities in the period from 2019 to 2024, in order to increase the volume of guarantee support in the framework of expanding the volume of lending to SMEs under the National Guarantee System (hereinafter – NGS).

In addition, a one-time contribution was made to the authorized capital of JSC "Corporation "SME"" in order to recapitalize JSC "SME Bank" so that to increase the volume of guarantee support in the framework of expanding the volume of lending to SMEs in accordance with the NGS in the amount of 5 billion rubles in 2019. Within the framework of the national project implementation, special attention is paid to the cost of leasing services for SMEs, but in certain sectors of the economy that are not related to agricultural production, transport, and trade. It is assumed that the amount of funding form the federal budget, the total amount of which for 2022–2024 will be 107.6 billion rubles. As we can see, the implementation of the national project involves the consistent implementation of state support measures that allow us to form a certain progressive dynamic in the development of both SMEs and the national economy as a whole.

The dynamics of changes in the number of SMEs is presented below.

**Table 1.** Dynamics of changes in SMEs, unit fraction

| Territory | 2019, un.frac. | 2020, un.frac. | 2021, un.frac. |
|---|---|---|---|
| Total Russian Federation | 1 | 1 | 1 |
| Central FD | 0,3105 | 0,3089 | 0,312 |
| North-West FD | 0,1174 | 0,117 | 0,1161 |
| South FD | 0,1168 | 0,1169 | 0,1163 |
| North-Caucasian FD | 0,0331 | 0,0338 | 0,0338 |
| Volga FD | 0,1784 | 0,1788 | 0,1774 |
| Ural FD | 0,0857 | 0,0861 | 0,0854 |
| Siberian FD | 0,106 | 0,1061 | 0,1056 |
| Far Eastern FD | 0,0519 | 0,0525 | 0,0532 |

(compiled on the basis of data published on the official website of the Federal State Statistics Service: https://rmsp.nalog.ru/)

According to Table 1 it can be noted that the share ratio of SMEs has remained virtually unchanged over the past three years, in which the largest share falls on the Central Federal District (CFD), a bit above 30 %.

Consider the dynamics of the turnover of small and medium-sized businesses in the Russian Federation for the period 2010–2019, which is reflected in Figures 2 and 3. According to the Figure 2, the variation of the turnover of medium enterprises is from 3522.1 to 9495.9 billion rubles. However, in 2019 there was a down trend compared to 2018 from 6622.0 to 6141.6 billion rubles. The most significant growth turnover was in the period 2015–2016 and increased from 4715.0 to 6761.4 billion rubles.

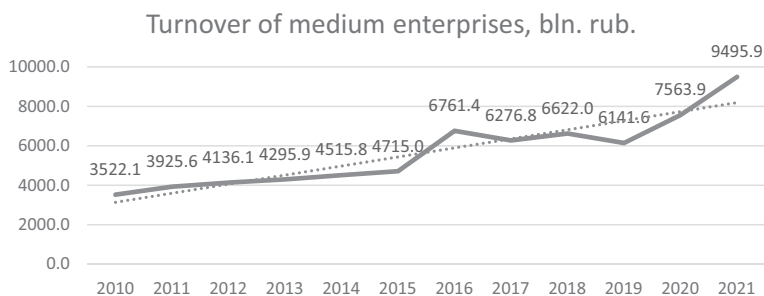

**Figure 2.** Dynamics of the turnover of medium enterprises in the Russian Federation, billion rubles

(Compiled by the authors based on data from the Federal State Statistics Service website)

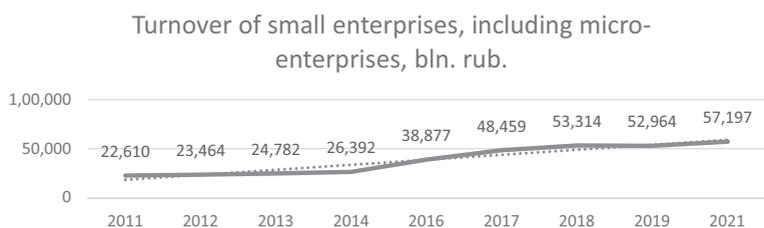

**Figure 3.** Dynamics of the turnover of small enterprises, including micro-enterprises, Russian Federation, billion rubles

(Compiled by the authors based on data from the Federal State Statistics Service website)

Figure 2 shows a qualitative leap in the indicator "turnover of small enterprises (incl. micro-enterprises)" for the period 2014–2016 from 26 392.218 billion rubles up to 38 877.027 billion rubles. respectively.

Figures 2 and 3 prove the effectiveness of implemented measures of state support for SMEs in the Russian Federation. It should be noted that during this period on the territory of the Russian Federation, the Ministry of Economic Development of the Russian Federation formed the "Forecast of the socio-economic development of the Russian Federation for 2013 and the planning period of 2014–2015". In this regard, according to the data of Figures 4–6, we can note that at that time the measures of state support for SMEs in the territory of the Russian Federation were quite effective.

## Analysis of the Source Data

The analysis used data for January–June 2021 for all constituent entities of the Russian Federation: the Central Federal District, the Northwestern Federal District, the Southern Federal District, the North Caucasian Federal District, the Volga Federal District, the Ural Federal District, the Siberian Federal District, Far Eastern Federal District. The research estimates the values of indicators by two groups of SMEs: small enterprises (excluding micro-enterprises), medium-sized enterprises.

The following indicators were used as variables:

– R&D share in the total volume for the subjects of the Russian Federation;
– fixed capital investments per capita in the Russian Federation.

Figure 4 illustrates the analysis of the relations between investments per capita and the turnover of small enterprises (excl. micro) in the Russian Federation.

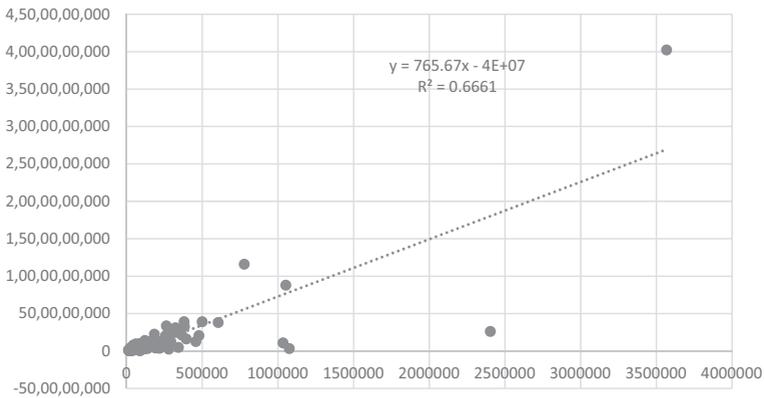

**Figure 4.** The relationship between investment per capita and the turnover of small enterprises in the Russian Federation, excluding micro-enterprises
(Compiled by authors)

Figure 5 shows the dependence between the indicator "fixed capital investments per capita" and the turnover of medium-sized enterprises.

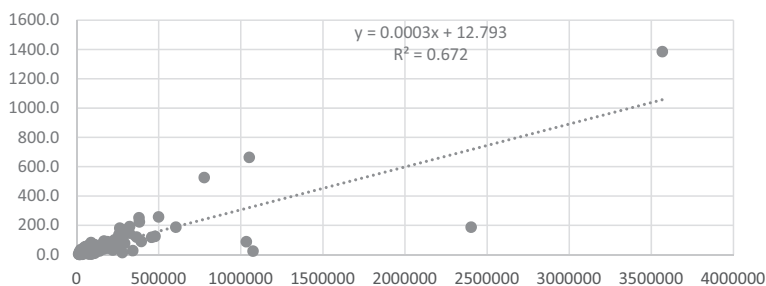

**Figure 5.** Dependence between investment per capita and the turnover of medium enterprises in the Russian Federation
(Compiled by authors)

The opposite trend is observed in relation to the influence of the volume of internal expenditures on R&D on the value of the turnover of SMEs, which is reflected in the Figures 7–9.

Figure 6 shows the changes in the share of R&D internal expenditures in GDP.

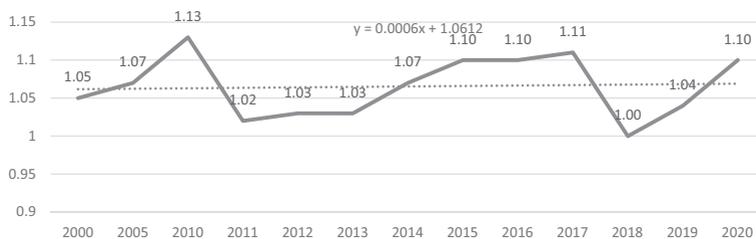

**Figure 6.** Dynamics of changes of the share of R&D internal expenditures in GDP, %
(Complied by authors)

According to Figure 6, the range of the indicator varies from 1.0 to 1.11 that proves stabilization. Figure 7 shows the dependence of the volume of internal expenditures on R&D on the turnover of small enterprises (excluding micro-enterprises).

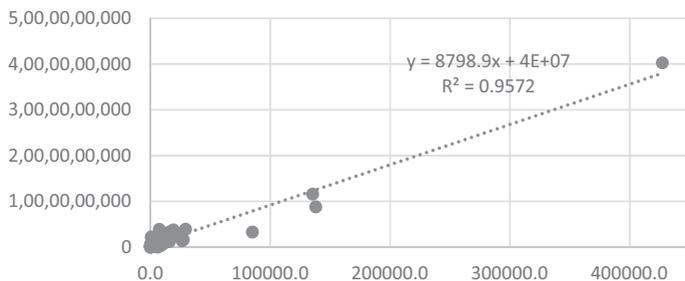

**Figure 7.** Dependence of internal R&D expenditures and the turnover of small enterprises in the Russian Federation

(Complied by authors)

If Moscow is excluded from the sample, then the value of the obtained regression is also characterized by a high level of determination coefficient, the obtained values of regression statistics, analysis of variance are presented in Figure 8.

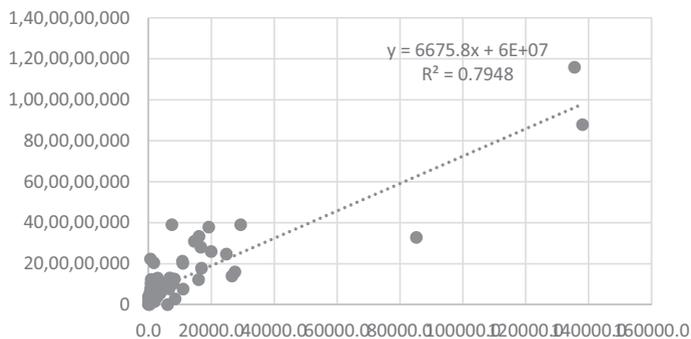

**Figure 8.** Dependence of the value of internal R&D expenditures and the turnover of small enterprises in the Russian Federation (except Moscow)

(Complied by authors)

Figure 9 reflects the impact of the volume of internal R&D expenditures on the volume of turnover of medium-sized enterprises, and under the above condition of excluding Moscow indicators from the calculations, the regression value acquires a higher coefficient of determination. The coefficient of dependence determination even under this condition is quite high.

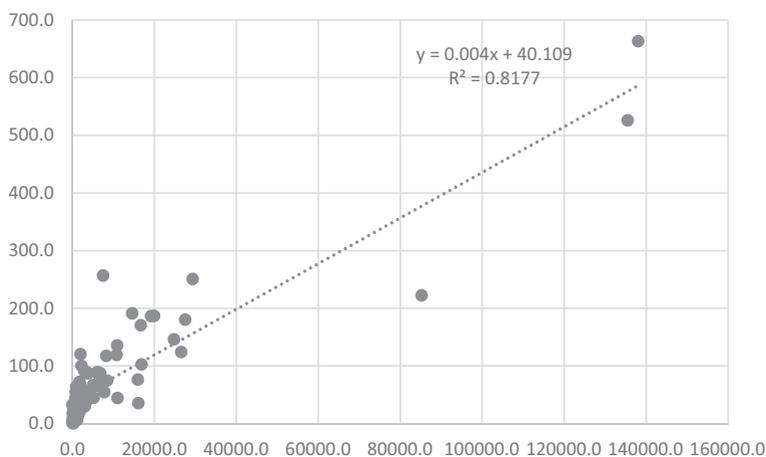

**Figure 9.** Dependence of the value of internal R&D costs and the turnover of medium-sized enterprises in the Russian Federation (except Moscow)

(Complied by authors)

According to the data obtained, it can be noted that in general, there is a dependence between the studied indicators in terms of their impact on the resulting indicator: the volume of the trade turnover of SMEs. In this case, we would like to note that the assumption is confirmed that, in general, the activities of SMEs directly depend on the level of development of the entire economy of the country as a whole, and the support measures implemented in all areas of the development of the national economy, in particular, in relation to SMEs. They will lead to the activation of small and medium-sized businesses. Since in the case of absence of any investment in fixed assets, or a decrease in R&D, in general, it will lead to a decrease in the volume of SME activity in general, which clearly allows us to say that the introduction of measures of state support and development of SME activities is positive.

In this regard, the implementation of the national project "Small and medium-sized businesses and support for individual entrepreneurial initiatives" involves the development of infrastructure support for small and medium-sized businesses. These initiatives are most clearly reflected in the territory of the Central and Volga Federal Districts, which in the future suggests becoming potential territories that ensure the economic growth of SMEs in these territories.

# References

Ashieim, B., Cooke, Ph., & Martin, R. (2009). Clusters and Regional Development: Critical Reflections and Explorations. *Economic Geography*, *84*(1), 109–112. https://doi.org/10.1111/j.1944-8287.2008.tb00394.x

Barkley D. L., & Henry M. S. (2005). Targeting Industry Clusters for Regional Economic Development: An Overview of the REDRL Approach. *Research Report Regional Economic Development Research Laboratory*, 01-2005-03. https://core.ac.uk/download/pdf/6360709.pdf

Bartlett, William. (2001). SME Development Policies in Different Stages of Transition. *n MOCT-MOST Economic Policy in Transitional Economics*, September 2001. https://www.researchgate.net/publication/227176058_SME_Development_Policies_in_Different_Stages_of_Transition

Bouri, A., Breij, M., Diop, M., Kempner, R., & Stevenson, K. (2011). *Report on Support to Smes in Developing Countries through Financial Intermediaries*. https://www.eib.org/attachments/dalberg_sme-briefing-paper.pdf

Decree of the President of the Russian Federation No. 13 of 16.01.2017. On Approval of the Fundamentals of the State Policy of Regional Development of the Russian Federation for the Period Up to 2025. (2017). *Collection of Legislation of the Russian Federation*, (4), article 637. http://www.pravo.gov.ru

Enright, M. J. (2003). Regional Clusters: What We Know and What We Should Know. *Innovation Clusters and Interregional Competition*, 99–129. https://doi.org/10.1007/978-3-540-24760-9_6

Herr, H., & Nettekoven, Z.-M. (2017). *The Role of Small and Medium-Sized Enterprises in Development What Can Be Learned from the German Experience?* https://library.fes.de/pdf-files/iez/14056.pdf

Hobohm, S. (2001). Small and Medium-Sized Enterprises in Economic Development: The Unido Experience. *Journal of Economic Cooperation Among Islamic Countries*, *22*(1), 1–42.

Hussain, M. N. (2000). *Linkages between SMEs and Large Industries for Increased Markets and Trade: An African Perspective*. Strategic Planning and Research Department. The African Development Bank. https://www.afdb.org/fileadmin/uploads/afdb/Documents/Publications/00157640-EN-ERP-53.PDF

Jamieson, D., Fettiplace, S., York, C., Lambourne, E., Braidford, P., & Stone, I. (2012). *Large Businesses and SMEs: Exploring How SMEs Interact with Large Businesses*. https://assets.publishing.service.gov.uk/government/uploads/system/uploads/attachment_data/file/34639/12-1196-exploring-how-smes-interact-with-large-businesses.pdf

Keskin, H., Sentürk, C., Sungur, O., & Kiris. H. M. (2010). The Importance of SMEs in Developing Economies. *2nd International Symposium on Sustainable Development*. Sarajevo. https://core.ac.uk/download/pdf/153446896.pdf

Ketels, C. C. (2017). *Mapping as Tool for Development*. Institute for Strategy and Competitiveness Harvard Business School. https://www.hbs.edu/

Kok, Jan de., Vroonhof, Paul., Verhoeven, Wim., Timmermans, Niek., Kwaak, Ton., Snijders, Jacqueline., & Westhof, Florieke. (2011). Do SMEs Create More and Better Jobs? *EIM Business & Policy Research Source*. https://ec.europa.eu/growth/sites/default/files/docs/body/do-smes-create-more-and-better-jobs_en.pdf

Marri, H. B., Nebhwani, M., & Sohag, R. A. (2011). Study of Government Support System in Smes: An Empirical Investigation. *Mehran University Research Journal of Engineering & Technology*, *30*(3), 435–446. http://publications.muet.edu.pk/research_papers/pdf/pdf131.pdf

McIntyre, Robert. (2001). *The Role of Small and Medium Enterprises in Transition: Growth and Entrepreneurship*. The United Nations University WIDER World Institute for Development Economics Research for Action 49. https://www.wider.unu.edu/sites/default/files/rfa49.pdf

Neagu, C. (2016). The Importance and Role of Small and Medium-Sized Businesses. *Theoretical and Applied Economics*, 3(608), 331–338.

Passport of the National Project "Small and Medium-Sized Entrepreneurship and Support for Individual Entrepreneurial Initiative". (2020). https://www.economy.gov.ru/material/directions/nacionalnyy_proekt_maloe_i_srednee_predprinimatelstvo_i_podderzhka_individualnoy_predprinimatelskoy_iniciativy/

Pech, M., & Vrchota, J. (2020). *Classification of Small- and Medium-Sized Enterprises Based on the Level of Industry 4.0 Implementation*. https://www.mdpi.com/2076-3417/10/15/5150/pdf

Porter, M. (1998). Clusters and New Economics of Competition. *Harvard Business Review*. http://clustermapping.us/sites/default/files/files/resource/Clusters_and_the_New_Economics_of_Competition.pdf

Porter, M. (2000). Location, Competition, and Economic Development: Local Clusters in a Global Economy. *Economic Development Quarterly*, *14*(1), 15–34.

Swords, J. (2013). Michael Porter's Cluster Theory as a Local and Regional Development Tool – the Rise and Fall of Cluster Policy in the UK. *Local Economy*, *28*(4), 367–381.

Tewari, Parth S., Skilling, David, Kumar, Pranav, & Wu, Zack. (2013). *Competitive Small and Medium Enterprises a Diagnostic to Help Design Smart SME Policy*. http://documents1.worldbank.org/curated/en/534521468331785470/pdf/825160WP0P148100Box379861B00PUBLIC0.pdf

Tsertseil, J. S. (2015). The Clusters and Special Economic Zone: The Improvement in the Development of the Region. *Journal of Global Economics*, *3*, 4.

Tsertseil J. S., & Kookueva, V. V. (2017). Development of Innovative Industrial Clusters: Problems, Tools and Prospects. *6th icCSBs 2017 the Annual International Conference on Cognitive – Social, and Behavioural Sciences*. The European Proceedings of Social & Behavioural Sciences EpSBS. ISSN: 2357-1330. https://doi.org/10.15405/epsbs.2017.11.15

Waqas Shair, Saem Hussain, and Muhammad Idrees

# Social Safety Net Programs and Food Insecurity in Pakistan

## 1. Introduction

Food security is an individual's access to adequate, safe, and nutritious food to meet the dietary need for an active and healthy life. In contrast, food insecurity concerns the limited or uncertain availability of food or individuals' access to nutritionally adequate and safe foods. According to Food and Agriculture Organization (FAO), the four dimensions of food security are availability, access, utilization, and stability. The availability of food is related to production level. The per capita food production has shown an upward trend since 1960 because the average growth of food production was almost 1 % more than the population growth. Population growth was 1.9 %, and growth of food production remained at 2.8 % globally over the last five decades (Byrnes & Bumb, 2017). Food access is subject to the purchasing power of an individual to afford an adequate level of food. Purchasing power is a positive function of a household's resources and a negative function of food prices. Food utilization is the food choice associated with the quality and quantity of the food. It is the metabolism of food by an individual and must be enough to meet the physiological requirement of the individual. Food utilization is affected by food preparation which depends on social and cultural norms. Food stability is related to the ability to obtain food over time. The unavailability of food may be seasonal, transitory, or chronic. The transitory food insecurity may be due to natural disasters, drought, war, internal conflict, job loss, and disability. On the other hand, chronic food insecurity is due to the presence of multidimensional deprivations. The dimensions of food security are effect by conflict, climate variability, and economic slowdowns.

## 2. Food Insecurity and Global Development Goals

Food insecurity always remains a heated debate among researchers due to its inclusion in the Millennium Development Goals (MDGs) in 2000. The eight goals and 21 targets of MDGs were designed to "shape the 21st century" with an intent to protect the human right (empowering women, reducing violence, and protecting property rights and freedom of speech), enhance human capital (education and health), and increase the living standard. Regarding food insecurity, the 1st MDG was related to "eradicating extreme poverty and hunger". Its target 1C was related to half the proportion of suffering from food insecurity by 2015. The MDGs were supplanted by Sustainable Development Goals (SDGs) in 2015. The SDGs were designed to be a "blueprint for achieving a better and more sustainable future for all" with an intent to protect planets, end poverty, and bring economic prosperity. The 2nd SDG is related to "Zero Hunger" by the end of 2030. The 2nd SDGs' primary targets are the prevalence of undernourishment, moderate or severe food insecurity, stunting, malnutrition, and anemia.

## 3. Global Overview of Food Insecurity

From the first decade and a half of the current century, the prevalence of hunger (severe food insecurity) has reduced from 15 % to 11 %[1]. While people suffering from hunger soared from 790 million in 2016 to 828 million in 2021, it implies that 1 in 10 people worldwide suffers from hunger, gone a day without eating (SDGs report, 2022)[2]. The situation is more vulnerable when looking into moderately or severely food-insecure people. Almost 1 in 3 (nearly 2.3 billion) people worldwide are moderately or severely food insecure, and an additional 350 million have been food insecure since the beginning of the pandemic. This surge is due to climate variability, pandemic, conflicts, economic shocks, growing inequalities, and food price hikes. The role of food price hikes is inescapable because global food prices surged almost 30 % in 2022 (SDGs report, 2022). These potential sources of rising food insecurity would keep the world off track of achieving zero hunger by 2030.

Globally, a 3.8 % increase in food insecurity from 2019 to 2020 added 320 million more people in one year (see Figure 1). The increase in food insecurity from 2019 to 2020 is 9.0 % in Latin America and the Caribbean, 5.4 points in Africa, 3.1 points in Asia, and 1.1 points in Northern America and Europe. The situation of food insecurity is worst in Africa due to the highest prevalence of food insecurity, 59.6 % of the population. Among these, almost 26 % are severely food insecure. The prevalence of food insecurity is relatively lower in Asia; nearly 25.8 % of people are food insecure. Nevertheless, it accounts for almost half the severely or moderately people of the world due to the larger population. It is important to note that the surge in food insecure people in 2020 is due to the

---

1   See: https://sdgs.un.org/goals/goal2, find in the tab of progress and info by choosing 2016.

2   See: https://unstats.un.org/sdgs/report/2022/

COVID-19 pandemic, which disrupted the supply chain globally and lowered purchasing power by raising food prices. Moreover, job losses due to COVID-19 also worsen the purchasing power of the individual, which in turn limits access to adequate food. The economic downturn followed by COVID-19 soared the number of unemployed people to 220 million, with more than 33 million jobs lost in 2020[3].

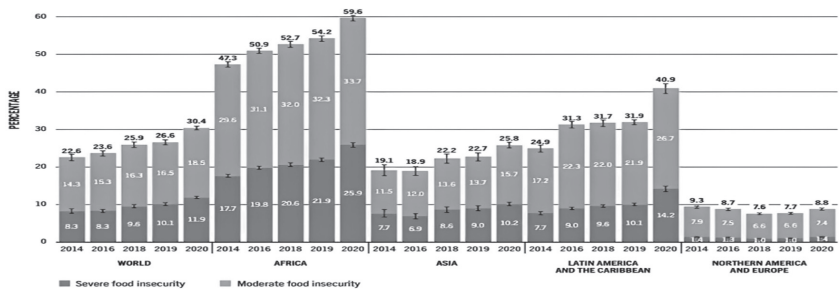

**Figure 1.** Severe or moderate food insecurity in the global and regional context
**Source:** FAO, IFAD, UNICEF, WFP and WHO, 2021

## 4. Global Overview of Social Safety Net Program

Social protection programs include social safety net/social assistance, social insurance, and labor market programs. The social safety net programs are non-contributory interventions designed to help individuals and household cope with chronic poverty, destitution, and vulnerability. The social safety net programs target the poor and vulnerable. It includes conditional and unconditional cash transfers, social pensions, food and in-kind transfers, school feeding programs, public works, and fee waivers. On the other hand, social insurance programs are contributory interventions designed to help individuals manage sudden changes in income because of old age, sickness, disability, or natural disaster. Individuals pay insurance premiums to be eligible for coverage or contribute a percentage of their earnings to a mandatory insurance scheme. It includes contributory old-age, survivor, and disability pensions; sick leave, maternity/paternity benefits; and health insurance coverage. The labor market programs can be contributory and non-contributory. The labor market programs aimed to improve chances of employment, earning, and income smoothing during unemployment. The importance of social protection programs has emerged over the last decade because the 3$^{rd}$ target of the SDG 1 is related to providing appropriate social protection systems to achieve "end poverty" by 2030. Moreover, access to adequate food can also be provided by cash transfer to the marginalized individual/household by designing the outcome-based SSN-program.

---

3   See: https://unstats.un.org/sdgs/report/2021/goal-08/

An economy's spending on SSPs is associated with a lower incidence of poverty and food hunger. The financial support by an economy to the vulnerable segment of society can help to maintain a minimum acceptable level of living. Moreover, SSPs can also serve as a recessionary hedge to cope with the economic downturn. Now the question arises of how much countries spend as a percentage of GDP on the SSN. According to the Atlas of Social protection indicators of resilience and equity (ASPIRE), the average spending of developing countries on SSN is almost 1.5 % of the GDP. In comparison, world spending on SSN as a percentage of GDP is 1.54 (see Figure 2a). On the other hand, Europe and Central Asia spend 2.2 % of GDP on SSN, which is more extensive when compared with other regions. However, it is essential to note that the South Asia region has the lowest SSN spending, with 0.9 % of GDP. Although Europe spent more than 2 % of its GDP on SSNPs, likewise its per capita spending was more than 253$. The higher per capita spending on SSNs allowed her to stall food insecurity at the lowest compared to other regions. The lowest per capita spending was in Sub-Sahara Africa, with an annual 16$, while globally, median SSN spending per capita was 80$ (see Figure 2b). It implies that a region with lower per capita SSN spending is experiencing higher food insecurity.

A common misconception regarding higher SSN spending is that countries with higher SSNs are often contented with fragility, conflict, and violence. Moreover, it is also argued that higher SSN spending is due to the inclusion of universal programs – funded mainly by donors – in the SSN portfolio of a country. Now a question arises: Do higher-income countries spend more on SSNs? The high-income countries spend almost 1.9 % of GDP on SSN, which is higher than the spending of low-, middle-, and upper-middle-income countries (see Figure 2c). Moreover, the expenditure of the OECD is 2.7 % which was higher when compared with other income group countries. However, high-income countries spend more on SSNs than others when the analysis was carried out using individual country observation. Then evidence was found on the presence of no relationship between GDP per capita level and SSN spending as a percentage of GDP (see Figure 2d). It was observed that countries with the same GDP per capita choose different spending levels on SSN, reflecting other policy preferences rather than economic conditions.

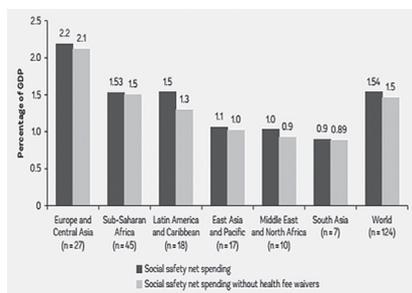

**Figure 2a.** Global and regional SSN spending
**Source:** World Bank, 2018

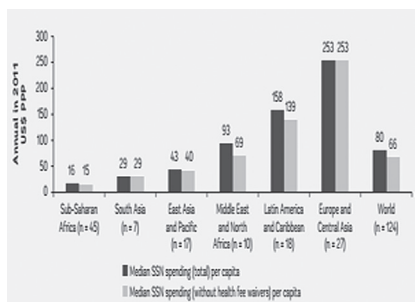

**Figure 2b.** Median SSN spending per capita
**Source:** World Bank, 2018

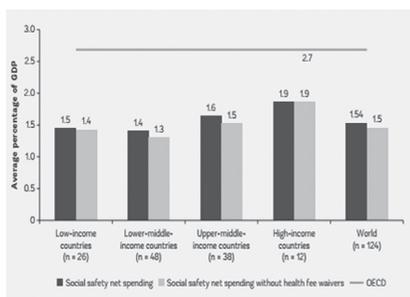

**Figure 2c.** SSN spending by country income groups
**Source:** World Bank, 2018

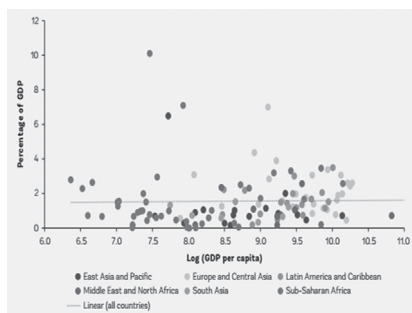

**Figure 2d.** SSN spending and income levels across region
**Source:** World Bank, 2018

## 5. Social Safety Net Program in Pakistan

Pakistan is striving to progress toward SDGs amid challenges of ensuring quality education, gender equality, skill development, health & sanitation, infrastructure development, and job creation. Pakistan is committed to alleviating poverty in line with the SDGs target Goal-1 "No Poverty" in all its manifestations by 2030. In this regard, SSNP, through redistribution of resources, can reduce poverty by reaching the poor and disadvantaged groups to maintain social harmony. Pakistan also takes advantage of SSNP and operates almost 30 SSNP (p. 11, World Bank, 2018). The two types of social safety programs are functional in Pakistan; one is budgetary, and the other is non-budgetary (p. 279, Economic

Survey of Pakistan, 2022). The budgeted SSN programs include the Benazir Income Support Program (BISP) and Pakistan Bait-ul-Mal (PBM). The non-budgeted SSNs are Zakat, Employees Old Age Benefits Institution (EOBI), Workers Welfare Fund (WWF), and Pakistan Poverty Alleviation Fund (PPAF).

Among these SSNs, BISP is a more comprehensive and well-organized unconditional cash transfer program that commenced in 2009 with 15.85 billion PKR and 1.76 million beneficiaries. BISP was implemented by focusing on poor women with an immediate objective of consumption smoothing and meeting the targets of SDGs to eradicate extreme and chronic poverty and empowerment of women. BISP provides paramount social safety to the poorest segments of society, and its carefully designed schemes add value to the overall mission of empowering women.

Since commencing of BISP in 2022, the cash transferred and beneficiaries have increased exponentially. The cash transfer increased almost 12-fold and reached 15.32 to 187.5 billion PKR (see Figure 3a). Moreover, beneficiaries increased 5-fold and outstretched to 9.11 million (5 % of the population).

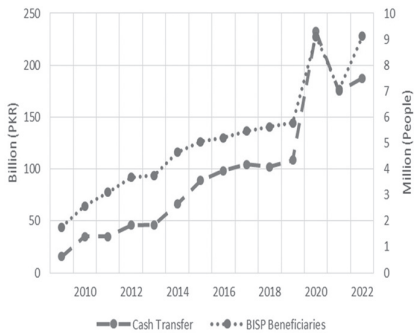

**Figure 3a.** BISP beneficiaries and cash transfer

**Source:** Economic Survey of Pakistan 2021–2022

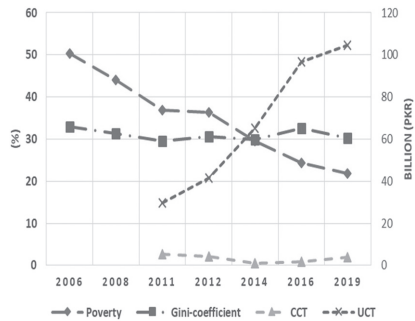

**Figure 3b.** Poverty, inequality, conditional, and unconditional cash transfer

**Source:** Economic Survey of Pakistan 2021–2022

Even though cash transfers from BISP and its beneficiaries are increased, the question arises of whether these transfers reduce poverty or income inequality. Figure 3b depicts poverty going down from 50 % in 2006 to 21 % in 2019. However, inequality remains stagnant because Gini-coefficient stalled from 0.3 to 0.33 scale point. It implies that the share of cash transfer supports the consumption of the bottom quantile group but does not raise their income share.

Though unconditional cash transfer is growing over time, conditional cash transfers have remained inert over the last decade in the range of 1 to 7 billion PKR. The descriptive at the macro level suggests that poverty is decreasing with increased cash transfers and BISP beneficiaries. However, household-level analysis is required to determine the food insecurity level in the BISP-recipient and non-recipient households.

## 6. Food Insecurity Methodology

As a measure of indicator, food insecurity refers to limited access to food, at the level of individuals or households, due to a lack of money or other resources. The severity of food insecurity can be measured using data collected with the food insecurity experience scale survey module (FIES-SM). The FIES-SM consists of eight questions[4] asking to self-report conditions and experiences typically associated with limited access to food. The severity of food insecurity can be quantified by assigning "1" for "yes" and "0" for the case of "no". Subsequently, apply the sophisticated statistical techniques based on the Rasch measurement model to figure out the food secure, moderate, and severe food insecure households based on the FIES reference scale erected by FAO.

After converting eight qualitative questions into eight dichotomous variables comprising "0" and "1" coding. Subsequently, apply sophisticated statistical techniques based on the Rasch measurement model to figure out the food secure, moderate, and severe food insecure households based on the FIES reference scale erected by FAO.

The FIES consider the individual/household mildly food insecure if they respond to compromising on food quality and quantity. Moderately food insecure concerns reducing food quantities or even skipping meals. Severely food insecure, for example, is experiencing hunger or going a day without eating. A household/individual is considered food secure if they respond "no" to all eight questions. Contrarily, a household/individual is considered mild food insecure for responding to "yes" for any of Q1–Q3, moderate food insecure for responding to "yes" for any of Q4–Q6, and severe food insecure for responding to "yes" for any of Q7–Q8.

The global recognition of food insecurity by using the FIES-SM is an outcome of the inclusion of the eight questions in the living standard survey of economies worldwide. Following the global recognition of FIES-SM, the Pakistan Bureau of Statistics (PBS) preliminary included eight sets of FIES-SM in the nationally representative survey Pakistan Social and Living Standard Measurement (PSLM) in the 2018–2019 round. The PSLM 2018–2019 round comprises a sample of 24,807 households.

---

4   See for eight questions: https://www.fao.org/in-action/voices-of-the-hungry/fies/en/

## 7. Food Insecurity in Pakistan

The spatial distribution in the map illustrating the food-insecure household across the administrative division of Pakistan is given in Figure 4a. In the sample, almost 36 % of households reported a prevalence of food insecurity. However, only three admin-divisions are less than 15 % insecure households, while other admin-divisions have a higher level of food-insecure households. The detailed assessment of food insecurity across the geographical distribution constitutes a powerful tool to help policymakers and program planners visualize which provinces or regions are most in need and, therefore, should be targeted for interventions aimed at guaranteeing them right to adequate food.

The greater disparities were observed in the expenditure per capita across the admin-divisions (see Figure 4b). Although food insecurity concerns access to food, the spatial distribution of expenditure per capita depicts that admin-divisions with higher expenditure represent lower food insecure households. Notably, 8 out of 29 admin-division's per capita expenditures are lower than the monthly poverty threshold. Therefore, these admin-divisions should be targeted for policy intervention to uplift the living standard of dwellers.

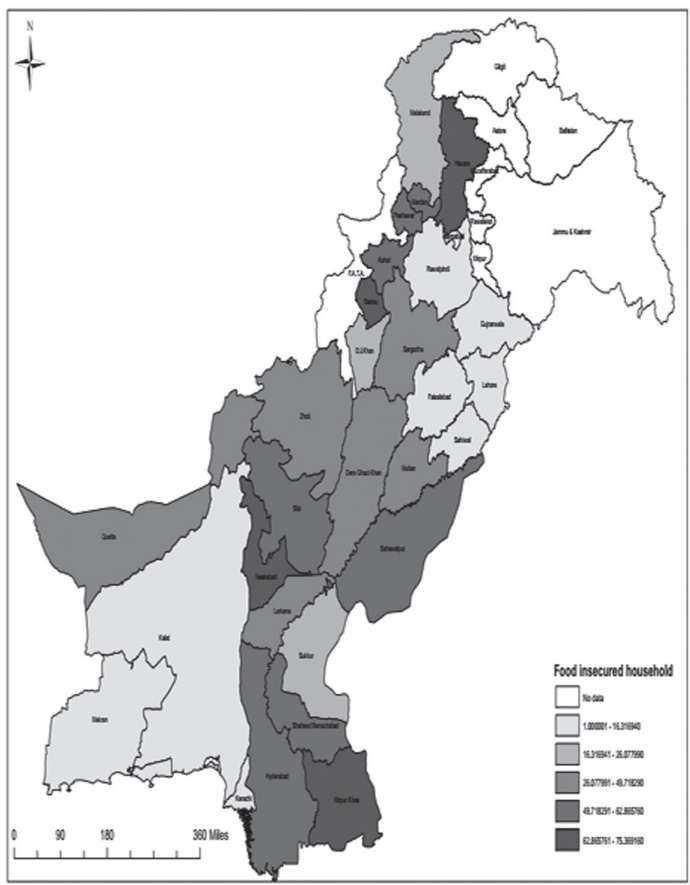

**Figure 4a.** Spatial pattern of food insecurity

**Source:** Authors' own estimation based on PSLM (2018–2019)

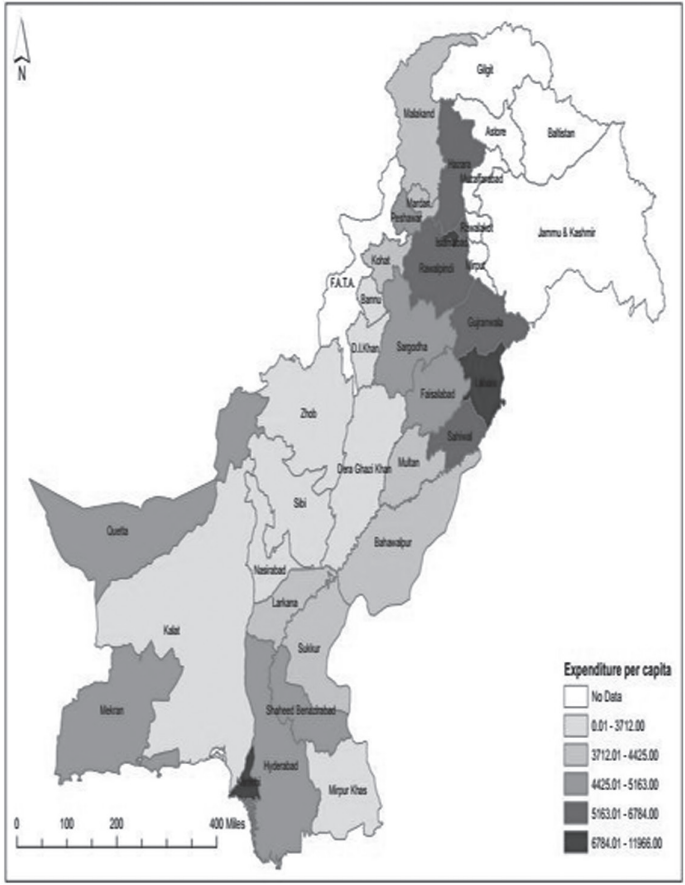

**Figure 4b.** Spatial pattern of Expenditure per capita
**Source:** Authors' own estimation based on PSLM (2018–2019)

## 8. Food Insecurity and SSN in Pakistan

The estimates based on the nationally representative data of the PSLM 2018–2019 survey, comprising a sample of 24,807, suggest that 36 % of households are food insecure in Pakistan. Almost 17 % of households are mild or marginally food insecure, which is not subject to policy targets. However, moderate food insecure households are 12 %, and severe food insecure households are 7 % (see Figure 5a). The severe or moderate were 26.6 % in the world and 22.7 % in Asia, while 19 % in Pakistan. Although severe or moderate food insecurity is relatively low in Pakistan, it is still higher than in other developing countries. After examining the food insecurity level in Pakistan, the food insecurity level in the BISP-recipient households will be examined.

Among Pakistan's different social safety net programs (SSNPs), the Benazir Income Support (BISP) is a comprehensive and extensive program aimed at alleviating poverty by empowering women. BISP is an unconditional cash transfer program that provides a stipend of 6000 PKR quarterly to vulnerable families. The question arises of the level of food insecurity in the BISP-recipient households. The estimates of the PSLM survey suggest that almost 37 % BISP-recipients are food insecure, 23 % mildly and 14 % severely food insecure (see Figure 5a). It intimates that amongst BISP-recipients, the greater proportion of the households are food insecure.

In the sample, almost 80 % of non-poor households are food secured, while only 64 % of the households are food secured in the whole sample. In the analysis, the poverty line is defined based on the cost of basic need (CBN) approach, which was 3757.85 PKR in 2019 (see economic survey of Pakistan, 2021–2022, page. 276). The greater proportion of food-secure households is the non-poor sample, while the lower proportion in the poor indicates that poverty is a potential source of food insecurity. The absence of poverty increases the prevalence of food security by 14 %. In the sample, almost one-third of the poor households are moderate or severe food insecure, which needs policy intervention because, in the sample, nearly 43.2 % of the households are poor (see Figure 5b).

It is important to note that food insecurity is not limited to poor households but also affects non-poor households. For example, in the sample, almost 9 % of the non-poor households are severe or moderate food insecure. However, it implies that non-poor households are also subject to policy focus to uplift against the transitory or chronic income shock which causes food insecurity.

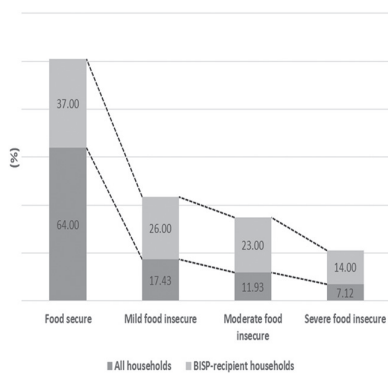

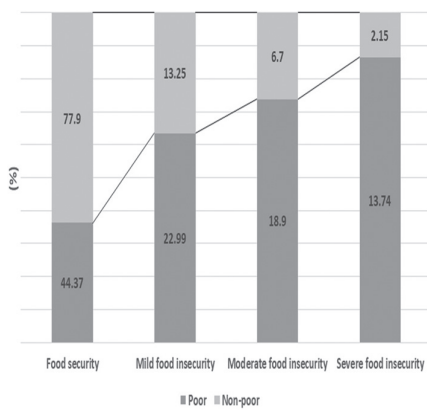

**Figure 5a.** Severity of food insecurity across all the households and BISP-recipient households

**Source:** Authors' own estimations based on PSLM (2018–2019)

**Figure 5b.** Severity of food insecurity across the poor and non-poor households

**Source:** Authors' own estimations based on PSLM (2018–2019)

Although BISP is a pro-poor SSN program, the non-poor household also reported receiving a stipend from the BISP program. The food insecurity level in poor BISP-recipient households is 68 %, while 47 % in the non-poor BISP-recipient (see Figure 6a). Likewise, the prevalence of mild, moderate, and severe food insecurity is higher in the poor-recipient household than in the non-poor-recipient household.

Regarding the income of the different household groups, the income per capita of the non-poor household is three-fold the income of the poor household (see Figure 6b). The per capita income of food secured household is 7105 PKR, which is almost two-fold the income of the food-insecure household. The average per capita income of the BISP recipient food-secure household is almost 896 PKR more than the BISP recipient food insecure household. It implies that adding 896 PKR per capita in the cash transfer to BISP recipient food insecure households can make the 63 % BISP recipient household food insecure to food secured given the identical endowment and characteristics across the household group.

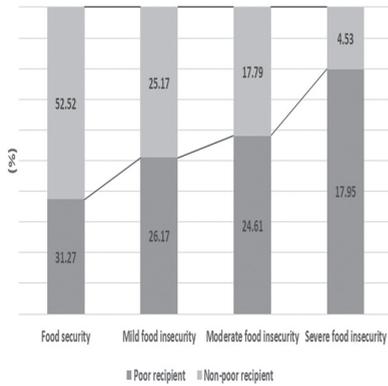

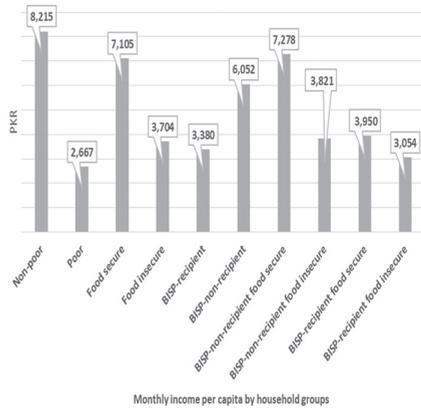

**Figure 6a.** Severity of food insecurity across poor and non-poor BISP-recipient households

**Source:** Authors' own estimations based on PSLM (2018–2019)

**Figure 6b.** Income per capita across household group

**Source:** Authors' own estimations based on PSLM (2018–2019)

## 9. Conclusion

This study has analyzed the social safety net program and the prevalence of food insecurity in Pakistan. The study's primary findings are that food insecurity's prevalence and severity varies across geographical location, poor or non-poor household status, and BISP-recipient or non-recipient households. It is obvious that per capita income is a major determinant of the prevalence of food insecurity in the household. The per capita income shows greater disparities across the poor and non-poor households, food secure and insecure households, and BISP-recipient and non-recipient households.

Although poverty always remains a hot debate among development economists, the importance of access to food is inescapable. Pakistan strives to "end poverty" by 2030, but policies to soften food insecurity are also required. The current SSNP is aimed to ease poverty and empower women under the famous political slogan of "food, clothing, and shelter". But food security–oriented policies are also required because poverty is 43.2 % in Pakistan while food insecurity is 36 %.

Here are some recommendations propounded for the effective policy implementation to ease the prevalence of food insecurity in Pakistan:

- Generally, social protection programs consist of social safety nets/social assistance, social insurance, and labor market program. In Pakistan, SSN and social insurance programs, to some extent, are functional. Still, measures are required to activate active and passive labor market programs to improve and smoothen the income of individuals.
- Food insecurity is not just due to the incidence of poverty. Some non-poor households are also experiencing food insecurity. Therefore, the beneficiaries of any program related to "zero hunger" are not only the poor but also the non-poor.
- BISP is an unconditional cash transfer program that gives liberty regarding spending. However, a conditional cash transfer program is required to implement access to food effectively. A successful example of conditional cash transfer is the model of Latin America to track the reduction in poverty and food insecurity for MDGs.
- Some poor and even food-insecure households are excluded from the consumption of SSNP. The steps need to be undertaken by the authorities for inclusion in the social safety net with no one being left behind. Especially focusing the poor quantile because 22 % of households from the poorest quantile receive cash transfers which are lower than in other lower-middle-income countries, while in China, it is 81 % (World Bank, 2018, p. 36).
- The SSN programs of Pakistan need to be evaluated, because from 2008 to 2013, the poverty reduction due to SSN was 3.16 and 0.67 % in income inequality. However, poverty and income inequality reduction globally are 2 and 8 %.
- The annual spending of Pakistan on SSN is 0.58 % of the GDP. This share needs to be standardized because world SSN spending is 1.5 % of the GDP and 0.9 % in South Asia. Moreover, monthly transfer through SSNs needs to be rationalized because the monthly transfer for Pakistan is 48$ on purchasing power parity (PPP), which is lower than the average of the world's 79$.

## References

Byrnes, B. H., & Bumb, B. L. (2017). Population Growth, Food Production and Nutrient Requirements. In Z. Rengel (Ed.), *Nutrient Use in Crop Production*, pp. 1–27. Boca Raton: CRC Press.

FAO, IFAD, UNICEF, WFP & WHO. (2021). The State of Food Security and Nutrition in the World 2021. In *Transforming Food Systems for Food Security, Improved Nutrition and Affordable Healthy Diets for All*. Rome: FAO. https://doi.org/10.4060/cb4474en

World Bank. (2018). *The State of Social Safety Nets 2018*. Washington, DC: World Bank. https://doi.org/10.1596/978-1-4648-1254-5

Nida Turegun

# History and Techniques of Managerial Accounting

## 1. Introduction

The degree to which management accounting is evolving has been extensively debated over recent years. Until the early part of the 20th century, managerial accounting had not improved and had lost its importance in order to inform the decisions of managers (Johnson & Kaplan, 1987). A variety of groundbreaking managerial accounting approaches have been developed across a range of industries since then, and probably in response to these criticisms.

Activity-based practices, strategic management accounting, and balanced scorecard are the most significant contributions. These practices are meant to encourage up-to-date technology and management practices, such as total quality management and just-in-time, and to find a strategic edge to face the challenge of international rivalry. These modern practices have allegedly affected the whole managerial accounting process and moved its emphasis from a mere role in cost determination and finance reporting to an advanced role in generating value through enhanced resource usage. It was also alleged that the world in which management accounting is conducted has evolved dramatically – with changes in information technology, more dynamic economies, different corporate systems, and modern management methods.

This chapter provides a synopsis of the managerial accounting to outline its evolution and stress its relevance for companies and management. The history of management accounting was therefore studied in order to provide a clearer explanation of the development of managerial accounting activities across the world with their peculiar characteristics.

## 2. History of Managerial Accounting

Cost accounting is regarded generally as managerial accounting. According to accounting historians, cost accounting is a result of the industrial upsurge. Cost accounting was applied in the 1800s with the emergence of industrialized cotton textile plants in United Kingdom and the United States (Ahmad, 2014). This opinion was aligned with Garner (1947). He noticed that cost accounting practices originated just after the 18th century with the upsurge of manufacturing process in the industrial revolution (Garner, 1947). The earliest mark of cost accounting uncovered was "job order costing" of the wool carding in Italy. Moreover, Francesco di Marco Datini, an Italian merchant born in Prato, kept double entry in 1390. The archives display evidence of accrual accounting, depreciation, and job costing. Likewise, in 1531 Raffello di Francesco de'Medici, an Italian cloth manufacturer, provided cost accounting evidences in his accounts (Ahmad, 2014).

As a consequence of increased industrialization and the growing size of companies, cost accounting began to grow further throughout the 19th century and into the middle of the 20th century. The areas of industrial engineering and managerial accounting evolved in parallel in the early decades of the 20th century. Industrial engineers established production management techniques throughout this era that included standards for material inputs, labor, and machine time, to which real outcomes could be measured. Therefore, this evolution contributed to the use of standard costing systems, which are still commonly used by manufacturing firms' planning and control (Caplan, 2006).

The word "cost accounting" began to shift to "managerial accounting" later in the 20th century, but the roots of managerial accounting practices started to develop in the 1920s. However, it was the 1960s that managerial accounting was started to be used commonly (Horngren, 1975). In addition, the rise of service sectors such as financial institutions and the proliferation of governmental and quasi-state bodies further encouraged the emergence of the managerial accounting in the middle and later parts of the 20th century (Kamal, 2015). The standard cost principles and costing methods adapted to a production process had to be changed to fit different organizations. The expression of cost accounting no longer properly represents an organization's accounting position while these improvements were in view. Consequently, the expression of management or managerial accounting has been increasingly embraced.

## 3. Techniques of Managerial Accounting

The International Federation of Accountants (IFAC) had divided the development process of managerial accounting into four timeline groups in 1998 (IFAC, 1998). The first timeline takes place before the 1950s, the second takes place from 1950 to 1979, the third takes place from 1980 to 1989, and the last one place in the 1990s.

## 3.1.  Before the 1950s

The IFAC labels managerial accounting before 1950 as a technical movement essential for the pursuit of structural goals. The primary emphasis was on the determination of the cost of the product. Production technology, with goods going through a number of different methods, was relatively straightforward. The cost of material and labor was easily recognizable, and manufacturing processes were generally ruled by manual activities alacrity. Direct labor then offered a natural basis for allocating overheads for each commodity. Budgets and the financial control of manufacturing activities complemented the emphasis on the prices of the goods (IFAC, 1998). In the 1900s, managers began to give importance to capital productivity and performance. At that time, the development of Du Pont techniques made the measurement of capital output smoother and paid special attention to the execution of investment returns. This knowledge has allowed managers to assign new investments between financial activities and to fund new capital needs (Kamal, 2015).

## 3.2.  1950 to 1979

During the 1950s and 1960s, managerial accounting focused on supplying knowledge for planning and control activities. Managerial accounting is viewed in this process as a management practice, albeit in a position for employees. It is also viewed as personnel support for line management by leveraging tools such as decision analysis and responsibility accounting. Management controls are directed toward production and internal management rather than strategic and environmental concerns (IFAC, 1998). The following can be explained by significant trends in managerial accounting from the 1950s (Hagerty, 1997; Smith, 1999):

- Advances of cost and managerial accounting in the 1950s can be named as cumulative sum charts, discount cash flows, optimum transfer pricing, and total quality management.
- Advances of cost and managerial accounting in the 1960s can be named as decision tree, computer technology, critical path scheduling, opportunity cost budgeting, management by objectives, and zero-base budgeting.
- Advances of cost and managerial accounting in the 1970s can be named as Just-in-time scheduling, portfolio management, information economics, strategic business units, agency theory, experience curves, materials resource planning, product repositioning, matrix organization, and diversification.

### 3.3.  1980 to 1989

Increased international economic rivalry in the early 1980s and the global crisis in the 1970s following the oil price crash placed the developed economies of Western countries at risk. The rapid technical growth that had an impact on many areas of the industry was followed by intensified competition. The use of technology and computer-driven systems, for instance, in many cases increased efficiency and lowered costs. Additionally, computing innovations, particularly the advancement of personal computers, have obviously altered the type and quantity of data that managers can access. Therefore, in successful management, the layout, implementation, and analysis of information systems became quite important (Kader & Luther, 2004). Briefly, advances of cost and managerial accounting in the 1970s can be named as activity-based costing, value-added management, target costing, theory of constraints, benchmarking, private labels, and vertical integration.

### 3.4.  1990s

The 1990s saw major uncertainties and remarkable developments in production and IT infrastructure worldwide (Ashton et al., 1995). The growth of the digital network and related technology, for instance, culminated in the appearance of ecommerce, which further heightened the challenge of international competition. Management accountants have concentrated on the development or formation of value through optimal utilization of resources. This was done by using technologies to verify consumer value factors, shareholder value, and corporate innovation. The chart below demonstrates the four phases of managerial accounting development and features of managerial accounting practices (IFAC, 1998).

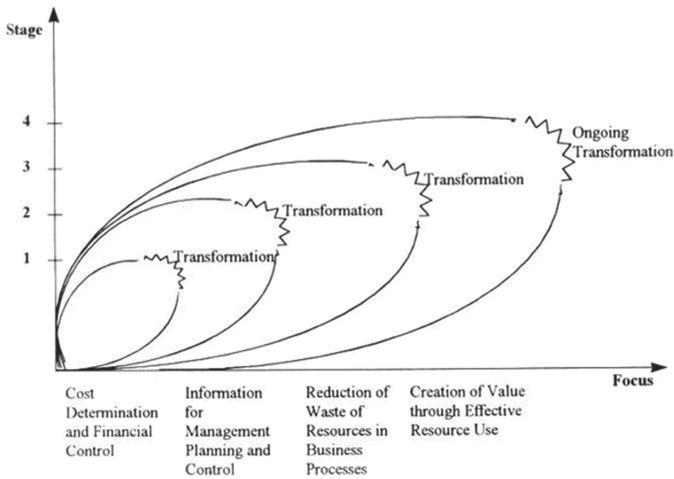

**Figure 1.** Evolution of management accounting
**Source:** (IFAC 1998) p. 85

Figure 1 demonstrates the four phases of managerial accounting development mentioned by the IFAC (1998). The transformation of emphasis away from the provision of information and resource processing, in the form of waste reduction (Step 3) and value development, is a major gap between Stage 2 and Stages 3 and 4 (Stage 4). The emphasis on the provision of knowledge in Stage 2 is not absent, though, but is refigured in Stages 3 and 4. Accompanied by other corporate tools, information becomes a resource; there is a clearer emphasis on eliminating duplication and exploiting resources for the production of profit. Managerial accounting is thus used as an integral aspect of the management process in Phases 3 and 4, as present information becomes directly available to management and as the difference between personnel and line management turn out to be indistinct (IFAC, 1998).

Subsequently, advances of cost and managerial accounting in the 1990s can be named as quality functional deployment, time-based competition, outsourcing, core competencies, gain sharing, business process reengineering, and learning organization (Hagerty, 1997; Smith, 1999). In addition, advances of cost and managerial accounting in 2000s can be named as activity-based costing, local information system, activity management, balanced scorecard, activity-based management, target costing, life cycle costing, and strategic management accounting. The development of managerial accounting techniques is displayed in Figure 2 as follows:

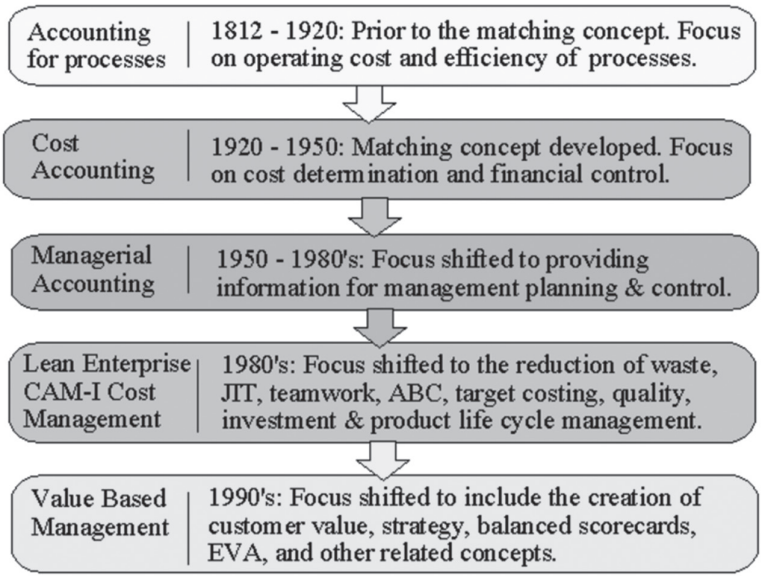

**Figure 2.** The evolution of management accounting practice
**Source:** Martin, 2006

## 4. Development of Managerial Accounting around the World

### 4.1. United States

Managerial accounting systems first developed in the United States. In the nineteenth and first quarters of the 20th century, the majority of the costs and managerial accounting practices were formed. In the 19th century cost accounting was used as a method of measuring subordinate managers' productivity. In addition, during the 19th century, internal accounting techniques were designed for measuring costs, throughput, and working capital (Kamal, 2015).

In the first half of 20th century, matching concept was developed. During that time, the design of Du Pont's managerial accounting practices allowed the assessment of the output of capital, bringing considerable emphasis to the implementation of return on investment (Kaplan, 1984). In the second half of 20th century, over 30 common accounting methods for cost and managerial have been developed. Some of the methods can be counted as discount cash flows, optimum transfer pricing, total quality management, computer technology, cumulative sum charts, opportunity cost budgeting, decision tree, zero-base budgeting, critical path scheduling, information economics, management by objectives, just-in-time scheduling, agency theory, strategic business units, portfolio management, diversification, product repositioning, materials resource planning, experience curves, and matrix organization (Askarany et al., 2007).

Starting from the 1980s, new management account techniques have been introduced and the emphasis moved to waste reduction, theory of constraints, teamwork, just in time private labels, benchmarking, target costing, value-added management, activity-based costing, vertical integration, investment, quality, and product life cycle management (IFAC, 1998). In the 1990s, the emphasis of the managerial accountants moved to value generation or development by the productive use of capital. This was to be done by the use of innovations that tested consumer value measures, shareholder value, and organizational ingenuity (Ashton et al., 1995).

## 4.2. Europe

The use of managerial accounting practices differs across European countries. The scope of the use of activity-based costing, which seems to be pointed to as an indication of the pursuit of new strategies, seems to be primarily strong in the United Kingdom and Belgium, and weak in Denmark and Germany. Activity-based costing method's cost reduction and control activities have proven especially successful in United Kingdom. On the other hand, activity-based costing system is not a costing instrument in France. In France, Sweden, Greece, and Germen, a large majority advocates full costing. The common method in Denmark and Finland is the variable costing method with an important usage in Italy and Spain. Tax law in Finland includes an inventory measurement on a variable cost basis. This was critical as the managerial accounts were promoted by variable costing. Only the United Kingdom has a formal institution devoted to management accounting called the "Chartered Institute of Management Accountants" (Kamal, 2015).

## 4.3. China

Management accounting practices were mainly used in companies starting from the 1970s as the economics restructuring policy was undertaken by the Chinese government. (Islam & Kantor, 2005). A responsibility cost accounting framework has been successfully adopted by Zhoulu Fertilizer Ltd. in China. Due to market competition, the prices are set and the responsibilities of different responsibility centers are then specified by the company. The most common practices for Chinese firms have been the contribution margin and cost-volume-profit analysis from all the western managerial accounting practices. Both methods have developed easy but effective instruments for Chinese administrators to assess the impact that various organizational choices would have on revenue and costs. (Kamal, 2015).

## 4.4. Japan

Management accounting practices started in the 19th century as the Japanese social system was significantly altered around the mid-19th century. The Western type of accounting built on double-entry system was presented by Yukichi Fukuzawa. He introduced Western bookkeeping in the 1870s. It was not adopted from the West but was "imported" from Japan alone. However, in the 19th century the Western type of accounting introduced into Japan was distributed over management as basic knowledge. Awareness of the Western accounting methods in modern Japan is generally and quickly being applied across society in the education system, but in practical they were widely adopted only for some time afterwards (Kudo & Kudo, 2015).

## 4.5. Russia

Prior to 1915, the managerial accounting was prepared by basic accounting and planning approaches. Cost and planning analysis were developed between 1915 and 1930. In this era, accounting of semi-finished goods, auxiliary and integrated production, overhead cost absorption, work in progress, and fixed and variable costs were used. Between 1931 and 1971, the hard-controlled managerial accounting scheme took place, and between 1971 and 1986, the rough regulated scheme took place. The economic measurement system's managerial accounting activities took place from 1987 to 1991. Each organization provides the production costs under this scheme and the effects of its operation to ensure sustainability of production. From 1991 on, businesses started using managerial accounting approaches when they became economically independent and approaches of convergence have begun. Commonly used technique in Russia is absorption costing. Furthermore, in comparison to other Western approaches, quality and inventory control techniques are commonly employed (Sokolov & Bikmukhametova, 2015).

## 4.6. Australia

In Australia, conventional management accounting techniques such as analysis for budgeting for planning financial position, capital budgeting, performance evaluation using return on investment are generally used rather than the new ones such as activity-based costing ranked, activity-based management, product life cycle analysis, and target costing (Askarany et al., 2007).

## 4.7. Thailand

Thai accounting principles were taken from the United States accounting standard in the 1970s and 1980s. The "Thai Accounting Standards" was replaced by "International Accounting Standards" which was developed by the "Federation of Accounting Professions" (Boonyanet & Komaratat, 2008). Thai corporate or technical institutes use Western managerial accounting standards and modified western language manuals. Thai major corporations are still in the conventional stage of managerial accounting practices: using cost variance analysis, budgetary control, and standard costing. Many use the same product cost information as for inventory assessment in preparing financial statements for decision-making purposes (Boonyanet & Komaratat, 2008).

## 4.7. Africa

Three professional bodies are affiliated for managerial accountants in Africa. These are The Institute of Management Accountants, The Chartered Institute of Management Accountants, and The International Federation of Accountants. The United Kingdom has largely affected managerial accounting in South Africa. In 1953 a branch organization of the Institute of Cost and Works Accountants was founded and a center was opened in Johannesburg at the beginning of the 1960s. The headquarters of The Chartered Institute of Management Accountants (CIMA) in Johannesburg shortly needed a CIMA officer at full time. There is currently only one CIMA center in Gauteng with 1200 practitioners. The CIMA office in Johannesburg has been unable to provide analysis on managerial accounting history in South Africa. Details can only be collected by means of personal interviews with longstanding CIMA South African members and copies of the correspondence held by some of these members (Waweru et al., 2004).

## 4.8. Canada

The origins of Canadian accounts lie in Toronto and Montreal's financial centers. Their services began in the 1840s primarily because of bankruptcies, but became popular as the economic development and stocks grew higher. In 1917, the position of fiscal accountants was created. The Government of Canada introduced the Income Tax Act, forcing corporations both to retain their accounts and to file tax returns.

In 1946, the Chartered Accountants Institute of Canada started to create their own country's accounting regulatory bulletins. Two years later, these claims were combined into the Canadian Institute of Chartered Accountant's Handbook, a loose-leaf binder. The Canadian Securities Managers who favored the implementation of U.S. laws, "Generally Accepted Accounting Principles", released National Policy Statement No. 27 in the Canadian Handbook in 1972. Later in 1975, all companies and corporations were legally bound to follow the "Generally Accepted Accounting Principles" under the Canada Business Company Act (Lee, 2018). Managerial accounting became a standardized occupation in the early 1940s.

### 4.9. Turkey

Turkish style of bookkeeping starts at 1326 in the Ottoman Empire. In 1850, Commercial Code required organizations to adapt to double-entry accounting method and introduced bookkeeping with two books: Inventory and Journal. (Guvemli & Guvemli, 2015). The first Turkish accounting book introduces American School of accounting, printed in 1913–14. In the Republic Era, accounting was more influenced with the tax reform of 1949 than any other regulation, because Tax Procedural Law of the time influenced the structure of record keeping and preparation of financial statements (Gucenme & Poroy Arsoy, 2006).

Between 1925 and 1975 management accounting was significantly influenced by the automation of production and advancements in Information Technology (IT) and global competition (Demir, 2008). During the 1990s, uniform accounting principles were finally embraced in 1992 with the Uniform Account Plan. Moreover, the Commercial Code of 2011 brings about the Turkish Accounting Standards, which are based on international accounting standards (Guvemli & Guvemli, 2015). Organizations are required to prepare year-end financial statements by Turkish Accounting Standards, which is a significant step taken toward the standardization of financial statements of organizations in Turkey.

## 5. Conclusion

Managerial accounting is in its early years. In the past, it played a secondary role in "public statements" and is still only a by-product of financial reporting in many businesses. Nevertheless, developments in the last 20 years have contributed to the growth of managerial accounting and been generally acknowledged as a distinct area of expertise. In the 1980s and 1990s the number of cost and managerial accounting advances has been above than 1960s and 1970s, and it is clear that there has been an explosion of managerial accounting advances in the literature over the past 20 years.

In fact, nothing new has been discovered in management accounting over the past 50 years. In the last 50 years, certain roles have remained untouched, but now they are said to have switched from practical to technical obligation. The trend really is to make management accounting work more competent. Management styles evolved considerably in the last quarter of the 20th century. The challenges such as the climate or the pace of technologies have changed. The ability to decide easily and to be an effective collaborator has also fundamentally improved. Management was once a system of order and control where basically each decision was taken at the top. There is more decentralized reporting and decision-making now. Businesses face various decisions and obstacles today, and they need managerial accountants to enter and work at any level of a squad.

## References

Ahmad, K. (2014). The Adoption of Management Accounting Practices in Malaysian Small and Medium-sized Enterprises. *Asian Social Science*, *10*(2), 236.

Ashton, D. J., Hopper, T., & Scapens, R. W. (1995). The Changing Nature of Issues in Management Accounting. In *Issues in Management Accounting*, pp. 1–20. Hertfordshire: Prentice Hall.

Askarany, D., Smith, M., & Yazdifar, H. (2007). Technological Innovations, Activity Based Costing and Satisfaction. *Journal of Accounting-Business & Management*, *14*, 53–63.

Boonyanet, W., & Komaratat, D. (2008). Diversification of Management Accounting Practices in the Thai Listed Companies. *Chulalongkorn Business Review*, *30*(1–2), 116–134.

Caplan, D. (2006). Management Accounting Concepts and Techniques. *Accounting and Law Faculty Books*, *1*. Retrived from http://scholarsarchive. library.albany.edu/accounting_fac_books/1

Demir, V. (2008). Yonetim muhasebesindeki degisim ve degisimi etkileyen faktorler. *Muhasebe ve Denetime Bakis*, *8*(26), 51–70.

Garner, S. P. (1947). Historical Development of Cost Accounting. *The Accounting Review, 22*(4), 385–389.

Gucenme, U., & Poroy Arsoy, A. (2006). Turkiye'de cumhuriyet doneminde muhasebe egitimi. *Mali Cozum*, *76,* 308–327.

Guvemli, O., & Guvemli, B. (2015). Turk ticaret kanunlarının Turk muhasebe dusuncesinin gelismesindeki etkileri. *Muftav Dergisi*, *8*, 26–50.

Hagerty, M. R. (1997). A Powerful Tool for Diagnosis and Strategy. *Journal of Management Consulting*, *9*, 16–25.

Horngren, C. T. (1975). Management Accounting: Where Are We?. In *Management Accounting and Control*, University of Wisconsin, Madison, 9–20.

IFAC. (1998). *International Management Accounting Practice Statement: Management Accounting Concepts*. New York.

Islam, M., & Kantor, J. (2005). The Development of Quality Management Accounting Practices in China. *Managerial Auditing Journal*, *20*(7), 707–724.

Johnson, H. T., & Kaplan, R. S. (1987). The Rise and Fall of Management Accounting. *Management Accounting, 68*(7), 22–39.

Kader, M. A., & Luther, R. (2004, October). *An Empirical Investigation of the Evolution of Management Accounting Practices*. WP No. 04/06.

Kamal, S. (2015). Historical Evolution of Management Accounting. *The Cost and Management*, *43*(4), 12–19.

Kaplan, R. S. (1984). The Evolution of Management Accounting. *The Accounting Review*, *59*(3), 390–418

Kudo, E., & KUDO, E. (2015). Accounting Knowledge and Merchant Education in Japan: An Historical and Comparative Study. 西南学院大学商学論集, *62*(2), 1–28.

Lee, A. S. (2018). *Exploring the History and Trends of Accounting in Canada and the United States*. Thesis, University at Albany, State University of New York.

Martin, J. R. (2006). *Evolution of Management Accounting Graphic*. Retrived 12.01.2021 from https://maaw.info/EvolutionOfMAGraphic.htm

Smith, M. (1999). *Management Accounting for Competitive Advantage*. LBC Information Services.

Sokolov, A., & Bikmukhametova, C. (2015). Cost Accounting in Russia-Historical aspects. *Asian Social Science*, *11*(11), 385–390.

Waweru, N. M., Hoque, Z., & Uliana, E. (2004). Management Accounting Change in South Africa: Case Studies from Retail Services. *Accounting, Auditing & Accountability Journal, 17*(5), 675–704.

Alexandru Trifu

# Flexibility of Labor Markets Supporting the Development

## 1. Introduction

We want to analyze the manifestations and actions that are produced rapidly in the labor market (labor force), and we will try to shed some light on the most rigid market and how it becomes increasingly flexible. We find the term *flexibility* specifically related to the labor market. *The labor market flexibility* refers to the willingness and ability of the workforce to respond to changes in the market conditions, including changes in the labor demand and the wage rate (the amount of basic wages paid to a worker per unit of time/hours or per day, or per unit of production if they work on a single piece flow basis). The labor market flexibility is an important aspect of how labor markets work to adjust supply to the demand. The labor market flexibility is central to the supply side of the macroeconomy and its overall performance in achieving macroeconomic objectives (www.economicsonline.co.uk, 2020).

In practice, employers should consider several aspects that justify the implementation of flexible approaches, such as (Dinu, 2020):

a.  the flexibility must be introduced for a reason, namely for the employer to identify those business problems to be solved;
b.  the establishment through internal policies of the forms and limits of flexibility, as well as the defining of the minimum requirements that are not negotiable and that the employer must implement so that, in the end, the benefits of flexibility are mutual;
c.  the establishment of the functions that will be the decision-makers for the implementation of forms of flexibility – these functions should be those that closely manage the activity of the employees and that can decide, in real terms, whether a certain form of flexibility is compatible with that activity or not;
d.  it is important that employees are informed about all the aspects and implications of a flexible formula that they wish to access;
e.  last but not least, it is necessary to document the implementation of the flexibility form applicable for each individual employee; in the absence of documentation or in case of poor documentation of the way in which the employee and the employer must collaborate and of the applicable rules during the implementation of the respective form of flexibility, real issues can generate in their relationship.

A variety of work flexibility is the so-called *on call* (on request). Although not regulated in certain legislations, this type of arrangement is often used both in the IT field and also in the field of service provision (electricity, gas, communications) in order to resolve problems/breakdowns that may occur at any time (both during the day and the night) and which require the urgent intervention of the employee.

Another approach to work flexibility is the mobility of the employees, which can be an attractive factor for them in making the decision to join a company. The flexibility through mobility offers employees the opportunity to transition to the workplace internally, within the same company, and apply for positions that they consider more attractive and more compatible with the training acquired, and also externally, through posting to other countries and even the definitive transition to a job abroad. We also must follow the interests of entrepreneurs/CEOs, who want well-trained employees, that is, people with adequate skills and knowledge, which are stable and which, through these qualities and the derived productivity, also ensure the development of the company/corporation and, through training, the development of the related economy. The characteristics of the goods that circulate on this market, the labor force, and the relationships between the economic agents that participate in the exchange relations provide the labor market a series of particularities (see also N. Dobrotă et al., 1995).

a. *First of all*, it has the characteristic of being the most **rigid** and for this reason, the most **sensitive** and fragile. Not only does it influence but also in many situations, it conditions the economic and social balance and the correspondence between the equilibrium level of goods and services and the level of full employment, levels that a natural mechanism does not coincide. The society cannot accept that some of its members do not have a permanent job and an appropriate income.

b. *Second*, the labor market is **the most organized and regulated market**. It involves an institutionalized framework within the limits of the law, where the demand is adjusted with the labor supply, strategies and policies for employment and use of labor are drafted, and the social partners (the union, employers, the state) play precisely circumscribed roles.

c. *Third*, from the viewpoint of the functioning mechanisms, the labor market is an imperfect market and this is because:

- *the pure market mechanisms (salary, marginal productivity, competition) operate in a framework regulated and accepted by economic agents;*
- *the salary ceased to be the only lever for regulating the volume of employment and the efficient use of work.* The salary and working conditions – the general atmosphere of activity becomes essential for individual results, as well as for the profitability and solvency of economic entities that evolve on specific markets.

This was the case regarding the labor market during the first decade after the collapse of the socialist system and the political economy in general.

As a summary, we can state that *the flexibility is specific to the labor market*. In the other cases we can talk about sensitivity, adaptability, and quick responses to the challenges of the economic and natural environment. Thus, a key element of the labor market flexibility is the flexibility to adjust wages to balance supply and demand. There are several types of wage flexibility, including the relative wage flexibility – which deals with the adjustment of wage rates across sectors of an economy or across regions and the real wage flexibility, the flexibility of real wages (nominal wages adjusted for inflation) to adjust to economic shocks.

Also referring to skills and training, the multi-skilled workers may be able to adjust their work patterns or workload to suit changing demand conditions. The training and training subsidies can similarly improve the labor mobility. Concerning the flexibility from the firm's point of view, a flexible labor market means that firms will have more freedom to hire workers when the demand increases and also to fire them when the demand decreases. If the workforce is better informed about job vacancies or promotion opportunities, workers can respond more effectively to changes in the firms' demands. The labor market is more flexible when there is a higher proportion of part-time work compared to full-time work. Flexibility is also improved when temporary contracts can be used. The multiple new and interconnected global crises, including inflation (especially in energy and food prices), the financial turmoil, the potential debt crises, and the global supply chain disruption – exacerbated by the war in Ukraine – mean that there is an increasing risk of further deterioration of hours worked in 2022, as well as a wider impact on global labor markets in the coming months (www.ilo.org/news/2022).

A large and widening divergence between richer and poorer economies continues to characterize the recovery. While high-income countries saw a rebound in hours worked, the low- and lower-middle-income economies suffered setbacks during the first quarter of the year, with a difference of 3.6 % and 5.7 %, compared to the reference value before the crisis. These divergent trends are likely to worsen in the second quarter of the current year. In some developing countries, the governments are increasingly constrained by a lack of fiscal space and the debt sustainability challenges, while businesses face economic and financial uncertainties and workers continue to lack sufficient access to social protection. More than two years after the start of the pandemic, many people in the labor market are still suffering from the impact:

a. The labor income has not yet recovered for most workers;
b. The gender gap in the hours worked also increased during the pandemic;
c. The sudden increase in job vacancies in advanced economies in the late 2021 and in the early 2022 led to a tightening of labor markets, with a growing number of available jobs relative to job seekers;
d. Driven by disruptions in production and trade, exacerbated by the crisis in Ukraine.

## 2. Methodology

In this chapter, we first conducted a review of the existing literature and tactics on the HR function and the interpersonal relations in the workplace, while planning to conduct a qualitative analysis of those working in the field about their actual experiences. Thanks to the analysis of the specialized literature and the experiences in this field, including our own experience, we offer a discussion about positive and negative techniques in the work force and their causes, the impact of new methods and activities at the workplace, and some good practices for professionals and managers in human resources so that they can effectively prevent any negative incidents to take place in their departments and organizations. We used *the synthesis* to separate the current trends and characteristics of the labor market, which is vital for the functioning of entities, their development and the national economies, and finding the most appropriate ways to reach *a modus in rebus*, both for entrepreneurs/CEOs and for workers.

## 3. Trends on the Labor Market as Support for Development

The most radical and continuous change that people experience is the so-called "Great Resignation". According to surveys, 44 % of employees are "job seekers". This means that almost half of employees are looking for a new job or they plan to do so soon. Unemployment rates skyrocketed at the start of the pandemic when many businesses closed. Seemingly arbitrarily, many people found themselves out of work, depending on whether they were declared "essential" workers or "unessential" workers. Higher wages, inflation, and more flexible work arrangements are all reasons why people are looking for new jobs today. There is no sign of this slowing down – 4.3 million people left their jobs in January. In 2021, nearly 48 million people quit their jobs. The Great Resignation can be considered the result of many factors converging and manifesting in the global workforce. Because it shows no signs of stopping, the Great Resignation may just become the new standard of workers leaving their jobs whenever they feel like it.

The Great Resignation is determined by the fact that employees want more freedom and independence. Many jobs have been cut during the pandemic, and workers now expect that to continue. The rise of flexible working hours means that entrepreneurial employees can build a business and start generating cash flow or take up a side business to earn extra income. Building a business is a great way for workers to give up the 9-to-5 interval and avoid the dreaded return to the office. Returning to the office can mean less quality time at home, more time commuting, and more time enriching others. Workers want to avoid spending hours in the car when they could be spending quality time with their friends and family. For employers, the writing is on the wall. To retain talent, they will need to find ways to give employees more freedom and independence.

Another factor changing the landscape is the increased automation and the wider adoption of technology. The automation and technology adoption does not only mean that more people are using digital calendars. It also means that employees use artificial intelligence, neural networks, and web micro-services to help them do their jobs. The emergence of the Internet has been called the information age, and today, more than ever, this has become true.

Today's workers need to understand the data and technologies available to leverage it for a business. As cloud computing and remote work become more common, these technologies will become more important for employees to understand (Bocetta, 2022). *The cloud computing* is the on-demand availability of computer system resources, especially the data storage (storage in cloud) and computing power, without active management directly by use. The relationship between what is known as human development and the economic development is a two-way relationship because each reflects negatively and positively on the other, that the economic growth occurs through the improvement of human capabilities and the achievement of the desired growth reflected in the development because this expands the options related to the human resources, especially for individuals in general (Ahmad Omar, 2020). The relationship can be extrapolated beyond the example in the analyzed document.

*The human capital* development is one of the most important factors impacting the economic growth in the national economy or society. The formation of the human capital depends not only on education and training but also on the amount of health and social services that work to build and maintain the human capital, as health and education are the main elements that determine the composition of production and export growth. The higher the level of education attained by the labor force, the higher the total productivity, because the more educated, the more likely they are to innovate and thus impact the overall productivity. In order to demonstrate that, first of all, the development of the employee/person employed or who is looking for a job must be ensured, we want to relate to an aspect that should not be neglected and which is also the substance of the post-graduate activities of the Alumni type: the compatibility rate between the nature of the job held and the basic training of the employee or candidate. There are frequent cases in which some jobs are occupied by workers who have other basic training, but who, through their knowledge, through their ability to adapt and learn, were employed. In the 15–34 age groups, according to some experimental statistics published by Eurostat, we are not doing well in this chapter. People who work in a different field from an educational viewpoint are considered to be people with a horizontal skills mismatch. (Popa, 2022).

In this regard, we believe that the most important thing for one's personal development and, subsequently, for the economic development is that *of a higher matching rate between the desired/practiced job and the training/education of the employee/applicant*. If we talk about the impact of the labor force on the general development of an economy, we found it suggestive to briefly present the example of Switzerland in this regard.

Switzerland is not only a developed country but also rich. *The wealth* measures the value of all valuable assets held by an individual, community, company, or country. The wealth is determined by taking the total market value of all physical and intangible assets owned, then subtracting all liabilities. At its core, wealth is the accumulation of limited resources (www.investopedia.com/terms). Switzerland ranks 3rd place in the list of the richest countries in the world, with a GDP of $93,520/inhabitant. It is no surprise to anyone, due to the fact that Swiss banks are the best in the whole world (Salins, Urban, 2022) But the wealth of this country is not only limited to the banking system; it must be emphasized that Switzerland is a member of EFTA.

The Swiss are also exceptional manufacturers; thus, the products such as watches, Swiss army knives, musical instruments, and chocolates manufactured by them are extremely famous. At the same time, their pharmaceutical and chemical industries are recognized worldwide. And last but not least, the tourism, especially the winter tourism, has a substantial contribution to the gross domestic product (Bercea, 2022). And in the issue that interests us is related to the facts that, from the total population, almost a quarter are immigrants.

On December 16, 2016, the Swiss Parliament adopted the Federal Act on Foreign Nationals implementing the result of the 2014 referendum in a manner that limited its effect, paving the way for the start of the normalization of the EU-Swiss relations. The Swiss government emphasized that Switzerland and the EU will still continue to be close partners and suggested the establishment of a structured dialogue to discuss common priorities for further cooperation (Lopez & Rakstelyte, 2021).

The Swiss work hard, and they are punctual and efficient. Through the measures related to the Federal Government, even taking into account the well-known neutrality of the Swiss, they are trying to find solutions so that the local workforce, also attracted by the migration phenomenon, can ensure the development of the economy – the exploitation of all natural, material, and human resources to create the necessary well-being. But the issue we need to highlight is that of the continued functioning of national economies. In this equation, the workforce has a decisive role. The economic-social situation we are in now is complex and worrying, not only because of the war in Ukraine and the gloomy forecasts regarding the supply of gas and oil for the coming winter but also because of the global warming which causes unexpected and intense destructive phenomena. Economists warn about the imminent entry into recession.

*A recession* is broadly defined as a severe, widespread, and prolonged decline in the economic activity, but a general rule of thumb is two consecutive quarters of decline in Gross Domestic Product (GDP) is the measure of a region's or a country's economic output. Related to our problem there is the need *to manage the inflation*.

Two U.S. jobs–related reports for July and August 2022 showed the country added 528,000 and 315,000 jobs indicating a still robust labor market despite the pressures. But as this is a figure that may lag behind a decline in the GDP, all eyes will be on the reports in the coming months for signs of trouble. Moving on to Europe, the big fear is for the coming quarters where the fiscal measures to manage inflation and the strangled gas supplies from Russia could impact the European economy. Another component that could lead our society into a recession is the constant talk about it that could lead to lower consumer confidence, as well as the gloomy predictions and the requests of the European states, especially, to reduce the consumption of electricity and gas (Ni Chulain, 2022).

## 4. Conclusions

The workers are the ones making the demands today. Surveys show that employees have more bargaining power than they have had before. With more and more people demanding flexible work environments and better wages, with the automation and technology dominating almost every work field, it's no wonder that employees are changing jobs. As the Boomers continue to retire, more changes are to be expected because the Millennials and the Zoomers become the dominant workforce in countries such as US.

A particular aspect to be followed up and which was already dealt with is related to ensuring the best possible compatibility rate between the job chosen, or the field in which the activity is carried out, and the basic training or knowledge that the person in question has. The developed countries are the countries where the people are considered to be "happy" there is a very high matching degree between the nature of the job and the training of the person in question.

various motivations ↓

Appropriate knowledge and skills  +  stability at work − − − − → a sound, profitable, reliable entity

climate of understanding and support ↑

We believe that in order to ensure the personal development of the employees and to also bring new value to the national/regional economy level, thorough varied knowledge, skills and the ability to communicate and understand, and jobs with longer stability are needed (something that is also up to the employer), so as to reach a beneficial consensus both for the employee and for the firm/company.

## References

Ahmad Omar, Dina. (2020). *Inter-Relationship between Economic Development and Human Development-Analytical Study of Selected Arab Countries*. Universidad del Zulia. www.redalyc.org/journal/, Retrieved September 4, 2022.

Bercea, Andrei. (2022). *(…) Care sunt cele may bogate țări (Which are the richest countries)-update 2021*. www.financer.com/ro/blog/, Retrieved September 4, 2022.

Bocetta, Sam. (2022). *7 Workforce Trends Workers Can Expect in 2022*. www.fee.org/articles/, Retrieved September 5, 2022.

Dinu, Gabriela. (2020). *Flexibilitatea pieței muncii, de la realitate la necessitate (Labor Market Flexibility, from Reality to Necessity)*. www.hrmanageronline.ro, Retrieved August 31, 2022.

Dobrotă, Niță et al. (1995). *Economie politică (Political Economy)*. București: Ed. Economică.

Lopez, M. A., & Rakstelyte, Ausra. (2021) *Spațiul Economic European (SEE), Elveția și Nordul*. www.europarl.europa.eu/facts/, Retrieved September 3, 2022.

Ni Chulain, Aisling. (2022). (…) *Economists Are Warning the Next Downturn Is Looming*. www.euronews.com, Retrieved August 31, 2022.

Popa, Dan (2022) *România este țara cu cea mai ridicată a nepotrivirii jobului cu domeniul în care ai studiat (Romania is the country with the highest rate of job mismatch with the field in which you studied)*, www.economie.hotnews.ro retrieved September 10th, 2022.

 www.economicsonline.co.uk, Retrieved August 28, 2022.

Salins, Veronique, & Urban, Sila. (2022). *Fostering a Strong Labor Market to Support the Recovery and Sustain Growth in Switzerland*. OECD Economics Department Working Papers no 1720. www.oecd.org, Retrieved September 2, 2022.

Salins, Veronique, & Urban, Sila. (2022). *ILO: Labor Market Recovery Goes into Reverse*. www.ilo.org/news/, Retrieved September 1, 2022.

Salins, Veronique, & Urban, Sila. www.investopedia.com/terms/, Retrieved September 1, 2022.

Luís Barbosa

# Society 5.0 – Exploring the Concept of Nature Citizenship in a Human-Centered Society Powered by Technology

## 1. Introduction

In terms of development, we are experiencing the 4th Industrial Revolution, also known as the digital age. According to Klaus Schwab (2015), there are three reasons why today's changes are not just an extension of the 3rd Industrial Revolution, but rather the arrival of a fourth and independent one: speed, scale, and system impact. The speed of current breakthroughs has no historical precedent. Compared to previous industrial revolutions, the 4th Industrial Revolution is evolving exponentially rather than linearly. Moreover, it is transforming nearly every industry in every country. And the breadth and depth of these changes herald the transformation of entire systems of production, management, and governance. The possibilities of billions of people connected via mobile devices with unprecedented processing power, storage capacity, and access to knowledge are limitless (Schwab, 2018).

The changes are so real and drastic for today's and tomorrow's society that "Peoples Living Lab—A Laboratory for Future Society" is the theme of the 2025 World Expo, which will be held in Osaka, Japan. The theme and concept are anchored in the Society 5.0 project launched by the Japanese government in 2017. This is a human-centered project that aims to "merge physical and digital spaces" with the ultimate goal of creating a super smart society where "life is more meaningful and enjoyable" (Medina-Borja, 2017, p. 235). According to the vision of the project's creators, "the interaction between people and technology should be used to create a sustainable, vibrant, and livable world where people are at the centre" (Medina-Borja, 2017, p. 235). Conceptually, the idea points out the way to a better future as it seeks to solve important social challenges.

In addition, as a global society, we must also solve important environmental problems. Humanity's impact on the Planet has never been more damaging than it is today, despite countless initiatives to reduce our negative footprint. These impacts have sparked discussion of a new geological age: the Anthropocene. The Anthropocene defines Earth's most recent geologic period as human-influenced or anthropogenic, based on overwhelming global evidence that atmospheric, geologic, hydrologic, biospheric, and other Earth system processes are now being altered by humans. Along with other scientists, Nobel laureate Paul Crutzen, who coined the term, argued that human-induced changes in the Earth system were so profound and of such long duration that it was possible to speak of a new epoch in Earth history. Although other people thought along the same lines as Crutzen, he combined his fame as a Nobel laureate with foresight and conciseness. Crutzen's idea of the Anthropocene broke through the false separation of the human and natural spheres in a provocative way. Crutzen pioneered a new view of nature and of ourselves (Hoffman & Jennings, 2015; Schwaegerl, 2021; Mendes, 2020).

Since 2019, we have been working on the concept of nature citizenship. In particular, we advocate that natural ecosystems must be protected by each country's Constitution and have the right to a legal identity or citizenship. We also advocate that every human being must become a guardian of nature and reconnect with natural ecosystems.

As humans and technology become more inseparable, we believe it is important to learn more about the Society 5.0 (S5.0) project on an academic level to understand how it envisions nature and how the concept of natural citizenship might fit into this project.

## 2. Society 5.0

According to the Japanese government, Society 5.0 is a people-centered society that balances economic progress with solving social problems through a system that strongly integrates cyberspace and physical space. Society 5.0 was proposed in the 5th Science and Technology Basic Plan[1] as a future society that Japan should strive for. It follows the hunting society (Society 1.0), the agricultural society (Society 2.0), the industrial society (Society 3.0), and the information society (Society 4.0) (Government) as contextualized in Figure 1:

---

1    Report on "The 5th Science and Technology Basic Plan Council for Science, Technology and Innovation", Cabinet Office, Government of Japan (December 18, 2015). Retrieved from: https://www8.cao.go.jp/cstp/kihonkeikaku/5basicplan_en.pdf. Last accessed 20.09.2022.

| | Society 1.0 | Society 2.0 | Society 3.0 | Society 4.0 | Society 5.0 |
|---|---|---|---|---|---|
| Society | Hunter-gatherer | Agrarian | Industrial | Information | Super smart |
| Productive approach | Capture/Gather | Manufacture | Mechanization | ICT | Merging of cyberspace and physical space |
| Material | Stone・Soil | Metal | Plastic | Semiconductor | Material 5.0* |
| Transport | Foot | Ox, horse | Motor car, boat, plane | Multimobility | Autonomous driving |
| Form of settlement | Nomadic, small settlement | Fortified city | Linear (industrial) city | Network city | Autonomous decentralized city |
| City ideals | Viability | Defensiveness | Functionality | Profitability | Humanity |

**Figure 1.** Contextualizing Society 5.0. categories

**Source:** Produced by authors. Research conducted by the University of Tokyo's Material Innovation Research Center

Proponents defend that the social reforms (innovation) in Society 5.0 will create a forward-looking society that overcomes the existing sense of stagnation, a society whose members respect each other, across generations, and a society in which every individual can lead an active and enjoyable life (Yano, Dai, Masuda, & Kishimoto, 2020).

Operationally, the big step is that in the previous information society, the common practice was to collect information via the network and have people analyze it. However, in Society 5.0, people, things, and systems are all connected in cyberspace, and optimal results achieved through AI that exceeds the capabilities of humans are fed back into physical space. This process brings new value to industry and society in a way that was not possible before (Deguchi et al., 2020).

Industry 4.0[2] calls for an industrial revolution that focuses on manufacturing, but says nothing about how such a revolution can affect the general public

---

2   Industrie 4.0 was a national strategic initiative led by the Ministry of Education and Research (BMBF) and the Ministry for Economic Affairs and Energy (BMWI). To deliberate on the initiative, a working group was formed consisting of actors from government as well as from businesses and universities. The working group was led by Henning Kagermann, the former chairman of SAP SE and president of the German (Kagerman, 2013). In contrast, as the concept of a people-centered society shows, Society 5.0 focuses heavily on the public impact of technology and the need to create a better society. The vision of Society 5.0 encompasses a course of reform to produce an inclusive society that meets diverse needs and preferences. This important differentiating aspect of Society 5.0 was mentioned in an address by late Prime Minister Shinzo Abe to German Chancellor Angela Merkel during the 2017 CeBIT conference in Hannover. After hearing Abe's statements about Society 5.0, Merkel expressed her strong support for this vision (Deguchi et al., 2020).

The project will promote economic development and the solution to social problems. Society 5.0 will also bring economic benefits to individuals by "providing the necessary goods and services to the people who need them, at the right time and in the right quantity". Society 5.0 will "facilitate human prosperity" (Government)—supported by improved legal regulations and education that enable dynamic engagement of all citizens in the new economy and society made possible by new technologies (Kendairen, 2016).

The realization of such a society will not be without difficulties, however, and Japan intends to face them head on, aiming to be the first country in the world to rise to the challenges in order to present a model for the future society (Government).

Despite the challenges, Society 5.0 will produce a human-centered society, a society centered on each individual human being rather than a future controlled and monitored by AI and robots (Government).

Society 5.0 will be increasingly diverse (Government); but this diversity does not only include the inclusion of people from different backgrounds. Thanks to its profoundly technologized, cyber-physical nature, Society 5.0 will be able to incorporate into its social structures and dynamics beings not previously found in the world's societies. (Gladden, 2019).

---

Academy of Science and Engineering (ACATECH). In April 2013, the working group issued its recommendations in a report titled "Recommendations for implementing the strategic initiative INDUSTRIE 4.0". As a proper noun, Industrie 4.0 denotes a uniquely German initiative, but the underlying concept—to deploy IoT in manufacturing—has gained global traction. Retrieved from: Recommendations for implementing the strategic initiative INDUSTRIE 4.0, Securing the future of German manufacturing industry, Final report of the Industrie 4.0 Working Group, April 2013. Last accessed 20.09.2022.

Although Society 5.0 originated in Japan, it does not only serve the prosperity of a single country (Fukuyama, 2018). So far, the project has grown and it is consistent enough to be the main theme of a world exhibition.

## 3. Nature Citizenship

In examining major development concepts from the perspective of environmentally sustainable development—including the Anthropocene, biomimicry, ecology, ecosophy, plant revolution, etc. and major documents such as the Paris Agreement on Climate Change, the Earth Charter, and the proposed content for the United Nations Declaration of Earth Rights, among others—we have found that most ongoing actions focus on climate change and that conservation programs tend to be in the legal realm, where rare sanctions are imposed after serious damage to nature and natural ecosystems has been committed.

Moreover, after studying several examples, one can see that the mere designation of "Protected Natural Area" (including World Heritage Sites) of a river, a forest, a coral reef, etc. is not sufficient to protect these ecosystems from harmful human interference (Barbosa & Bogalheiro, 2021).

The debate about whether nature has rights was sparked by Professor Christopher D. Stone. Originally published in 1972, "Should trees have standing?" was an essential tool for the then burgeoning environmental movement and sparked a worldwide debate about the legal rights of nature, aided by a recent US Supreme Court decision in Sierra Club v Morton (Tănăsescu, 2022). Central to Stone's book is the compelling argument that the environment should be granted legal rights (Stone, 2010). The movement has grown, including new variants such as spiritual. In terms of visibility of the issue, one of the most important milestones was the fact that the people of Ecuador voted in 2008 to include the rights of nature (Pacha Mama) in the country's Constitution (Tănăsescu, 2022; Barbosa & Bogalheiro, 2021). Other countries, such as Bolivia and Panama, have followed suit in subsequent years. On the other hand, some cultures, especially indigenous cultures and countries like Bhutan, show that this vision is not new in human history. In this remote Asian country, culturally, there is a holistic relationship between citizens and nature so much that the Constitution provides in its Article 5 that every citizen is a custodian of the kingdom's natural resources and environment (Barbosa & Bogalheiro, 2021).

In recent years, the recognition of rights to natural systems has increased significantly For instance, in New Zealand, where the Whanganui River became the first watercourse to be recognized as a legal entity in 2017[3], the "Mar Menor" in Spain (in 2022)[4] was recognized as the first ecosystem with legal personality in Europe.

Nevertheless, we understand that there is a lack of comprehensive and global action that effectively protects nature. A new theoretical framework needs to be considered – the concept of citizenship applied to nature – and new mechanisms to motivate its implementation by global actors.

Our vision is that this is the right impulse to bring together the opportunities created by the new digital networks for the benefit of nature.

This study is part of a broader study on Nature Citizenship, a conceptual approach we have been working on since 2019. Nature Citizenship advocates giving natural systems an identity and granting them legal status. It calls for the inclusion of nature citizenship in the constitutions of countries. It advocates that a citizen can constitutionally be a custodian of natural systems: collects and publicizes natural systems that have already been granted natural rights by law; promotes the reunification of humans and nature. It is based on the concepts of regeneration and the new ecology. It differs from environmental citizenship because the latter refers to education for respect for the environment.

How will it be implemented? It will be by creating a technological hub/digital network (possibly using blockchain technology) that includes SSIs[5] and using this hub as a communication, interaction, and dissemination point for Nature Citizenship.

So, in addition to promoting the holistic vision between people and nature, our aim is to work toward the incorporation of the rights of nature into national and regional law in all countries. In parallel, the main purpose of creating a digital network is to establish a global communication channel to disseminate the concept and support-related initiatives.

---

3   How New Zealand's Whanganui River is legally a person. Retrieved from: https://edition.cnn.com/2020/12/11/asia/whanganui-river-new-zealand-intl-hnk-dst/index.html#:~:text=In%202017%2C%20that%20fight%20finally,to%20speak%20on%20its%20behalf. Last accessed 01.10.2022.

4   Spain makes history by giving personhood status to salt-water lagoon, thanks to 600,000 citizens. Retrieved from: https://www.euronews.com/green/2022/09/22/spain-gives-personhood-status-to-mar-menor-salt-water-lagoon-in-european-first. Last accessed 01.10.2022.

5   SSI = Self-Sovereign Identity.

## 4. Data Analyses and Results

To understand how the creators of S5.0 envision nature in the context of the project, we have studied the full report with particular attention to its concept:

*Society 5.0 Will Bring About a Human-centered Society*

*In society up to now, a priority has generally been placed on social, economic, and organizational systems with the result that gaps have arisen in products and services that individuals receive based on individual abilities and other reasons. In contrast, Society 5.0 achieves advanced convergence between cyberspace and physical space, enabling AI-based on big data and robots to perform or support as an agent the work and adjustments that humans have done up to now. This frees humans from everyday cumbersome work and tasks that they are not particularly good at, and through the creation of new value, it enables the provision of only those products and services that are needed to the people that need them at the time they are needed, thereby optimizing the entire social and organizational system.*

*This is a society centered on each and every person and not a future controlled and monitored by AI and robots.*

*Achieving Society 5.0 with these attributes would enable not just Japan but the world as well to realize economic development while solving key social problems. It would also contribute to meeting the Sustainable Development Goals (SDGs) established by the United Nations.*

*Japan aims to become the first country in the world to achieve a human-centered society (Society 5.0) in which anyone can enjoy a high quality of life full of vigour. It intends to accomplish this by incorporating advanced technologies in diverse industries and social activities and fostering innovation to create new value.* (Government)

The data were retrieved from the internet *corpus latente* in English, with the Government Report[6] as the main source. Using qualitative methodology, after reading and observation, we applied the technique of content analysis. This technique allowed us to create the following "cloud" highlighting the key words of the concept, as presented in Figure 2:

---

6   *Op. Cit.*

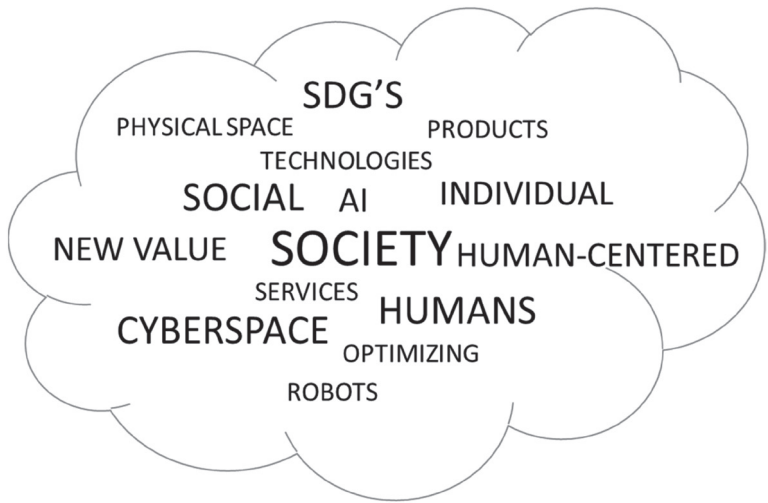

**Figure 2.** Keywords "cloud"

The results confirm that the focus is on society and social issues. The results allow to align with the opinion of Atsushi Deguchi and Kaori Karasawa (2020) that the literature on Society 5.0 is full of references to humanity, people, and individuals. For example, there are frequent expressions such as "more humanity", "respect for human dignity", a "human-centred society", "people-friendly", "more freedom for individuals", and a society tailored to "diverse human preferences". These phrases suggest that the happiness of individuals is central to the design of environments and institutions and that Society 5.0 must be designed to achieve this goal. The free and effective use of information, coupled with innovation in the design of environments and institutions, will liberate individuals from the constraints that prevent them from leading better lives. Once freed from these constraints, individuals can fulfil their desires and needs without undermining the sustainable development of society as a whole. Such a society is a happy society because the people in it are not only physically healthy but also mentally healthy when they lead a satisfying and meaningful daily life. This is the people-centered vision that Society 5.0 champions (Deguchi & Karasawa, 2020).

In terms of technology, S5.0 is not the logical extension of today's society; Society 5.0 is a revolutionary break with prevailing ideas and practices (Deguchi et al., 2020). The "cyber-physical convergence concept can thus help create a resource-efficient society. Cyber-models that meticulously replicate the real world can help us understand how best to use resources" (Matsuoka & Hirai, 2022), and this vision would definitely help mitigate or at least drastically reduce the impact of climate change.

However, as results demonstrate, nature and natural ecosystems are excluded from the keywords of the project.

## 5. Conclusion

Nature Citizenship is a comprehensive system in which society understands nature holistically; every citizen should feel and act as a guardian of nature and natural ecosystems.

In our opinion, digital technologies, including digital communication networks, are the ideal vehicle to operationalize nature citizenship by creating a digital identity, promoting the concept, and bringing governments together to incorporate nature citizenship into the national constitution.

When it comes to Society 5.0, as Atsushi Deguchi and Kaori Karasawa point out, "perhaps the most important thing is to keep in mind how technological innovation should steer society in a better direction, and to ensure that the principle of human-centered society is embedded in the minds and hearts of the actors and organizations involved in technology development and community development, as well as in the hearts and minds of engineers and every member of the public". In this context, we believe that for a truly new society, it is indispensable to include the "nature factor".

In terms of technology, Gladden (2019) has noted that Japanese attitudes to new technologies are markedly different from those of other countries, but in our perspective digital presence is now a global fact. As we envision it, a human-centered society that uses technology as an instrument for a healthier and happier life should be the ultimate purpose of each government and stakeholder. As a society, we are more and more affected by the negative human footprint on our planet. One of the goals of S5.0 is to save the planet through better use and distribution of resources, and it can become a key factor in the right direction. According to our findings, in S5.0 nature should be included in a more expressive way: as a people and nature-centered project.

World Expo 2025 will help promote the S5.0 concept and share it with the rest of the world – ideally it would take up Nature Citizenship as a new "flag".

## Acknowledgment

The chapter in this book is sponsored by CICANT & Universidade Lusófona do Porto, Porto, Portugal

# References

Barbosa, L., & Bogalheiro, M. (2021). Exploring the Path for Nature Citizenship in the Digital Era. *Global Communication Challenges (ESR)*. Nairobi: IAMCR. Retrieved from https://iamcr.org/node/16809

Deguchi, A., & Karasawa, K. (2020). Issues and Outlook. In *Society 5.0*. Singapore: Springer.

Fukuyama, M. (2018). Society 5.0: Aiming for a New Human-Centered Society. *Japan SPOTLIGHT*. Retrieved from https://www.jef.or.jp/journal/pdf/220th_Special_Article_02.pdf

Gladden, M. E. (2019). Who Will Be the Members of Society 5.0? Towards an Anthropology of Technologically Posthumanized Future Societies. *Social Sciences, 8*(5), 2–39.

Government, O. J. (n.d.). *Society 5.0*. Retrieved from Cabinet Office: https://www8.cao.go.jp/cstp/english/society5_0/index.html

Hoffman, A. J., & Jennings, P. D. (2015, March). Institutional Theory and the Natural Environment. *Organization & Environment, Special Issue: Review of the Literature on Organizations and Natural Environment: From the Past to the Future*, 8–31. Retrieved from https://www.jstor.org/stable/26164720

Kagerman, H. E. (2013). *Recommendations for Implementing the Strategic Initiative Industrie 4.0*. Frankfurt: ACATECH.

Kendairen. (2016). *Toward Realization of the New Economy and Society*. Japan Business Federation. Retrieved from https://www.keidanren.or.jp/en/policy/2016/029.html

Matsuoka, H., & Hirai, C. (2022). Habitat Innovation. *Society 5.0*, 25–42. https://www.researchgate.net/publication/341750546_Habitat_Innovation

Medina-Borja, A. (2017). Smart Human-Centered Service Systems of the Future. *Future Services & Societal Services in Society 5.0*, 235–239.

Mendes, J. (2020). O "Antropoceno" por Paul Crutzen & Eugene Stoermer. *Anthropocenica. Revista De Estudos Do Antropoceno E Ecocrítica, 1*, 113–116.

Schwab, K. (2018). *Shaping the Future of the Fourth Industrial Revolution*. London: Penguin Books.

Schwaegerl, C. (2021). The Anthropocene: Paul Crutzen's Epochal Legacy. *Anthropocene Magazine*. Retrieved from https://www.anthropocenemagazine.org/2021/02/the-anthropocene-paul-crutzens-epochal-legacy/

Stone, R. D. (2010). *Should Trees Have Standing?: Law, Morality, and the Environment.* Oxford: Oxford University Press.

Tănăsescu, M. (2022). *Understanding the Rights of Nature – A Critical Introduction*. Bielefeld: New Ecology, Transcript Verlag.

Yano, M., Dai, C., Masuda, K., & Kishimoto, Y. (2020). *Blockchain and Crypto Currency*, C. D. Makoto Yano (Ed.). Tokyo: Springer.

Adelina Milanova and Pavlinka Naydenova

# A Cultural Framework of Innovation Potential

## 1. Introduction

Defining the innovation potential is not unambiguous, and clarifying its essence in the organization is a complex process that needs in-depth and multifaceted research. The authors are of the opinion that the innovation potential of the organization is understood as a system of factors and conditions necessary for the implementation of innovation activity, and more specifically, as a possible but not realized ability to achieve innovation goals, consistent with both demand and with the resources of the organization (intellectual, material-technical, human, financial), situated in an appropriate infrastructure. The innovation potential of any organization depends on the specifics and scale of its activity, and the degree of use of the potential determines the organization's receptivity to innovation.

The innovation potential can be judged by the possibility of the organization to develop innovatively on the basis of new knowledge – the result of scientific research activity, with the aim of revealing its own competitive advantages. The concept of innovation potential corresponds to the concept of innovativeness, since the innovativeness of the organization is measured by the possibility of implementing technologically new or significantly improved products, processes, or combinations of products and processes, including organizational, marketing, and social.

Innovation should not be considered as an independent and one-time event, but as a process deeply realized and integrated into the organization's management system; therefore, innovation capability reflects basic aspects of management activity aimed at the distribution of strategic resources and the realization of knowledge that transforms it into a measure of the organization's competence to respond adequately to changes in the environment.

A key factor for realizing the innovation potential is the presence of an effective management mechanism, which is based on a certain culture of the organization. The characteristics of the two concepts – culture and innovation, confront them with a particular kind of contradiction: on one hand, culture is conservative, expressing traditional understandings and perceptions layered over the years, and on the other hand, innovation is an expression of change, bearers of transformations through innovations. The linking of these two concepts through categories from the scientific toolkit of social economic anthropology reflects the vision of the authors that the person with his value system realizes the mutual connection and mutual influence between them.

## 2. A Novel Vision for Analyzing Innovation Potential

According to the authors, the innovative potential of the organization is formed, developed, and realized as a result of human activities, attitudes, knowledge, and experience; it is a complex and multi-layered phenomenon, a reflection of the characteristics and specifics of human capital, whose carriers are the exponents of a certain value system, and it is manifested through the culture established in the organization.

The relevance of the current research problem is dictated by the understanding of the complex causal relationships that determine the behavior of individuals involved in innovation activity of the deep (invisible) roots of visible external manifestations of receptivity or resistance to changes.

The adequate cultural environment for the generation and implementation of innovation activity is identified with creativity and entrepreneurship, with a desire to take measurable risk, readiness for change through "long life learning", and the corresponding mobility. The dominant culture in the organization must contain those principles (rules or norms) on the basis of which the ability to successfully innovate is developed and uniqueness in relation to competitors in innovation activity is preserved. If innovation initiatives are not tailored to the specifics of the organization's culture, then the efforts made are doomed to failure or the innovation potential inherent in the organization is not being used.

Studying the conditions and incentives for the formation of innovative entrepreneurial attitudes can be extremely useful as a basis for creating the future uniqueness and competitiveness of the organization.

In principle, the manifestations of innovative thinking and behavior that stimulate the innovation potential in the organization could be systematized as follows (Rogers, 2003; Feldman, 2004; Lopesa, 2016):

- Institutionalization of changes (innovations) – establishment of rules and norms to provoke and support creative/innovative activity in the culture of the organization.
- Establishing an organizational-management policy that encourages both the search for maximum effect and allows certain risk in the process of innovations.
- Application of motivational mechanisms for continuous renewal.

- Management of the human capital, as a main factor for the manifestation of innovation potential, aimed at its optimal use.
- Formation of proactive behavior regarding changes and renewal in conditions of uncertainty/risk and uncertainty of expected results.

The policy of the organization that can provide comparative competitive advantages, according to the peculiarities of the cultural identity and the available innovation potential, as well as the practical steps that are taken, should be in the following directions:

- development of human capital – training aimed at developing initiative, promoting readiness for lifelong learning, stimulating skills for creative application, and spreading the acquired knowledge in practice;
- accumulation of intangible assets (own or acquired): patents, licenses, know-how, documentation, innovation programs, and plans for testing the results of research activities;
- getting to know good practices – striving for saturation with information about implemented innovation projects, about innovation results, and about the reasons determining the achieved success/failure or efficiency/inefficiency in their implementation.

The innovation potential of the organization is determined by the degree of its readiness to perform the tasks that lead to the achievement of foreseen innovation goals. The manifestation of the innovation potential of any organization can be stimulated by generating and managing the competences for innovation development and innovation activity taking into account the internal and external environment of the organization (Danchev, 2015; Kolev, 2017).

Some authors support the thesis that the Innovation Potential of any organization can be increased by improving and managing its competences for development and innovation activity in its field. Competences are considered a complex mastery of specific knowledge and skills, as well as flexible behavioral patterns, attitudes, and evaluative relationships. It is especially important that competent management stands out with a continuous and systematic search for appropriate guidelines and methods for the development and use of the organization's innovation potential (Naydenova, 2018; Milanova & Naydenova, 2013). The study of the circumstances that characterize the manifestation of the innovation potential highlights the strong influence of the available managerial knowledge and skills for the successful innovative development of the organization; the important role of market determinants and the need for institutional-normative support (Panteleeva, 2013).

Increasingly, along with the market requirements for optimizing innovative demand, it is necessary to implement and manage changes related to the cultural specifics of the organization, which would consolidate entrepreneurship, ingenuity, and efficiency of the innovation process. It is about changes in the stereotypes of thinking, decision-making, and their implementation, which is associated with the quality of human capital in the organization. Human capital and its effective management have long been central to advanced systems for determining the development of innovation potential. The trends are described in detail by Schultz (1961), Becker (1967), Armstrong (2004), and Tripon and Blaga (2011).

## 3. The Concept of Innovation Potential within Social Anthropological Discourse

In their previous research, the authors have always sought not only to establish facts and their subsequent analysis but, above all, to search for relevant answers to the question of why a given process or a specific phenomenon takes place and manifests itself precisely in the corresponding way, and not according to the expectations laid down in basic findings and assumptions. In this particular case, guided by the idea of seeking an in-depth explanation, the authors approach the study of some of the "invisible" motives for the behavior of human capital, placing the problem in an objective and unbiased social anthropological discourse. A specific object of research, which is the subject of the present study, is the innovation potential related to the educational sphere. Due to the more specific nature of this sphere, as well as to certain differences with the production one, our hypothesis is that the relationship between the manifestation of the national and the organizational cultural matrix is not strictly demarcated, and the focus will be precisely on the analysis of the national cultural dimensions and their manifestation in the academic environment. It has become a question (Hofstede, 2001; Milanova & Naydenova, 2013, 2021) that when talking about national culture, when identifying differences, we mainly address values, while in organizational cultures, the focus is on practices specific to each structure.

Regarding the study of the organizational/corporate culture in Bulgaria, it has been proven that the national cultural dimensions still dominate the organizational ones (Milanova & Naydenova, 2013; Milanova & Naydenova, 2017). In this sense, the specific organizational cultural dimensions that the authors adhere to in their previous studies, regardless of their clarity in defining them, are significantly dependent on what determines the national economic genotype. In parallel with this, when bringing out certain visions, the formation of human capital, as well as social capital, in particular corporate social capital, stands out Milanova & Naydenova, 2022), bearing in mind the larger relative share of the so-called exceptions to the general picture of the national cultural matrix and its extrapolation in a specific environment. This fact could lead to certain deviations in proving or rejecting some statements, but at the same time, our aspiration is to look for the deep reasons, in general, for the greater deviations found in the expected manifestation of the innovation potential in the researched area and more specifically about discrepancies in pre-set expectations. In the study, we adhere to Geert Hofstede's system of indicators and indices (Hofstede, 2001), which enables an operational comparison of the cultures of different countries in order to assess the possibilities and directions for joint activities or to borrow certain practices and policies in different spheres (https://www.hofstede-insights.com/country-comparison).

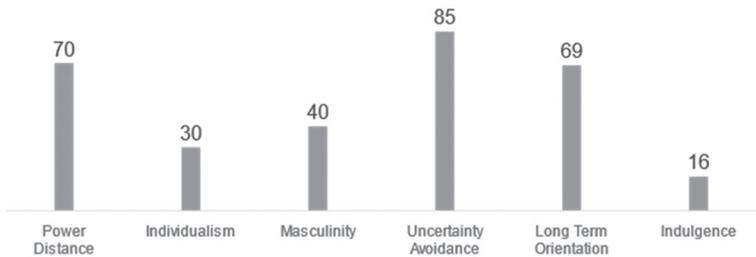

**Figure 1.**  National cultural dimensions – Bulgaria
**Source:** https://www.hofstede-insights.com/country-comparison

Based on these social-anthropological parameters (we operate mainly with the first four dimensions and respectively with their value indicators), as well as on the selected methodology, the authors derive the following hypothetical dependencies:

Innovations in this case are accepted if they are promoted by certain authorities.

The innovation potential could be deployed and developed if there is a so-called "blessing" on the imposed and perceived as indisputable authority in the relevant community, which in our case is precisely the specific educational institution.

Innovative attitudes are refracted through the values of the internal group, without these values claiming to be adequate to modern ones.

Innovation potential could not be seen as a universal asset but is refracted through the prism of the moral and sometimes ideological ideas, values, and imperatives of the inner group.

Innovative attitudes depend on the degree of ambition and assertiveness and can be strongly influenced by social emphases in some cultures, especially if possible social problems are predicted.

Innovation potential may exist, but its deployment is largely dependent on the above attitudes. It is possible that the social narrative is just an excuse for not making more of an effort, but it is definitely leading in softer cultures.

Innovative attitudes are realized as far as the corresponding threshold of tolerable stress in different cultures is concerned. Innovation itself is an object of interest in cultures that do not presume an uncharacteristic departure from the comfort zone when undertaking innovative projects that involve difficult-to-predict risk.

Innovation potential is an important factor for growth in all cultures, but in this case it is in the position of a real bet regarding the different combination of specific national cultural dimensions.

The innovation potential, the implementation of innovations, and innovation attitudes are to a significant extent dependent on the established cultural dimensions, the specific manifestation of which is tailored to the specific indices. The perception of the new, respectively, the attitudes of real risk-taking, and innovation are mainly associated with the avoidance of uncertainty but also with the fact that how much a given culture is more individualistic. (Hofstede, 2001; Minkov, 2002). Taking responsibility for carrying out certain activities, whether routine or innovative, is directly related to power distance and individualism, in their inverse proportionality. Teamwork, which is especially necessary in the realization of the available innovation potential, is addressed to individualism and the avoidance of uncertainty. Of course, one should not neglect the presence of two main focuses in the cultural matrices, namely defensiveness and independence, which, as can be seen above, were generated during the definition of the cultural profile, but at the same time, in subsequent stages, are tied to no less extent to the avoidance of uncertainty, and naturally to power distance. As already explained, foci such as ambition or social commitment are primarily related to the firmness of a given culture, so they should not be universalized, but rather perceived as additional motivating factors to realize the innovation potential in a specific organization.

Basically, economic analyses include the rational decisions of economic agents within the institutional definition. Thus, these agents /individuals are defined as economic beings who seek to develop their own benefit /satisfaction and act in a rational way. But the substantial differences in economic choice and economic development are explained not only by technological and/or organizational innovation and resource protection but also to a significant extent by the specific values of the people mentioned. The cultural conditions in which economic activity takes place are directly related to economic results – this is actually the main postulate in Max Weber's analysis (Weber, 2009).

## 4. Main Findings, Results, and Discussion

The research was carried out in two stages – the first one included the actual conducting of in-depth interviews and the construction and corresponding interviewing in the focus groups. In-depth interviews were conducted, and the main expert segments were defined: teachers in secondary and higher education – 20 teachers from higher educational institutions / universities in Sofia, and 8 teachers from secondary schools in Sofia, of which 6 are public and 2 are private. The main areas forming the thematic blocks, in the framework of which problem questions are formulated and specifying questions are asked, are defined as follows:

1. Innovations are closely related to their promotion by a relevant authority, and in this sense the innovation potential could unfold and develop if it is perceived and tolerated by the imposed authority, which can be both a certain person and a specific educational group or institution.
2. Innovative attitudes are refracted through the values of the internal group, without these values claiming to be adequate to modern ones. It is essential to understand and explain the phenomenon of specific refraction through the prism of the moral and sometimes ideological notions, values, and imperatives of the ingroup and its formal or informal leader.
3. Innovative attitudes in certain cultures are strongly influenced by prosocial emphases. In certain cases, this social narrative justifies not making more of an effort, and it is definitely dominant in softer cultures. Ultimately, the innovation potential may exist, but its concrete realization is more or less dependent on the interrelationships described above.
4. There is a threshold for the realization of innovative attitudes determined by the degree of stress tolerance. In principle, innovation is an object of real interest in cultures where leaving the comfort zone when undertaking innovative projects is not associated with an uncertain and hard-to-predict risk.

The second stage of the team's research is a review of the reports made and an analysis of the specifics and trends leading to the drawing of concrete conclusions that will be our basis in conceptualizing the problem of innovation potential and the corresponding organizational culture.

It becomes clear that, according to the majority of the interviewees, the following basic assumptions are formed:

– In principle, innovation potential exists – 27 of the interviewees confirm.
– The available innovation potential is relevant for the implementation of positive changes, intensifying and dynamizing the innovation capacity of the respective organization – 20 of the interviewees confirm, 7 believe that the innovation potential is not relevant for the implementation of changes, and 1 rejects the thesis of its existence altogether.
– The role of national cultural dimensions prevails, which consciously or unconsciously can change the right direction for the deployment of innovation potential – 17 confirm, 6 hesitate, and 5 do not accept this statement in principle.
– An important part of the good management of educational institutions is the competent compliance with national specifics – 18 confirm, 5 hesitate, and 5 do not accept this statement in principle.
– It will become more and more difficult to change the mentality and expectations of the young generations, in view of the incorrect assessment of the training methods, already from secondary education /and earlier/, which are a direct reflection of the family values, built as a function of the imposed social cultural characteristics and parameters – 18 confirm, 6 accept relatively, and 4 do not accept this statement.
– In Bulgaria, any idea of lobbying in a field such as education would be not only ineffective but also conflicting – 28 /the entire set of interviewed experts/ accept the statement.

## 5. Conclusion

As a result of the conducted research, located in a specific social anthropological discourse, the following conclusions stand out:

1. In the educational space, the microclimate is quite strongly influenced by the manifestation of national cultural dimensions, which applies equally to both the teaching staff and the students. The influence of the dimensions individualism / collectivism, power distance – egalitarian / non-egalitarian culture, uncertainty avoidance – stressed culture / relaxed culture, masculinity (hardness) / femininity (softness) of culture is significant.

2. The existence of innovation potential is extremely important for positive dynamics in educational processes, both in secondary and higher education.

Innovative features are perceived as essential aspects in educational strategies. The emphasis is on combining innovation potential with high standards. An essential focus is national and international collaboration in academia to realize the mission of outstanding commitment to knowledge-generating research oriented toward optimal decisions about science applications, without neglecting critical reflection on the effects of innovation. The idea of the innovative combination of scientific and practical knowledge is based on partnerships, where the specific social cultural environment and its requirements, attitudes, and abilities that led to certain conditions, constructs, and effects must be taken into account.

3. The research carried out in a social cultural discourse not only leads to the clarification of the reasons that led to a kind of blocking of the available innovation potential and, respectively, to the impossibility of its deployment, but also sends specific messages.

The specifics of the social cultural environment do not necessarily mean that it is negative or positive; rather, it is interpreted as compatible or incompatible with certain innovative proposals for which, for example, the potential turns out to be available, but insufficient innovation capacity is found.

When developing relevant policies and expecting a positive result, it is necessary to carefully assess the specific social cultural context and assume its essential role in order to avoid irrelevant declarativeness and ineffectiveness.

## References

Armstrong, M. (2004). *Narachnik za upravlenie na choveshkite resursi*. Burgaz: Delfin Press.

Becker, G. S. (1967). *Human Capital and the Personal Distribution of Income: An Analytical Approach*. Ann Arbor: Institute of Public Administration.

Danchev, A. (2015). Evolution of the Idea of Soft Social Infrastructure: The Latest Achievements. In: *Management and Sustainable Development* 2/2015 (51) http://oldweb.ltu.bg/jmsd/files/articles/51/51-01_A_Danchev.pdf

Feldman, M. (2004). *The Significance of Innovation*. Swedish Institute for Growth Policy Studies

Hofstede, G. (2001). *Kulturi i organizatsii. Software na uma*, 373p. Sofia: Klasika i Stil. https://www.hofstede-insights.com/country-comparison

Kolev, B. (2017). *Economic Culture,* 430p. Sofia: UNSS Publishing House (in Bulgarian)

Lopesa, A. (2016). Innovation Management: A Systematic Literature Analysis of the Innovation Management Evolution. *Brazilian Journal of Operations & Production, Management*, *13*( 1), 16–30

Milanova, A. (2008). Organizational Culture and Mobilization of Knowledge. In Chobanova, R. (Ed.), *Demand for Knowledge in the Process of European Economic Integration*, pp. 201–213. Sofia: Bulgarian Academy of Sciences.

Milanova, A., & Naydenova, P. (2013). *Human Capital Management in a Corporate Environment. Social-Anthropological Determination and Motivation,* 115p. Sofia: Academic Publishing House "Prof. Marin Drinov".

Milanova, A., & Naydenova, P. (2017). Corporate Social Capital – Reality and Pragmatic Definiteness. In *Improving the Competitiveness of Enterprises and National Economies,* pp. 155–177. Niš: University of Niš.

Milanova, A., & Naydenova, P. (2022). *Korporativniat sotsialen capital kato upravlensko predisvikatelstvo,* 204p. Sofia: Academic Publishing House "Prof. Marin Drinov",

Minkov, M. (2002). *Zashto sme razlichni?* Sofia: Klasika i Stil.

Naydenova, P. (2018). Motivation of Human Capital in the Business Organization. *Economic Studies*, *6*, 103–114

Panteleeva. (2013). *Upravlenuie na inovatsiite v industrialnoto predpriatie*. Svishtov: AI D. Tsenov.

Rogers, E. (2003). *Diffusion of Innovations*, 5th ed. New York: The Free Press.

Schultz, T. W. (1961). Investment in Human Capital. *American Economic Review*, *51*, 1–17.

Tripon and Blaga. (2011). Stimulation of the Innovative Potential in Online Lifelong Training of Human Resources. *Acta Marisiensis, Seria Technologica*, *8*(2), 262–267, Verlag: De Gruyter, Poland

Weber, M. (2009). *The Theory of Social and Economic Organization*. https://www.perlego.com/book/779024/

Livia Ilie

# The 5 I's for Academic Leadership in the 21st Century

## 1. Introduction

Universities are millennial institutions that have played a major role in the world through educating the citizens, creating knowledge, and driving changes at technological, economic, and social level. Nowadays they have to face tremendous challenges brought by digital transformation, political developments, and environmental, social, and economic major crises. These challenges have to become opportunities for universities to innovate both in teaching and research.

Over time, higher education institutions went through many academic revolutions and their role evolved, becoming more complex, adding new missions to their traditional ones. From the initial mission of preserving the knowledge and transmitting it to new generations, universities had to add the mission of generating knowledge through pure and applied research. Moreover, in the knowledge society, universities have to create knowledge and to put it at use becoming more relevant to the society through their economic and social impact. In the last decades, universities had to shift from their focus on teaching and research to a third mission, generally labeled as "contribution to society". As synthetized by Compagnucci and Spigarelli (2020), the third mission is reflected in already established concepts like technology transfer, entrepreneurial university, and triple helix model and is referring to a large variety of activities that help transfer knowledge to organizations and communities becoming real engines that contribute to economic, social, and cultural development; promote innovation and social welfare; and contribute to the education of highly skilled human capital for a knowledge-based economy, demonstrating both the efficient use of the public funding and its accountability. Universities had to abandon their ivory tower and become part of the general society, developing their strategies in dialogue with governments, industries, and society at large.

Even if the central responsibility in the current context remains the teaching – educating people to live and work in a highly technological world, and the research missions – Altbach (2008) argues about the complex role of universities: academic role (general education providing skills in logic, critical thinking, and writing, as well as specialized training for a large variety of professionals, ensuring access and equity for individuals); preservation and dissemination of knowledge (universities remain repositories and organizers of knowledge, and scientific heritage and provide free dissemination through libraries, courses, online solutions); intellectual centers (creative institutions with teachers and researchers that get involved through debates and analysis in the life of their communities); international institutions (through ICT and informal networks universities are involved in international exchange of ideas, data, and knowledge, and international research and exchange programs for students and staff); engines of economic development (direct economic impact through job creation, investment and expenditures made by staff and students, stimulation of tech-based industries in the local communities by providing skilled graduates and scientific innovation).

The European University Association (EUA) has published in February 2021 its vision for 2030 for "Universities without walls". EUA identifies a series of trends and challenges that is shaping the strategies of universities, out of which we mention:

– **Technological developments**. Universities are responsible for producing knowledge for new technologies and social innovation, and to transfer it and study the impact on our lives and on the society at large and at the same time to train graduates in order to be able to adapt to new labor market requirements. Universities themselves have to change the way they work due to digitalization and new technologies.
– **Underfunding of HEI**. Funds available for higher education and research differ across region. In a highly competitive market and increased economic crisis, universities struggle for resources and are forced to become more entrepreneurial in their approach.
– **Demographic changes and social disparities**. Universities have to adapt to cope with these challenges and design their study programs to address a more diverse student population. Key concerns become lifelong learning, access, equity, and inclusion.

We can add some other relevant challenges that universities are facing globally, being in a position to balance between contradictory trends, as synthetized by Mitchel (2010):

– Universities should keep a balance between the **need for mass education** – that comes from reasons of equity and economic efficiency, as a result of the fact that the knowledge-based society requires new skills and higher qualification of the workforce – and **the need for excellence in education** in the context of fierce competition between universities.

– The internationalization and globalization processes of the last decades led to both **increased cooperation** between universities and **fierce competition** between them in the battle to attract students and funds. An important aspect of the issue of internationalization is the implementation of the Lisbon Strategy and the Bologna Process, which gradually led to the creation of a European Area of Higher Education and to facilitating the mobility of students, teachers, and researchers within the European Union.

Based on their analysis, EUA envision for 2030 *universities without walls*: these means "open and engaged in society while retaining their core values…responsible, autonomous and free, with different institutional profiles, but united in their missions of learning and teaching, research, innovation and culture in service to society."

## 2. The 5 I's

In the current international context, we can no longer talk about education and research outside **interdisciplinarity, innovation,** and **internationalization**. These dimensions affect both fundamental missions of universities: teaching and research.

### 2.1. Interdisciplinarity

Teaching-learning can no longer be done on narrow disciplines, without connecting them with other disciplines. In the information age, students and graduates must be better equipped to find, understand, and use information. In the cognitive process, the higher hierarchical level has to be addressed: analysis, synthesis, evaluation, and even creation of new information. Rapid advancement of technology and artificial intelligence, increased complexity and uncertainty, brings a shift in the skills students need to master. The skills of generating the dots (that come from narrow specialization) are losing value in favor of skills of connecting the dots (mastered by generalists) (Mansharamani, 2020).

In addition, according to the European framework, universities must be able to develop competences for life, i.e. a multifunctional package of knowledge, skills, and attitudes necessary for personal fulfilment and development, social inclusion, active citizenship, and employability in the knowledge-based society. This implies a greater emphasis on transversal skills: communication in general (online communication in particular), teamwork, social responsibility, and ethics. Students must also be prepared to be able to react and innovate in a constantly changing world. It requires the development of new skills. The skills of anticipation, quick reaction, and communication in inter- and multicultural environments become essential (Ilie & Bondrea, 2016).

In their article, Lattuca et al. (2004) bring the arguments of different researchers in what concerns the role of interdisciplinarity in better preparing students for work and citizenship by developing high-order cognitive skills like critical thinking, problem-solving, and using of multiple perspectives in analyzing complex situations. Interdisciplinarity, by integrating disciplinary perspectives, provides students and researchers with opportunities to connect knowledge in order to solve complex problems and create innovative solutions in many fields. The challenges of the current complex environment that researchers have to address have multiple dimensions and implications so that an inter-, trans-, and multidisciplinary approach is needed.

In its extensive literature review on interdisciplinarity, Chettiparamb (2007) provides an overview of the concept and its teaching implications, highlighting the need to boost interdisciplinarity in both teaching and research. We retain here some points in favor of interdisciplinarity that are cited in the above-mentioned article from Nissani (1997): creativity requires interdisciplinary knowledge; many research topics as well as intellectual, social, and practical problems need interdisciplinary approaches; interdisciplinarity offers more flexibility in research and helps breach communication gaps by mobilizing its intellectual resources.

## 2.2. Innovation

Knowledge and **innovation** are key drivers of development. Universities must innovate both in the scientific fields in which they do research and also in the field of education through the development of innovative study programs and new teaching methods adapted to new generations. Innovation must manifest even in the administrative area through identification and implementation of flexible structures, digitization of processes, implementation of new methods of motivating and stimulating the performance of teaching-, research-, and administrative staff. In the current context, innovation requires interdisciplinary research.

Universities have to play a central role in the innovation ecosystem. The development of knowledge and skills has to be in partnership with all stakeholders. Universities will have to support innovation in education and research by promoting an entrepreneurial mindset (EUA, 2021).

Universities are not only responsible for technological innovation. The universities' third mission is mostly understood around knowledge and technology transfer. But universities also have to integrate social innovation in order to address the current societal challenges. Bayuo et al. (2020) published an extensive literature review on broadening the "scope of university engagement in social innovation and inclusion".

## 2.3.  Internationalization

Considering the technological developments and globalization that have allowed the free movement of information, talents, and capitals, universities cannot afford to be unconnected from international flows of information and knowledge; international networks of research and education; international flows of students, researchers, and teachers; and international financial grants for education and research. The **international dimension** of the university must be manifested both in the alignment of the contents and study programs to international standards, as well as the ability to participate in international projects and research teams. Learners and graduates have to be prepared to work in a globalized market place and to solve problems that have global concerns.

Internationalization is not a new strategy for universities; by contrary, it has characterized the higher education since its beginning. Still, due to globalization and internationalization and a higher demand for foreign languages and multi-cultural competences, internationalization in higher education received more focus in the last decades, and became an issue of institutional and public policy (Bergan, 2010). As mentioned by Bergan (2010), the current context of internationalization is different in many aspects. It is no longer about educating an elite and professionals for few domains. It is about mass-education and convergent competences. All sectors of the economy need higher education, not only in terms of training and re-training appropriate workforce but also for developing and transferring new knowledge that is needed for economic and social progress. Research cannot be conducted anymore locally, but internationally. Big global issues like climate change, human rights, economic, social and health crisis can be addressed only through international cooperation. To make the Earth a better place for future generation, higher education should educate professionals and citizens capable of addressing global issues, and this cannot be done unless HEI internationalize.

When designing internationalization strategies, universities have to take into consideration more than the mobility of staff and students, international research networks, knowledge exchange, and international teaching initiatives. Considerable attention has to be given also to what is called "internationalization at home". In her article, Robson (2017) is mentioning that there are real benefits of studying abroad, mainly the development of transversal skills that will increase students' employability. Still, more comprehensive internationalization strategies have to include internationalization at home in order to ensure that, on one hand, the gains of mobility experiences are maintained and further exploited; on the other hand, it is about creating equity opportunities for students that are not benefiting from an international mobility. In this context, important attention should be given by universities to internationalization of curriculum. The internationalization of curriculum refers both to content and teaching and assessment processes having in mind that universities are more and more accountable to produce employable graduates in a global economy. Nowadays, employers require skills and intercultural competences, a global mind-set necessary to live and work in fluid and interconnected settings. (Robson, 2017).

## 2.4. Infrastructure

In all their existence, universities were paying attention on continuously improving their teaching and research infrastructure. Well-equipped classrooms, libraries, and laboratories were priorities for their investments. In order to meet the challenges of the 21st century, universities have to further develop their **infrastructure**, both the physical and the digital one. In the context of ICT developments, mobility of talents, globalization, internationalization of education, and research, it is absolutely necessary to have the appropriate infrastructure to allow for both types of interaction.

The labor market is asking for new skills and ever-changing competences. In the information era, digital skills become compulsory. In addition, the rapid changes in all sectors and on all dimensions are imposing permanent upskilling and reskilling of the workforce in a lifelong learning approach. Universities have to ensure an appropriate digital infrastructure in order to deal with these challenges.

Moreover, COVID-19 pandemic and the newly energy crisis are making it even more obvious that universities have to develop a hybrid infrastructure providing to their staff, students, and partners both a physical (remains essential for dialogue and social interaction) and a virtual space (improves access to education and research, enhances cooperation and innovation). The COVID-19 pandemic put a lot of pressure on the HEI to invest in their digital infrastructure in order to develop distance and blended teaching and learning. As mentioned also in the EUA vision for 2030 (2021), the learning and research environments have to be "designed to accommodate different needs" for a diverse academic community and to "allow for flexible and blended approaches".

## 2.5. Impact

**Impact** is what defines the HEI of the 21st century. Globally, universities are under pressure in order to justify public investment. It can be considered a matter of ethics: what contribution do universities make to the well-being of society who invests in them? In this context, it is very important to emphasize the impact that universities are producing in society through education and research. Both teaching activity, by educating and professionally training of students, and research-innovation activity must have an impact in society by developing technologies and increasing the competitiveness of the economies. The relationship with the labor market is the example of the immediate impact, which takes the form of providing the employable specialists for all sectors. It is not, however, the only form of impact, because universities are not a mere supplier for the various needs of third parties. A university is training people of culture and art, consciences actively engaged in promoting a climate of peace, with an attitude that is environment friendly, and in creating a better and more beautiful world for all.

Over the centuries, universities were essential for the economic and social progress. In their extensive study (data analyzed from 15000 universities in 78 countries), Valero and Van Reenen (2019) have identified several channels through which higher education institutions have an impact on growth. (1) *Greater supply of human capital*: universities are producers of human capital (measured in years of schooling). Human capital is important for growth and development. The skilled workforce is more productive than the unskilled one. In consequence, the communities and regions where universities were developed registered increased GDP per capita. (2) *Innovation*: universities have a direct but also an indirect impact on the innovative capacity of the region. Researchers produce innovations that is transferred to the local organizations, or innovation is produced in cooperation with local businesses. Moreover, the skilled graduates will be in the position to innovate for their organizations, not only in terms of technology but also in developing innovative solutions and implementing better managerial practices. (3) *Support for democratic values*: universities provide a platform for dialogue and debates of ideas being drivers of democracy, producing a human capital that can strengthen and improve institutions. (4) *Demand effects*: universities create jobs, invest in their teaching and research infrastructure, and boost tax revenues through increased consumption of goods and services in the local community with a direct impact on the GDP.

As poles of learning and knowledge, universities are essential for the sustainability of the socio-economic competitiveness of regions and countries.

Each euro invested in education and research brings an important return on investment and has a multiplication effect in the economic, employment, and innovation dynamics (Saúde et al., 2015).

Siegfried et al. (2007) mention studies that demonstrate the impact of universities in the form of innovation and technology transfer, better quality of life in the region and improved public service, and contribution to local culture. But higher education should not only look for economic competitiveness and efficiency, but it should also address the public values of global social movements, as well as react to the challenges of cultural divisions and conflicts. Moreover it should not abandon the intrinsic motivations of academics and students in basic research and the transmission of universal and democratic values (Serrano-Velarde, 2010).

Still, universities deserve support not only for their economic benefits. Stronger societies become those where universities are not only engines for the knowledge society, but they also aim to fulfil humanistic and cultural goals of individuals and communities (Altbach, 2008).

## 3. Conclusions

The ability of universities to adapt to ever-changing demands depends on, but it is not just about increasing their autonomy and adequate funding. The universities need to develop their institutional capacity to define their strategies and differentiate themselves in a very competitive market. Academic excellence requires both appropriate governance able to balance between academic and administrative functions and strong leadership to facilitate coherence and unity between the various academic departments in order to respond to the need of interdisciplinarity in both education and research. In addition, innovation requires inter- and multidisciplinary research. Although universities tend to have a conservative approach, a strategic approach in this regard must consider adapting the curriculum and research themes to this trend. Such developments also require university reorganizations and a new type of relationship between various departments and scholars.

*Interdisciplinarity*, *innovation,* and *internationalization* of HEI impose important investment in appropriate *infrastructure* and in reorganizing the internal processes. Moreover, it becomes necessary for the continuous investment in people, teachers, and researchers, in order to be able to address the challenges and implement the changes. In the current context, very dynamic and with an open international competition, universities must become more entrepreneurial. The entrepreneurial approach is not questioning the academic role and purpose of universities, their fundamental mission. It just reinforces the idea that universities are living, functioning organizations in a continuously changing environment and in order to survive and, moreover, to be the elite organizations, trailblazers, they must be able to rethink themselves in order to have an ***impact*** in the society at large.

## References

Altbach, P. (2008). Chapter the Complex Roles of Universities in the Period of Globalization. In *Higher Education in the World 3: Higher Education: New Challenges and Emerging Roles for Human and Social Development* (GUNI Series on the Social Commitment of Universities), 3rd ed., pp. 5–14. Palgrave Macmillan.

Bayuo, B. B., Chaminade, C., & Göransson, B. (2020). Unpacking the Role of Universities in the Emergence, Development and Impact of Social Innovation – A Systematic Review of the Literature. *Technological Forecasting & Social Change*, *155*, 120030.

Bergan, S. (2010). Internationalisation of Higher Education: A Perspective Beyond Economics. In S. Bergan & R. Damian (Eds.), *Higher Education for Modern Societies – Competences and Values,* pp. 57–74. Council of Europe Publishing.

Chettiparamb, A. (2007). Interdisciplinarity: A Literature Review. *The Interdisciplinary Teaching and Learning Group*, Subject Centre for Languages, Linguistics and Area Studies, School of Humanities, University of Southampton.

Compagnucci, L., & Spigarelli, F. (2020). The Third Mission of the University: A Systematic Literature Review on Potentials and Constraints. *Technological Forecasting & Social Change, 161*, 120284.

EUA. (2021). *Universities Without Walls. A Vision for 2030*. https://www.eua.eu/downloads/publications/universities%20without%20walls%20%20a%20vision%20for%202030.pdf (accessed on September 2022).

Ilie, L., & Bondrea, I. (2016). Developing Students' Competences for the Future – Key Priority in the Economics of Universities. *International Conference Knowledge-Based Organization Volume*, *22*, 201–205.

Lattuca, L. R., Voigt, L. J. & Fath, K. Q. (2004). Does Interdisciplinarity Promote Learning? Theoretical Support and Researchable Questions. *The Review of Higher Education*, *28*(1), 23–48.

Mansharamani, V. (2020). https://www.cnbc.com/2020/06/15/harvard-yale-researcher-future-success-is-not-a-specific-skill-its-a-type-of-thinking.html (accessed on October 2022).

Michel, A. Main Challenges Facing the Future of Higher Education in Curaj, A. (coord.) (2010). *The For-Uni Blueprint A Blueprint for Organizing Foresight in Universities,* pp. 13–20. Bucharest: The Publishing House of the Romanian Academy.

Robson, S. (2017). Internationalization at Home: Internationalizing the University Experience of Staff and Students. *Educação*, Porto Alegre, *40*(3), 368–374

Saúde, S., Borralho, C., Feria, I., & Lopes, S. (2015). The Impact of Higher Education on Socioeconomic and Development Dynamics: Lessons from Six Study Cases. *Investigaciones de Economía de la Educación*, *10*, 887–905.

Serrano-Velarde, K. (2010). New Challenges to Higher Education: Managing the Complexities of a Globalized Society. In S. Bergan, & R. Damian (Eds.), *Higher Education for Modern Societies – Competences and Values*, pp. 39–48. Zaragoza: Council of Europe Publishing.

Siegfried, J. J., Sanderson, A. R., & McHenry, P. (2007). The Economic Impact of Colleges and Universities. *Economics of Education Review*, *26*, 546–558

Valero, A., & Van Reenen, J. (2019). The Economic Impact of Universities: Evidence from Across the Globe. *Economics of Education Review*, *68*, 53–67

Isabela-Anda Dragomir* and Brândușa-Oana Niculescu**

# Tradition and Transgression in English Language Teaching

## 1. Introduction

With the number of non-native or second language speakers of English outnumbering those of primary or native speakers, there is no doubt that English is the global lingua franca, being a vehicular language that connects people with different cultural and language backgrounds. Especially against the backdrop of globalization, defined as "economic, political, cultural, linguistic and environmental interconnections and flows that make many of the currently existing borders and boundaries irrelevant" (Steger, 2020, p. 10), the spread of English is viewed as a positive trend, a strong promoter of social networking, economic integration, education, technology, and modernity.

The teaching of English in a multicultural world has been a matter of continuous change and adaptation to the challenges of the current environment (economic, political, security). The constant search for resources, increased migration due to conflicts and wars in various parts of the world, and the amplified mobility of the educational and labor market are just a few of the underlying conditions that support and trigger continuous reinterpretations of ELT methodology, making it a domain that is invariably under "construction". Researchers and practitioners have been travelling back and forth between a wide array of methods and techniques of English language teaching, often starting from traditional approaches that are recycled and reinvented, while innovating and creating new avenues for skill development and language proficiency.

One of the most recent approaches would be the "alienation" from the traditional methods that emphasize grammatical, lexical, and phonetic forms, in favor of considering English a "contact language", used as a means of communication between people speaking different native languages (Firth, 2009). In this novel context, a characteristic feature of teaching English is the greater stress placed on function, which translates in using English as a means of intercultural and interlingual communication. The ultimate goal of learning the language is that EFL speakers be able to express their thoughts, effectively using the available forms and functions, regardless of the L1 influences.

## 2.  Tradition in Foreign Language Teaching and Learning

Traditional approaches in the teaching and learning of English as a foreign language can be best summarized in Almeida Filho's words: "In order to learn, students resort to conventional ways of learning typical of their region, ethnicity, social class, since tradition informs, in a naturalized, subconscious, and implicit manner, the way through which a new language should be learned" (1993, p. 13, as cited in Michelan de Azevedo and Lopes Piris, 2018, p. 419). In the relation between students and learning, resorting to traditional methods is a matter of socio-historical inheritance. Learning occurs based on a set of historically constructed rules that govern the manner in which knowledge is produced and controls the way in which learners relate to this knowledge. In foreign language teaching and learning, tradition is viewed as a transmission of customs from generation to generation.

Most traditional approaches are based on the learning of lists of words for comparisons between L1 and the target language, aimed at expanding vocabulary. Also, a variety of grammar exercises in context are used to consolidate syntactic and morphological knowledge about the language. Traditionally, students are advised to read texts aloud, memorize them, and recite various passages, translate texts, paraphrase, or answer comprehension questions. Methods such as Grammar Translation, Audio-lingual, and Community Language Learning pivot on experiencing language in use with focus on grammar and vocabulary in context. However, in most cases, this context is divorced from the multicultural reality that often exists in real-life educational environments nowadays.

With linguistic diversity on the rise around the world, English language instruction now happens in spaces that are increasingly becoming multilingual and which entail the addition of different cultures to the educational repertoire of students and teachers alike. The latter category is now expected to add novel pedagogical approaches to their teaching, by which diverse linguistic practices and cultural experiences are recognized and exploited as valuable resources (Bonnet &Siemund, 2018).

As Cenoz and Gorter point out in their article *Teaching English through pedagogical translanguaging* (2020), new approaches to English language teaching presuppose a shift from monolingual ideologies, by which teaching a language based on the isolation of the target culture is no longer an applicable course of action. The learners in this new paradigm are no longer monolingual speakers. They are characterized by multilingualism, and are typified by a new type of competence, which is far more complex and more qualitative (Cook, 1992). What makes their specific case even more interesting is that multilinguals, as opposed to the traditional monolinguals, have a different social and cultural experience, which definitely affects their foreign language learning strategies. Multilingual learners need to accommodate a wide array of elements that have a direct impact on their learning. They navigate between different types of discourses, interact in various kinds of communicative situations, on different topics, and often with different audiences than in the native language. Consequently, boundaries between languages become fluid and the "native" speaker's traditionally hegemonic profile is no longer the measuring standard of competence.

If traditionally, the main objective of foreign language teaching was to develop communication skills in the target language; more modern pedagogies advocate the development of a vast range of competences that supplement the linguistic ones: social, pragmatic, discursive, and cultural. Against this backdrop, traditional methods are no longer fully efficient and need to be complemented with "out-of-the-box" teaching and learning strategies and techniques that are better suited to accommodate the diverse linguistic and cultural profile of the students of today and especially of those of the future.

## 3. Transgressive Pedagogies in English Language Teaching

In the current context, typified by constant change and an increased need for re-adaptation, the most likely to succeed are those who are willing to attempt disruption, leaving behind the old ways and embracing new, creative, innovative approaches and following a path of subversion – replacing traditional practices with modern ones. The field of foreign language teaching and learning makes no exception from this tendency. More recently, teachers have been put in positions that challenge them out of their comfort zone, by prompting them to take risks and adapt to the current needs of their students. These courageous teachers, genuine "catalysts" for learning, inspiring individuals who leave deep meaningful footprints in the minds and souls of the learners, are celebrated by bell hooks in her book *Teaching to Transgress* (1994). She makes a point in praising those approaches that go "against and beyond boundaries", challenging tradition and confronting the constraints of mandated curricula, in favor of valuing the cultural experience of learning itself.

Against this backdrop, Culturally Responsive Teaching (CRT), defined as a way of "using the cultural knowledge, prior experiences, frames of reference, and performance styles of ethnically diverse students to make learning encounters more relevant to and effective for them" (Gay, 2000, p. 29), makes the most informed approach of choice when it comes to English language teaching and learning in a multicultural world. By embracing CRT, teachers are able to move their students through a process of culturally relevant understandings and engagement. Even with restricted subject-matter (a specific course in English, such as a Writing course, or ESP, for example, Military English), the novelty of CRT is that emphasis is placed on the process of learning the language, over the content, which becomes a mere scaffolding framework that sustains the building and development of skills. With teaching a foreign language, it so often happens that we might have to teach content that is not necessarily relevant for the students. The alternative is to supplement already-existing scripted syllabi and standardized curricula with content that becomes significant for the cultural identity of the students. In such a context, teachers assume the role of socio-cultural mediators between student individuality and academic spaces (Nieto, 2017). CRT can be successfully employed to connect students with themselves, with the teachers, and with the outside world, thus creating a learning community that recognizes who students are individually, how they interact with the real world, what learning styles they employ, and what prior knowledge they possess.

Starting from Prada and Nikula's (2018) discussion of transgression under the umbrella of translanguaging, that incorporates three secondary/additional "trans-" conceptualizations (transcending, transformative, and transdisciplinary), we argue that employing CRT in English language teaching mitigates all possible discrepancies that navigating between two (or more) different languages might generate.

The concept of *translanguaging,* which harnesses a great transformative power for the individual and for their own learning, can be defined as "multiple discursive practices in which bilinguals engage in order to make sense of their bilingual worlds" (García, 2009, p. 45). The approach draws on the utility of a linguistic repertoire, that incorporates all the linguistic resources available to the members of a particular community, translated as "every bit of language we accumulate" during social and cultural experiences (Blommaert &Backus, 2013, p. 28). In the context of CRT, the meaning and importance of these language "bits" are dictated by the relevance they have for the learner, and by their usefulness in developing abilities and skills required in different communicative occurrences. The ultimate goal of language learning is to use language to gain knowledge and be able to articulate one's thoughts and ideas; thus, language becomes a vehicle through which thinking is expressed and transformed. The process of translanguaging also presupposes transcendence, an iterative journey between different linguistic structures and systems, for the comprehension and use of which a new pedagogical strategy is needed. Transgression, in this context, boils down to operationalizing linguistic elements at the intersection of the learners' own cultural repertoires and, at the same time, opens new avenues through which students explore their own experiences that promote a positive attitude toward multilingualism. Given the fact that translanguaging connects cognition, linguistic experience, and social relations in a holistic manner, the dimension of transdisciplinarity is also an essential part of any discussion of transgression in the process of language learning.

Especially relevant for bi-/multilingual groups of learners, CRT aims to incorporate a more systematic and fluid understanding of the target language and the socio-historical context attached to it. Starting from the premise that languages do not operate in a vacuum, modern English language learning perspectives encourage teachers to see the classroom as an area populated by the students who occupy it, and who bring into this space their social, cultural, and linguistic practices and, ultimately, their own agenda.

In practical terms, culturally responsive or relevant teaching can be defined as a pedagogy that recognizes the importance of including students' cultural references in the process of learning (Muñiz, 2019). As opposed to traditional teaching, which focuses on teacher–student dynamics and standardized content, CRT harnesses the power of the learners' individual and cultural experiences that reflect the current social context (Burnham, 2020, July 31). As an example, while traditional teaching methods may cultivate classical knowledge of the arts, science, and literature (learning of the English language by studying renowned authors such as Shakespeare, Dickens, Salinger, or Poe), CRT aims to include contextual examinations of various authors, from different parts of the world, that illustrate an assorted repertoire of experimental learning opportunities belonging to a multicultural community that is representative for today's society.

By successfully integrating CRT into classroom instruction, teachers obtain important benefits for education: they strengthen the students' sense of identity, promote equality and inclusiveness in the classroom, actively engage students in the course material, and support critical thinking. Most CRT pedagogies pivot on a number of strategies that are directed at fostering powerful connections between content and the process of learning, root-tapping into the students' cultural background and prior experiences.

First and foremost, learning should be contextualized. Topics in the curriculum must be tied to the social communities the students belong to. A lesson in British or American culture and civilization needs relevant reflections in the present; a vocabulary list of words related to the topic of military equipment is more likely to be internalized if, for instance, it is addressed specifically to a group of infantry cadets that will use that equipment as part of their future training or missions.

Culturally responsive teaching also pivots on the activation of the students' prior knowledge. Since learners enter the classroom with diverse experiences, they should be encouraged to draw on their previous knowledge and share it with the community they are part of. One example would be teaching the art of argumentation which, in addition to observing a specific discursive format, heavily draws on the pertinence of the arguments themselves and the manner in which they are supported throughout the discourse. Expressing a point of view is undoubtedly one illustrative example of cultural differences, as some cultures tend to be more opinionated than others, which reflects in the manner learners construct an opinion and defend it. The role of CRT in this context would be to mitigate the different (culturally-biased) perspectives on a particular issue (e.g. slavery, gender issues) through the use of inclusive and tolerant language.

In the same vein, it is extremely important to support students to leverage their cultural capital. Especially with multicultural mixed classrooms, it is often the case that some students may not have a voice. Activating on these students' cultural capital, making them feel understood and respected, and encouraging them to express themselves even if they represent a less powerful culture are some of the mechanisms through which CRT promotes inclusion, equality, and tolerance, and fights the damaging effects of the systemic discrimination students might otherwise experience outside the classroom.

CRT goes beyond mere instruction, which is often an abstract, immaterial concept. Cultural responsiveness is also illustrated in the physical classroom setup: the choice of literature in the classroom library, the posters being displayed on walls, the materials on bulletin boards, the arrangement of students, the position of the teacher. These may seem small alterations to the traditional representation of a classroom environment, but, in the long run, they can transform a typical classroom into a more culturally responsive location.

Building relationships between students, between students and teachers, and between the school and the outside community is one of the long-term goals of education. Diversity can be bridged through a common cultural understanding, and one of the most powerful instruments to achieve this goal is the language. With English, in particular, being the lingua franca of the world, it is only natural to assume that language teaching through CRT unequivocally meets this objective.

By and large, the success of English language teaching and learning that fosters the application of culturally responsive pedagogies depends on a number of conditions that encapsulate the essence of the strategies discussed above: establish inclusion, develop positive attitudes, enhance meaning, and foster self-confidence.

## 4. Conclusion

Transgression in English language teaching is represented by any novel approach that provides a framework for the application of new methods and strategies of language instruction. The transformatory action being called into action here pivots on a shifting emphasis from the traditional monolingual education to the emergence of a new concept – that of multilingualism. Adopting transgressive pedagogies, for instance, culturally responsive teaching, in the English language classroom, represents an opportunity to align language instruction to the demands and challenges of the future. Supporting language learning and, more importantly, language awareness translates directly into building positive attitudes toward different and diverse cultures and promotes inclusion, tolerance, and equality through diversity. One of the main benefits of adopting CRT as a method of English language teaching is that the student – and implicitly what they know and how they know it – becomes the foundation of teaching interactions and curriculum. CRT recognizes and values the potential of culturallyrelated knowledge and skills as a resource that enhances development and proficiency. CRT focuses on positive intra- and inter-personal relationships, creating communities where diversity is embraced and harnessed into a vector of change and progress. Consequently, it becomes a validating mechanism that incorporates students' diverse knowledge and practice into the curriculum. CRT's comprehensive umbrella integrates the social, cultural, and historical dimensions of the individual, of the group, and of the community. Such a teaching strategy is also empowering and transformative, in that it changes the way in which learners perceive themselves and gives them the chance to access their own potential, building self-confidence, a sense of belonging, and, ultimately, life-enduring skills.

The future is definitely one of multiculturalism. In a world in which borders are constantly moving, and concepts such as nationality, race, gender, and sex are becoming increasingly fluid, we will all need to feel unique and integrated at the same time. Culturally sustaining pedagogies will represent the way ahead, reflecting the goal of not only responding to but also constantly nourishing linguistic and cultural diversity in education. We remain confident that the recipe of success will integrate a dual-track approach, in which teachers will support their students' native linguistic and cultural abilities while simultaneously offering access to the competences that are practiced and valued in the target/dominant culture.

## References

Blommaert, J., & Backus, A. (2013). Superdiverse Repertoires and the Individual. In I. Saint-Georges & J.-J. Weber (Eds.), *Multilingualism and Multimodality*, pp. 11–31. Rotterdam, The Netherlands: Sense Publishers.

Bonnet, A., & Siemund, P. (2018). Multilingualism and Foreign Language Education: A Synthesis of Linguistic and Educational Findings. In A. Bonnet, & P. Siemund (Eds.), *Foreign Language Education in Multilingual Classrooms*, pp. 1–29. Amsterdam, the Netherlands: John Benjamins Publishing Company.

Burnham, C. (2020, July 31). 5 Culturally Responsive Teaching Strategies. *Northeastern University*. Available at: https://www.northeastern.edu/graduate/blog/culturally-responsive-teaching-strategies/(Accessed on August 2022).

Cenoz, J., & Gorter, D. (2020). Teaching English through Pedagogical Translanguaging. *World Englishes and Translanguaging*, *39*(2), 300–311. Available at: https://onlinelibrary.wiley.com/doi/epdf/10.1111/weng.12462 (Accessed on July 2022).

Cook, V. J. (1992). Evidence for Multicompetence. *Language Learning*, *42*(4), 557–591.

Firth, A. (2009). The Lingua Franca Factor. *Intercultural Pragmatics, 6*(2), 147–170 (Accessed on July 2022).

García, O. (2009). Chapter 8 Education, Multilingualism and Translanguaging in the 21st Century. In T. Skutnabb-Kangas, R. Phillipson, A. Mohanty & M. Panda (Eds.), *Social Justice through Multilingual Education*, pp. 140–158. Bristol, Blue Ridge Summit: Multilingual Matters.

Gay, G. (2000). *Culturally Responsive Teaching: Theory, Research, and Practice*. New York: Teachers College Press.

Hooks, B. (1994). *Teaching to Transgress: Education as the Practice of Freedom*. New York, NY: Routledge. Available at: https://sites.utexas.edu/lsjcs/files/2018/02/Teaching-to-Transcend.pdf (Accessed on August 2022).

Muñiz, J. (2019). Culturally Responsive Teaching. A 50-State Survey of Teaching Standards. *New America*. Available at: https://www.newamerica.org/education-policy/reports/culturally-responsive-teaching/ (Accessed on August 2022).

Nieto, S. (2017). *Language, Culture, and Teaching: Critical Perspectives*, 3rd ed. New York: Routlege.

Prada, J., & Nikula, T. (2018). Introduction to the Special Issue: On the Transgressive Nature of Translanguaging Pedagogies. *EuroAmerican Journal of Applied Linguistics and Languages, Special Issue*, *5*(2), 1–7.

Steger, M. B. (2020). *Globalization: A Very Short Introduction*, 5th ed. Oxford: Oxford University Press.

Daniela Antonescu, Victor Platon, Ioana Cristina Florescu,
Andreea Constantinescu, and Florina Popa

# The Impact of Digitalization on Regional GDP: The Case of Romania

## 1. Introduction

Digital transformation helps to increase the quality of life and makes services more efficient so that the economy could function well. For digital technology to remain relevant, it must create added value. One relevant aspect for today's impact of digitalization is the network of transmission of stored information. Thus, in the latest years and with the help of data transmission networks, the computer field has experienced a constant / fulminant development at the global level. This chapter aims to analyze the relationship between regional GDP and a number of indicators specific to digitalization using panel regression models.

In order to achieve the proposed objectives, the literature dealing with digitization was consulted and specific data was created/processed at the level of the eight development regions from official statistical sources (Eurostat, 2022). Data processing was conducted on the basis of econometric modelling, which provides the opportunity to highlight how selected indicators can contribute to regional development.

## 2. Literature Review

Quantifying the influence of digital technologies on economic development is currently one of the major concerns of scientific studies and analysis. Following the literature review, several studies of interest to the subject were analyzed and presented below.

In the article of Hardy A.P. "*The role of the telephone in economic development*", the relationship between the telephone services and economic development has been analyzed. In this chapter, the author investigates the role of the telephone as a contributing agent to the economic development. The time series for 60 countries over a period of more than 13 years was used to quantify how the phone contributed to economic development. The analysis of dynamics and cross-correlation techniques showed that the phone contributed to economic development (Hardy, 1980).

Another relevant paper in terms of using modelling to analyze the influence of telecommunications on economic development is "*Economic impacts of mobile versus fixed broadband*" conducted by Thompson and Garbacz (2011). They used the stochastic border method in order to model the relationship between these two indicators. The article builds the stochastic frontier based on the Cobb-Douglas production function, obtaining results related to the efficiency of using the digital potential at the level of the analyzed countries (Thompson et al., 2011).

In 2022, Harald Edquist published the article entitled "*The economic impact of mobile broadband speed*". Based on the panel analysis of data from 116 countries, between 2014 and 2019, the paper investigates the association between broadband internet speed and labor productivity. The authors identified a significant and robust relationship when a one-year gap is introduced for the series defining mobile broadband infrastructure. The interpretation of the results shows that a 10 % increase in mobile broadband in the period *t-1* is associated with a 0.2 % increase in labor productivity in the period *t* (Edquist, 2022).

Some studies have looked at the impact of fixed/mobile broadband on economic development. Qiang and Rossotto (2009) found an association between fixed broadband access and the growth of GDP per capita in both developed and developing countries. Czernich et al. (2011) estimated that a 10 % increase in fixed broadband penetration results in an annual GDP per capita growth of 0.9–1.5 percentage points (Qiang et al., 2009).

## 3. Analysis of the Indicators Used

### 3.1. Regional Framework for Analysis

In Romania there are eight development regions (1. North-East, 2. South-East, 3. South Wallachia, 4. South-West Oltenia, 5. West, 6. North-West, 7. Center, and 8. Bucharest-Ilfov) as planning territorial subunits (Figure 1).

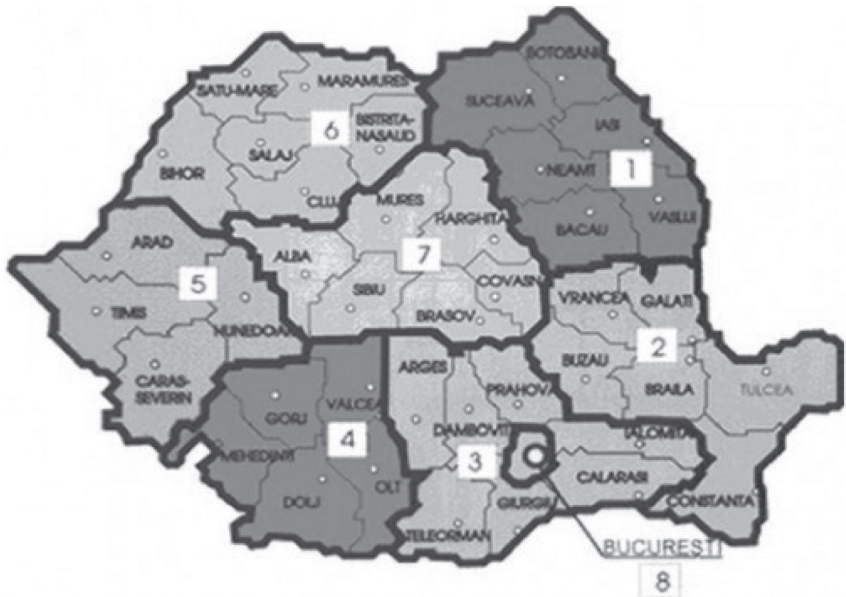

**Figure 1.** The Romanian Economic Development Regions

**Source:** Regional Development Plan of ADR Muntenia, https://www.adrmuntenia.ro/dezvoltare-regionala/static/2

The regions are no administrative units. Each of these regions has an Agency for Regional Development and its own Regional Development Plan.

The panel regression analysis will include data from the eight regions for the period 2010–2021. Gross domestic product will be the dependent variable, and the two independent regressors will be the broadband internet infrastructure and online commerce.

## 3.2. Gross Domestic Product, at Regional Level

In Romania, at regional level, the average value of GDP per capita in 2020 was 21,937.5 euros/inhab., 44.44 % higher than in 2014, but down from 2019 by -1.07 % (Graph 1). The analysis by regions shows a maximum value of GDP per capita in the Bucharest-Ilfov region of 49,200 euro/inhabitant, and a minimum in the North-East region of 13,600 euros/inhabitant. In evolution, the largest increase was recorded by the South Muntenia region (+58.5 %), followed by North-West (+52.7 %) and North-East (+49.5 %).

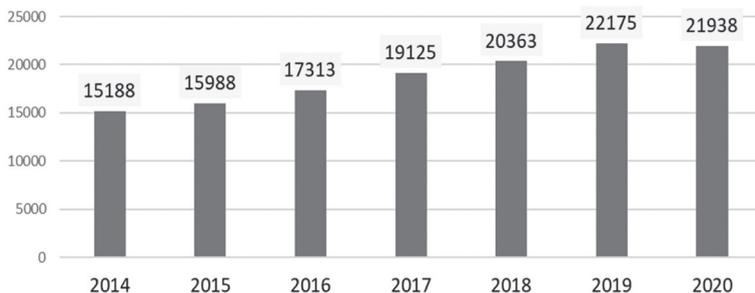

**Graph 1.** Regional GDP per capita (average) in Romania, 2014–2020
**Source:** Own computations with data from Eurostat, 2022.

### 3.3. Broadband Infrastructure in Romania

Relevant data in a study developed by ANCOM (National Authority for Management and Regulation in Communications) shows that in Romania there were 5.7 million fixed internet connections at the end of 2020 (+8 % annual evolution), of which 4.5 million (80 %) represented very high-speed connections (over 100 Mbps) (ANCOM, 2021).

On average, 7 out of 10 households have a fixed internet connection, with 8/10 being urban households and 6/10 rural households. This gap decreased during 2020, with the growth rate of the number of connections in rural areas (+ 16 %) being well above that in urban areas (+4 %) (EU DESI, 2022).

Between 2010 and 2021, over three quarters of households in Romania (78.2 %) had access to their home internet network. In urban areas, 84.8 % of households are connected to the Internet, and in rural areas only 69.7 % of the households.

At regional level, internet connection was more widespread among households in the Bucharest-Ilfov region (78.5 % of the population), followed by the Regions West (67.8 %) and North-West (67 %). The regions that represent the average broadband internet connection are: South-West (61 %), Center (60.8 %), and South (60.3 %). The smallest shares are in South-East (59.8 %) and north-east (55.8 %) regions.

| TIME | 2010 | 2015 | 2019 | 2020 | 2021 |
|------|------|------|------|------|------|
| **North-West** | 28 | 70 | 85 | 89 | 90 |
| **Center** | 23 | 65 | 80 | 82 | 90 |
| **North-East** | 17 | 57 | 77 | 77 | 87 |
| **South-East** | 23 | 57 | 77 | 79 | 84 |
| **South - Muntenia** | 23 | 61 | 79 | 82 | 86 |
| **Bucharest - Ilfov** | 33 | 80 | 91 | 92 | 94 |
| **South-West Oltenia** | 15 | 62 | 83 | 82 | 86 |
| **West** | 22 | 75 | 87 | 89 | 90 |

**Table 1.** Regional Broadband infrastructure in Romania, 2010–2021 (% in total housing)

**Source:** Own computations based on the data taken from Eurostat, 2022.

The connection types used to access the internet from home are 77.5 % in favor of fixed broadband connections (fixed broadband connections), followed by mobile broadband connections (66.3 %) and narrowband connections (13.6 %).

Households with broadband access are shown in the following table as a share of total households (Table 1). It can be seen that all regions have increased the value of this indicator, with a tendency to reduce territorial inequalities.

### 3.4. E-Commerce, at Regional Level

In 2020, Romania recorded a significant growth rate of e-commerce, even if only 45 % of Romanian Internet users bought online. The e-commerce market reached the value of 5.6 billion euros and marked an annual growth of 30 %, in the context of the pandemic and the previous modest performance of this sector, compared to the situation of e-commerce in other countries (Eurostat, 2022). The e-commerce market in Romania was estimated at 6.2 billion euros in 2021, with local sales accounting for half of the total achieved in Eastern Europe (Eurostat, 2022).

| Time | 2010 | 2015 | 2020 | 2021 |
|------|------|------|------|------|
| **North-West** | 2 | 14 | 40 | 37 |
| **Center** | 5 | 9 | 39 | 43 |
| **North-East** | 3 | 9 | 32 | 33 |
| **South-East** | 3 | 7 | 30 | 31 |
| **South  Muntenia** | 2 | 11 | 32 | 33 |
| **Bucharest - Ilfov** | 8 | 19 | 56 | 58 |
| **South-West Oltenia** | 3 | 9 | 36 | 35 |
| **West** | 4 | 8 | 39 | 41 |

**Table 2.** E-commerce in Romania's regions during 2010–2021 (%)
**Source:** Own computations based on the data taken from Eurostat, 2022.

The analysis of the e-commerce indicator by development regions in Romania, in 2021, shows that the first three places are occupied by the most developed regions, namely: Bucharest-Ilfov with 58 %, Center with 43 %, West with 41 %, and North-West with 37 % (Table 2).

## 4. Data and Methodology

The analysis is based on two independent variables – broadband infrastructure and online commerce – and is carried out at the level of development regions (NUTS 2) in Romania. The aim is to estimate the impact of broadband infrastructure on GDP per capita using regional data for the period 2010–2021.

The methodology presents the following phases:

1. Selection of indicators: GDP per capita at regional level (€/inhab.), the share of households using broadband infrastructure in total households, at regional level (%); individuals who have shopped online in total population, at regional level (%).
2. the data at the level of the eight development regions in Romania were taken from the Eurostat database for the period 2010–2021;
3. Modelling the GDP/capita – Broadband, E-commerce dependence using multiple panel regression (8 regions, over a period of 11 years) according to the equation:

$$\text{Yit} = C + a_1 {}^* \text{Xit} + a_2 {}^* \text{Yit} + \varepsilon_{it} \qquad (\text{Eq. 1})$$

4. Comparison of OLS model, model with fixed effects and model with random effects using AIC, $R^2$, MRSE, etc.; Hausman test.
5. Selecting the optimal model and discussing the results.
6. Analysis of regional heterogeneity.
7. Discussions, conclusions.

Modelling dependence between variables, based on econometric equations, provides the opportunity to highlight how the two indicators of the digital economy contribute to regional development.

## 4.1. Modelling the Relationship between GDP Per Capita Indicators and Those of Internet Infrastructure and Online Commerce, at Regional Level, in Romania

Modelling the dependence of GDP – Internet infrastructure and on-line Trade was achieved with the help of the Panel regression, in which GDP per capita is dependent variable and online commerce and broadband infrastructure are independent variables. The panel analysis is based on data published by Eurostat for the eight development regions of Romania, in the period 2012–2021.

The model presented below is a particular form, adapted to the purpose of this article.

## 4.2. The OLS Model (Ordinary Least Squares Model)

Following the application of the OLS method, the validity of the regression model is indicated by the (F-statistics) probability of 0.0000. This means that the model is valid and has all coefficients different from zero.

The probability related to the online commerce indicator is below the 5 % threshold, and that of broadband infrastructure exceeds this percentage.

According to the $R^2$ value, the influence of endogenous variables on exogenous variables is only 42 % and the probability attached to the broadband indicator is 0.2421 % (Table 3).

The equation for the OLS model is next:

**GDP_PER_CAPITA = 64.00942\*BROADBAND+ 377.7075**
$$\textbf{*E-COMMERCE + 6485.240 + } \boldsymbol{\varepsilon_i.} \qquad \text{(Eq. 2)}$$

For the above reasons, we believe that the model does not explain the relationship between variables well, and we will explore two other models: random effect panel and fixed effect panel.

### 4.3. The Random Effect Model

We will test if the panel model with random effects is appropriate. To test the model, we use the Housman Test with $H_0$ hypothesis: *The model with RE is appropriate*, using Chi-Sq statistics (Table 4).

It can be observed that the probability of the Chi-Sq value of 46.8740 is void. Under these conditions we reject $H_0$ and accept H1: The fixed effect model is appropriate. We will look at this model next.

### 4.4. Fixed Effects Model

According to the Hausman test, the model with Fixed Effects is suitable in our case. The estimates identify a new relationship between the selected variables (Table 5).

Broadband internet infrastructure and online commerce, as in the previous case, have a significant positive effect on GDP per capita. In the case of Online commerce, it can be seen that an increase of 1 % will lead to an increase in GDP per capita of 19,276 €/inhabitant. Broad-band Internet infrastructure also has a significantly positive relationship with GDP per capita; the influence being positive and of 3,904 €/inhab. at an increase of 1 % (Table 6).

All estimators are statistically significant at significance levels of 0.1, 0.05, and 0.01.

The overall fit of the model with FE is very good, with independent variables explaining a significant part of the variance of the dependent variable, as shown by the corrected $R^2$ (0.96) (Table 6).

The final equation of the model with FE is as follows:

**GDP_PER_CAPITA = 39.0437564071\*BROADBAND + 192.762233534\* E-COMMERCE + 11048.9379175 + [CX=F]**     (Eq. 3)

The data in Annex 5 show that, in the case of the model used, the probabilities related to the broadband, online commerce, and C coefficients are 0.0000, which shows that they are significant for the chosen significance thresholds of 5 %.

The value of the coefficients (Eq. 3) thus shows that with the increase of online trade by 1 %, the GDP per capita will increase by 19,276 €.

The broadband Internet infrastructure also has a significantly positive relationship with GDP per capita (39.04 €/inhab.) (Eq. 3). The high value of $R^2$ (0.96), shows a very strong influence of the two endogenous variables on the dependent variable.

It can be concluded that R-squared is not equal to zero, and the correlation between the dependent variable and the dependent variables is statistically significant. The probability corresponding to F-statistic is 0.0000 in all three models (Table 7), which indicates the validity of the regression models. We therefore reject the null hypothesis that all regression coefficients are equal to 0.

## 4.5. Heterogeneity at Regional Level

The fixed effect model allows the identification of regional heterogeneity, thus defining the individual regression equations for the eight regions.

By calculating the longitudinal effects, it is possible to calculate the intercept for the eight development regions concerned (Table 8) taking into account that the slope is the same.

Thus, as per the calculations, according to the constant term, Bucharest-Ilfov region ranks first. The intercept is 29,594 €/inhab., the highest value, because the Bucharest-Ilfov region is the most developed region of the country.

West region is in the second place, the fastest growing region between 2000 and 2007. Along with the Bucharest-Ilfov region, the West region has the fastest growth in the entire decade (EU funds, 2013), with a constant term of 11,459 €/inhab. In the third place is the Center region, the third region of the country in terms of level of development, which is also reflected in the constant term, 9727 €/inhab. This is followed by the North-West regions with 8690 €/inhab., South-East with 8830 €/inhab., South Muntenia with 7859 €/inhab., South-West with 7013 €/inhab., and last, North-East with 5218 €/inhab.

According to the general equation (Eq. 3), the following slopes were obtained: BROADBAND=39.04376 and E-COMMERCE=192.7622.

The general equations for the eight regions are derived from Eq. 3:

$$\text{GDP\_PER\_CAPITA (N-V)} = 8.690 + 39.04376*\text{BROADBAND} + 192.7622*\text{E-COMMERCE}$$

$$\text{GDP\_PER\_CAPITA (B-Ilf)} = 29.594 + 39.04376* \text{BROADBAND} + 192.7622* \text{E-COMMERCE}$$

$$\text{GDP\_PER\_CAPITA (N-E)} = 5.218 + 39.04376* \text{BROADBAND} + 192.7622* \text{E-COMMERCE}$$

$$\text{GDP\_PER\_CAPITA (C)} = 9.727 + 39.04376* \text{BROADBAND} + 192.7622* \text{E-COMMERCE}$$

$$GDP\_PER\_CAPITA \text{ (S-E)} = 8.830 + 39.04376^* \text{ BROADBAND} + 192.7622^* \text{ E-COMMERCE}$$

$$GDP\_PER\_CAPITA \text{ (S)} = 7.859 + 39.04376^* \text{ BROADBAND} + 192.7622^* \text{ E-COMMERCE}$$

$$GDP\_PER\_CAPITA \text{ (S-V)} = 7.013 + 39.04376^* \text{BROADBAND} + 192.7622^* \text{ E-COMMERCE}$$

$$GDP\_PER\_CAPITA \text{ (V)} = 11.459 + 39.04376^* \text{BROADBAND} + 192.7622^* \text{ E-COMMERCE}$$

The regional equations are useful to analyze each of the eight regions separately and make forecasts.

In the general equation (Eq. 3), if the dependent variables were null, then it would result in a value of GDP per capita of 11,048.94[1]€. In the case of Bucharest-Ilfov region, under the same conditions, a value of GDP/capita of 29,594€ would be obtained, the lowest value being that of the North-East region, where a value of 5,218€/inhab. would be recorded (Graph 2).

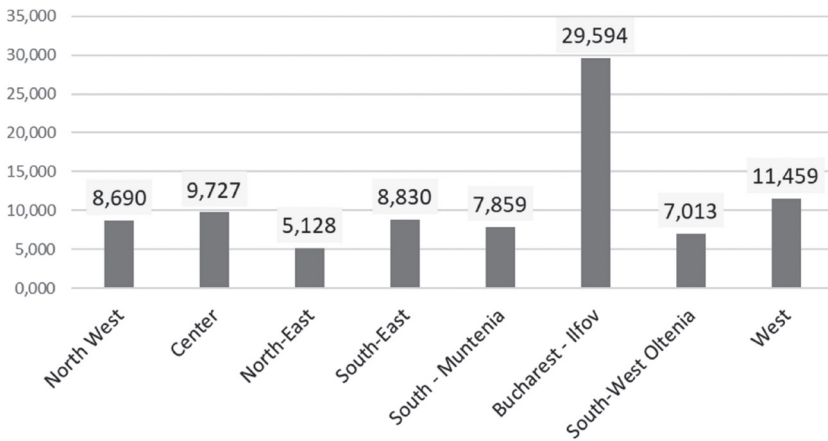

**Graph 2.** The intercept of the regional regression equations
**Source:** Own computations based on data taken from Eurostat, 2022.

---

1    There would be no broadband infrastructure and no online commerce.

If we take into account the value of the constant term in Eq. 3 (**11048.94 €/person**) which represents an average value of the eight regions, we note that the Bucharest-Ilfov region has a value of 2.6 times higher. Above the average is only the West region with a value of 1.03 times higher (approximately equal to the average). The rest of the regions are below average.

## 5. Conclusions

Digitalization is an essential element for the current structure of society, given the COVID-19 pandemic and geo-political challenges.

The digital analysis in Romania shows that, in 2010–2021, over three quarters of households (78.2 %) had access to the internet network from home; in urban areas, 84.8 % of households are connected to the Internet, and in rural areas only 69.7 %. In territorial aspect, internet connection was more widespread among households in the Bucharest-Ilfov region (78.5 % of the population), followed by the West (67.8 %) and North-West (67 %) regions. The regions representing the average broadband internet connection are South West (61 %), Center (60.8 %) and South (60.3 %).

The e-commerce market in Romania was estimated at 6.2 billion euros in 2021, with local sales accounting for half of the total achieved in Eastern Europe. The regional analysis of the average values of e-commerce, during 2010–2021, shows that the first three places are occupied by the most developed regions, namely: Bucharest-Ilfov with 26.5 %, North-West with 17 %, and Center with 16.4 %. They are joined by the West region with an average of 14.9 %.

As regards the econometric model used to highlight the relationship between the three proposed indicators (GDP, broad-band indicators and online trade), it demonstrates a conclusive picture of the indicators' influences on regional development. The resulting equations show the direct and positive link between broadband infrastructure and e-commerce with GDP per capita, proving that any increase among independent variables will lead to an increase among dependent ones.

From the analysis of the panel regression model, it turned out that the Fixed Effects model is the most suitable to illustrate the dependence of regional GDP on the two regressors (internet infrastructure and online commerce). The values of the coefficients in the fixed effects model equation are statistically significant and have a positive influence on GDP. For example, with the growth of online trade by 1 %, the GDP per capita will increase by 19,276 euros. In the case of broadband Internet infrastructure, a significantly positive relationship with GDP per capita can also be observed. The high value of $R^2$ shows that the two endogenous variables explain 96 % of the variation of the dependent variable.

The Fixed Effect model allows the identification of regional heterogeneity, thus defining the individual regression equations for the eight regions.

By calculating the longitudinal effects, the constant terms for the eight development regions considered can be calculated by taking into account that the slope is the same for each region.

From the general equation, independent equations can be deducted, specific to the eight regions considered. With the help of the eight resulting equations, the regional GDP can be forecasted according to the regressors considered (broadband internet infrastructure and online trade).

If we take into account the value of the intercept in Eq. 3 (**11048.94**), which represents the average value of the eight regions, we note that the Bucharest-Ilfov region has a value of 2.6 times higher. Above the average is only the West region with a value of 1.03 times higher (approximately equal to the average). The rest of the regions are below average.

From a regional approach, digitalization can have an important impact on the level of territorial development and on the reduction of economic and social inequalities.

## References

Czernich N., Falck O., Kretschmer T., Woessmann L., 2011, Broadband Infrastructure and Economic Growthm, The Economic Journal, Vol. 121, Issue 552 https://doi.org/10.1111/j.1468-0297.2011.02420.x

Edquist, H. (2022). *The Economic Impact of Mobile Broadband Speed*. https://www.sciencedirect.com/science/article/pii/S0308596122000532

EU Funds. (2013). https://ec.europa.eu/regional_policy/en/policy/evaluations/ec/2007-2013/

EU Funds EU DESI. (2022). https://digital-strategy.ec.europa.eu/en/policies/desi

EU Funds Eurostat. (2022). *Households that Have Broadband Access by NUTS 2 Regions*. https://ec.europa.eu/eurostat/databrowser/view/tgs00048/default/table?lang=en

EU Funds Regional Development Plan of ADR Sud – Muntenia. https://www.adrmuntenia.ro/dezvoltare-regionala/static/2, 2007.

Hardy, A. P. (1980, December). The Role of the Telephone in Economic Development. *Telecommunications Policy*, *4*(4), 278–286. https://www.sciencedirect.com/science/article/abs/pii/0308596180900440

Qiang C., & Rossotto C. (2009). *Economic Impacts of Broadband, World Bank Information and Communications for Development 2009: Extending Reach and Increasing Impact.*

Thompson, H. G., & Garbacz, C. (2011). Economic Impacts of Mobile Versus Fixed Broadband. *Telecommunications Policy*, *35*, 999–1009.

ANCOM. (2021) *Traficul de internet a crescut cu peste 50% in 2020, ancom.ro*

# Appendix

**Table 3.** OLS model

| Dependent Variable: GDP_PER_CAPITA | | | | |
|---|---|---|---|---|
| Method: Panel Least Squares | | | | |
| Sample: 2010 2021 | | | | |
| Periods included: 12 | | | | |
| Cross-sections included: 8 | | | | |
| Total Panel (balanced) observations: 96 | | | | |
| Variable | Coefficient | Std. Error | t-Statistic | Prob. |
| BROADBAND | 64.00942 | 54.36765 | 1.177344 | 0.2421 |
| E-COMMERCE | 377.7075 | 91.94749 | 4.107860 | 0.0001 |
| C | 6485.240 | 2586.557 | 2.507287 | 0.0139 |
| R-squared | 0.421127 | Mean dependent var | | 16637.50 |
| Adjusted R-squared | 0.408678 | S.D. dependent var | | 9066.633 |
| S.E. of regression | 6972.013 | Akaike info criterion | | 20.56795 |
| Sum squared resid | 4.52E+09 | Schwarz criterion | | 20.64808 |
| Log likelihood | -984.2615 | Hannan-Quinn criter. | | 20.60034 |
| F-statistic | 33.82844 | Durbin-Watson stat | | 0.066978 |
| Prob(F-statistic) | 0.000000 | | | |

**Source:** Own computations in EViews (Eurostat data, 2022)

**Table 4.** EGLS panel method with random effects

| Dependent Variable: GDP_PER_CAPITA | | | | |
|---|---|---|---|---|
| Method: Panel EGLS (Cross-section random effects) | | | | |
| Sample: 2010 2021 | | | | |
| Periods included: 12 | | | | |
| Cross-sections included: 8 | | | | |
| Total panel (balanced) observations: 96 | | | | |
| Swamy and Arora estimator of component variances | | | | |
| Variable | Coefficient | Std. Error | t-Statistic | Prob. |
| BROADBAND | 32.56717 | 12.88289 | 2.527939 | 0.0132 |
| E-COMMERCE | 247.1348 | 21.96992 | 11.24878 | 0.0000 |
| C | 10589.90 | 1145.267 | 9.246671 | 0.0000 |
| | Effects Specification | | | |
| | | | S.D. | Rho |
| Cross-section random | | | 2730.607 | 0.7420 |
| Idiosyncratic random | | | 1610.254 | 0.2580 |
| | Weighted Statistics | | | |
| R-squared | 0.761754 | Mean dependent var | | 2792.089 |
| Adjusted R-squared | 0.756631 | S.D. dependent var | | 3974.305 |
| S.E. of regression | 1960.624 | Sum squared resid | | 3.57E+08 |
| F-statistic | 148.6765 | Durbin-Watson stat | | 0.330481 |
| Prob(F-statistic) | 0.000000 | | | |
| | Unweighted Statistics | | | |
| R-squared | 0.361823 | Mean dependent var | | 16637.50 |
| Sum squared resid | 4.98E+09 | Durbin-Watson stat | | 0.023706 |

**Source:** Own computations in EViews (Eurostat data, 2022)

**Table 5.** Hausman test for the model with RE

| Correlated Random Effects - Hausman Test | | | | |
|---|---|---|---|---|
| Equation: EQ_OLS_RANDOM_EFFECTS | | | | |
| Test cross-section random effects | | | | |
| Test Summary | | Chi-Sq. Statistic | Chi-Sq. d.f. | Prob. |
| Cross-section random | | 46.874074 | 2 | 0.0000 |
| Cross-section random effects test comparisons: | | | | |
| | | | | |
| Variable | Fixed | Random | Var(Diff.) | Prob. |
| BROADBAND | 31.558049 | 32.567165 | 0.253945 | 0.0452 |
| E-COMMERCE | 242.832628 | 247.134798 | 0.999435 | 0.0000 |

**Source:** Own computations in EViews (Eurostat data, 2022)

**Table 6.** Fixed effects EGLS Panel Model

| Dependent Variable: GDP_PER_CAPITA | | | | |
|---|---|---|---|---|
| Method: Panel EGLS (Cross-section weights) | | | | |
| Sample: 2010 2021 | | | | |
| Periods included: 12 | | | | |
| Cross-sections included: 8 | | | | |
| Total panel (balanced) observations: 96 | | | | |
| Linear estimation after one-step weighting matrix | | | | |
| Variable | Coefficient | Std. Error | t-Statistic | Prob. |
| BROADBAND | 39.04376 | 6.896786 | 5.661152 | 0.0000 |
| E-COMMERCE | 192.7622 | 12.52349 | 15.39205 | 0.0000 |
| C | 11048.94 | 312.6908 | 35.33503 | 0.0000 |
| | Effects Specification | | | |
| Cross-section fixed (dummy variables) | | | | |
| | Weighted Statistics | | | |
| R-squared | 0.964452 | Mean dependent var | | 23603.01 |
| Adjusted R-squared | 0.960732 | S.D. dependent var | | 7741.394 |
| S.E. of regression | 1389.480 | Sum squared resid | | 1.66E+08 |
| F-statistic | 259.2516 | Durbin-Watson stat | | 0.944994 |
| Prob(F-statistic) | 0.000000 | | | |
| | Unweighted Statistics | | | |
| R-squared | 0.968575 | Mean dependent var | | 16637.50 |
| Sum squared resid | 2.45E+08 | Durbin-Watson stat | | 0.353874 |

**Source:** Own computations in EViews (Eurostat data, 2022)

**Table 7.** Comparative analysis between the three models (OLS, fixed effects, and random effects)

| Indicator/Coefficient | Model OLS | Model EA | Model EF |
|---|---|---|---|
| BROADBAND | 64.00942 | 32.56717 | 39.04376 |
| (the associated probability) | (0.2421) | (0.0000) | (0.0000) |
| E-COMMERCE | 377.7075 | 247.1348 | 192.7622 |
| (the associated probability) | (0.0001) | (0.0000) | (0.0000) |
| C | 6485.240 | 10589.90 | 11048.94 |
| (associated probability) | (0.0139) | (0.0000) | (0.0000) |
| R-squared | 0.421127 | 0.761754 | 0.964452 |
| Adjusted R-squared | 0.421127 | 0.756631 | 0.960732 |
| S.E. of regression | 6972.013 | 1960.624 | 1389.480 |
| F-statistic | 33.82844 | 148.6765 | 259.2516 |
| Prob(F-statistic) | 0.000000 | 0.000000 | 0.000000 |

**Source:** Own computations in EViews (Eurostat data, 2022)